\documentclass[useAMS,amssymb,epsfig,usegraphicx,usenatbib,twocolumn]{emulateapj}

\usepackage[usenames,dvips]{color}
\usepackage{epsfig}

\newif\ifAMStwofonts

\makeatother

\shorttitle{Growth of bulges in disc galaxies}
\shortauthors{S. Sachdeva, K. Saha and H.P. Singh}

\begin{document}

\title{Growth of bulges in disc galaxies since $z\sim1$}
\author{Sonali Sachdeva and Kanak Saha}
\affil{Inter-University Centre for Astronomy and Astrophysics, Pune 411007, India}

\and 

\author{Harinder P. Singh}
\affil{Department of Physics and Astrophysics, University of Delhi, Delhi 110007, India}

\begin{abstract}
We investigate the growth of bulges in bright ($M_B<-20$) disc galaxies since $z\sim1$, in rest-frame {\it B} and {\it I}-band, using images from HST ACS and WFC3 in GOODS-South for high redshifts ($0.4<z<1.0$) and SDSS for local ($0.02<z<0.05$). The growth history has been traced by performing two-component bulge-disc decomposition and further classifying the bulges into pseudos and classicals using Kormendy relation. We have about $27$\% pseudo and $40$\% classical bulges in our sample. Classical bulges are brighter than pseudo, in both rest-bands, at all redshifts probed here; in fact since $z\sim 0.77$, classical are about $\sim 1$~mag brighter than pseudo bulges. Both bulges have witnessed substantial growth, more than half of their present day stellar mass has been gained since $z\sim1$. Their host discs have grown concurrently, becoming progressively brighter in rest-frame {\it I}-band.

The high redshift host discs of both pseudo and classical bulges are found to be equally clumpy in rest-frame {\it B}-band. In the same band, we found that the growth of classical bulges is accompanied by fading of their host discs - which might be an indication of secular processes in action. However, both host disc as well as the bulge have grown substantially in terms of stellar mass. Our analysis suggests that, clump migration and secular processes alone can not account for the bulge growth, since $z\sim1$, accretion and minor mergers would be required. 
\end{abstract}



\section{Introduction}
\label{sec:intro}

Bulges are lately known to be structurally related to the processes witnessed by their host disc galaxies. The primary process understood to be the cause of formation of bulges is major mergers, where, violently changing gravitational potential transforms newly formed discs to ellipsoids \citep{BarnesandHernquist1992} and host discs assemble around it from the surviving matter \citep{BarnesandHernquist1996,Bournaudetal2005,SpringelandHernquist2005,Naabetal2006,Hopkinsetal2009}. Evidence suggests that due to diminishing frequency, this process probably has not played a significant role in bulge growth for the past half age of the Universe \citep{Linetal2004,LopezSanjuanetal2009,Bridgeetal2010,Blucketal2009,Blucketal2012,Kaviraj2014}.
During this time, bulge growth is discerned to have been facilitated by multiple minor mergers and continuous matter accretion that displaces stars from their disc orbits \citep{SteinmetzandNavarro2002,Parryetal2009,Hopkinsetal2010,Oseretal2010}. This is supported by the observation of a large number of satellite fractions and simulations \citep{VelazquezandWhite1999,NaabandBurkert2003,Bournaudetal2007a,Youngeretal2007,Jogeeetal2009}. In addition, disc galaxies are expected to internally grow a bulge with time as they evolve to reach a more stable configuration \citep{BinneyandTremaine1987,Athanassoula2005,Combes2008,Kormendy2008,Sellwood2014}. This process of internal secular evolution attempts to explain the presence of bulges with disc like properties \citep{KormendyandKennicutt2004,Peletieretal2007,FisherandDrory2008,Gadotti2009}. Another bulge formation process that has lately gained evidence through observations and simulations is clump coalescence \citep{Bournaudetal2007b,Genzeletal2011,ForsterSchreiberetal2011}, where, massive clumps formed in volatile high redshift discs fall towards the centre, also facilitating inflow of cold gas and causing star-bursts, leading to bulge formation\citep{Elmegreen2007,DekelandKrumholz2013,Ceverinoetal2015}.

Unraveling the relative role of these bulge growth mechanisms, i.e, understanding which of these mechanisms was dominant at which stage in the formation/growth of a particular bulge type, is imperative to comprehend present-day disc galaxies which differ to a large extent in terms of the bulges they host. Empirically, useful insight can be obtained by tracking the growth of bulges inside disc galaxies as a function of redshift. Although extensive studies have been carried out in terms of benchmarking structural and kinematic properties of the bulges across optical-IR wavelength in the local Universe \citep{Gadotti2009,FisherandDrory2010,Simardetal2011,LacknerandGunn2012,Kelvinetal2012,Meertetal2015}, comprehensive studies exploiting multi-band photometry in the high redshift Universe remain inadequate. Using CANDELS data, \citet{Bruceetal2014} and \citet{Langetal2014} decomposed massive galaxies in {\it H}-band (infrared), from $1<z<3$ and $0.5<z<2.5$, respectively. In their decomposition, the bulge was already fixed to have a de-Vaucouleur's profile (S\'ersic index $n$ fixed at 4). As a consequence, their focus was to analyze the affect of change in the dominance of this component on morphology distribution and fraction of passive/quenched galaxies with redshift. \citet{MargalefBentaboletal2016} do allow the bulge component to be fitted freely, however, since their study is also based at very high redshifts ($1<z<3$), it is limited to analyzing the evolving fraction of one and two-component galaxies with time. 

In the intermediate redshift range, $0.1 < z < 1.3$, \citet{DominguezPalmeroandBalcells2009} examined a sample of prominent bulge galaxies in GOODS-N using HST/ACS data. Since their aim was to study the relation of bulge colours with central surface brightness, they extracted these two parameters from the central aperture. \citet{Tascaetal2014} using HST/ACS data, studied the broad division of rest-frame {\it B}-band light into bulges (including ellipticals) and discs, for 3266 galaxies at $0.7<z<0.9$, and compared it with the local distribution. Apparently, a thorough investigation of the evolution of bulge and disc properties in the redshift range $0 < z <1$, accounting for the light emitted by both young and older stellar populations (as well as dust), i.e., simultaneously in optical and infrared, is lacking. This ($0<z<1$) marks an important phase in the evolutionary history of disc galaxies. Before $z\sim1$, i.e., in the redshift range $1 < z <1.5$, disc galaxies have formed and matured to an extent that they resemble 'Hubble types' with noticeable bulges\citep{Delgado-Serranoetal2010,Buitragoetal2013,Mortlocketal2013,HuertasCompanyetal2016,MargalefBentaboletal2016}.
Thereafter, i.e, since $z=1$, the present-day disc galaxies have gained more than half of their stellar mass and size \citep{Trujilloetal2011,Marchesinietal2014,Ownsworthetal2014}. Thus, the period from $z=1$ to $z=0$, covering more than half the age of the Universe, is crucial in terms of tracing bulge growth inside disc galaxies.

In this paper, we undertake this examination, starting from $z=1$ to the present epoch, both in rest-frame {\it B} (0.37 - 0.49 $\mu$m) and {\it I}-band (0.78 - 0.85 $\mu$m), using HST ACS and 3D-HST data of Chandra Deep Field South (CDF-S), at the intermediate redshift range ($1.0 - 0.4$) and a similar sample at the local range ($0.02 - 0.05$) from SDSS. We perform 2-component bulge-disc decomposition in both rest-frame optical and infrared band, to derive bulge properties which are further classified into classicals and pseudos following Kormendy relation, as suggested by \citet{Gadotti2009}. In the rest of the paper, we investigate the physical properties of these bulges as seen in optical and infrared as a function of redshift and comment on the possible physical mechanisms that might have caused the diversity in the appearance of the bulges.
 
The paper is organized as follows: in Section~\ref{sec:data} we describe the data sample, 
decomposition process and bulge classification; in Section~\ref{sec:bulgegrowth}, we present the growth history of both types of bulges in disc galaxies; in Section~\ref{sec:luminosity} we compare the luminosity distribution of the bulges with their host discs. Section~\ref{sec:discuss} presents a discussion on issues related to bulge growth. Throughout the paper, we consider a flat $\Lambda$-dominated CDM universe with $\Omega_{\Lambda}$=0.73, $\Omega_m$=0.27, H$_o$=71 km sec$^{-1}$ Mpc$^{-1}$. All magnitudes are quoted in the AB system and stellar masses are based on \citet{Chabrier2003} IMF.


\section{Data and analysis}
\label{sec:data}
Our galaxy images are from three sources catering to the two redshift ranges, 0.4-1.0 and 0.02-0.05, probed in this study. First source is GOODS survey in CDFS done using HST-ACS\footnote{Based on observations obtained with the NASA/ESA HST, which is operated by the Association of Universities for Research in Astronomy, Inc.(AURA) under NASA contract NAS 5-26555.} \citep{Giavaliscoetal2004}. The images are taken from it's {\it V} (F606W), {\it i} (F775W) and {\it z} (F850LP) filters, for galaxy-sources having redshifts 0.4-0.6, 0.6-0.8 and 0.8-1.0 respectively, to obtain their rest-frame {\it B}-band images. In \citet{Sachdeva2013} we performed single S\'ersic component fitting on all the detected galaxies and used the derived global S\'ersic index ($n_g<2.5$) to separate a magnitude limited ($M_B<-20$) disc dominated sample. The completeness of this sample and the selection cuts made to ensure that are described in detail in that work. 

The second source is CANDELS and 3D-HST Treasury Program\footnote{Based on observations taken by the 3D-HST Treasury Program (GO 12177 and 12328) with the NASA/ESA HST} which employs {\it HST}'s Wide Field Camera 3 (WFC3) \citep{Groginetal2011,Koekemoeretal2011,Skeltonetal2014}. We match the RA Dec of our optical sample with those provided in the 3D-HST photometric catalog \citep{Skeltonetal2014} to a maximum distance of 0.5 arc-seconds which strikes a balance such that maximum number of sources are matched and there is no case of double matching. Then according to the x and y centroids provided in the catalog, we take cutouts from the {\it J} (F125W), {\it JH} (F140W) and {\it H} (F160W) filter images, for galaxy-sources having redshifts 0.4-0.6, 0.6-0.8 and 0.8-1.0 respectively, to obtain their rest-frame {\it I}-band images. 

The third source is NASA Sloan Atlas\footnote{http://www.nsatlas.org. Funding for the NASA-Sloan Atlas has been provided by the NASA Astrophysics Data Analysis Program (08-ADP08-0072) and the NSF (AST-1211644).} \citep{Blantonetal2011} compiled using SDSS DR-8 \citep{Eisensteinetal2011}, for redshift range 0.02-0.05. The images are taken from its {\it g} and {\it z} filters, as the range corresponds best to rest-frame {\it B} and {\it I}-band respectively. The selection of the sources in this local sample was done using the NYU-VAGC (New York University-Value Added Galaxy Catalog catalog) \citep{Blantonetal2005a}, which has a specially prepared catalog for low redshift sources \citep{Blantonetal2005b}. Selection process for this local sample is detailed in \citet{Sachdeva2013}.

Thus, we obtain rest-frame {\it B} and {\it I}-band images of disc dominated sample of 570 galaxies in the intermediate (0.4-1.0) and local (0.02-0.05) redshift-ranges. For images in the intermediate redshift range, 10 arcsec cutout is extracted and for images in the local range, 3 arcmin cutout is extracted, which ensures that an average galaxy size covers $\sim$60\% of the total area. The cutouts are final in the sense of flat-fielding, bias subtraction, cosmic ray removal etc. An important concern is to decontaminate, i.e, removing the neighbouring sources around the galaxy. The neighbouring sources in each cutout are recognized through SExtractor's \citep{BertinandArnouts1996} segmentation map. Decontamination is achieved using IRAF (Image Reduction and Analysis Facility) {\it imedit} task, which creates a circular annulus of a chosen radius surrounding the selected neighbouring source and replaces the pixel values of that source with average value of the pixels inside the annulus, detailed in \citet{Sachdevaetal2015}. After finalizing the cutouts, we fit elliptical isophotes at successively increasing radii using the {\it ellipse} task of {\it IRAF} and obtain the radial intensity profile for each galaxy. For 5 galaxies, no comprehensible profile was obtained, i.e, the task did not converge to provide meaningful isophotal values, which occurs when there is degeneracy of bright pixels around the centre. Since, it was not viable to fit any parametric functions to such profiles, they were removed from the analysis. 

\begin{figure*}
{\includegraphics[width=18.0cm]{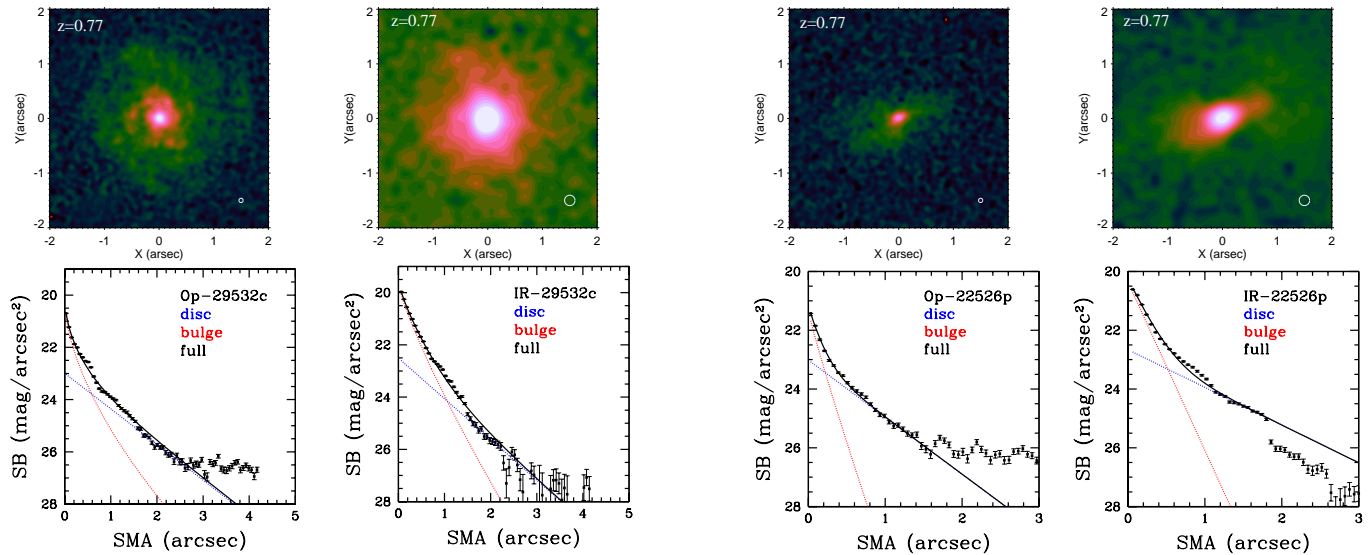}}
\caption{The rest-frame optical ({\it B}-band) and infrared ({\it I}-band) images and their corresponding SB profiles, for two galaxies (ID:27512 and 23183) is shown. The profiles have been decomposed into bulge (red dotted line) and disc (blue dotted line) parts. The total intensity obtained by adding the two functions is marked as "full" and shown as solid black line. The white circles on the images depict the PSF FWHM (circle's diameter), for optical images it is 0.077" and for infrared it is 0.19".}
\label{fig:profiles}
\end{figure*}

\subsection{1D bulge-disc decomposition}
\label{sec:decomposition}

Using the obtained radial intensity profiles, in both rest-frame {\it B} and {\it I}-band, of our sample of 565 galaxies, we perform one dimensional bulge-disc decomposition on each profile, individually, employing a S\'ersic bulge and an exponential disc. This is accomplished through the following steps. First, we visually inspect the full intensity profile of each galaxy, and select the part where intensity falls linearly on log-normal plot. On this selected portion of the profile, we fit exponential disc profile given by,

\begin{equation} 
I_d(R) = I_0 \exp(-R/R_d),
\end{equation}

where $I_0$ and $R_d$ are the central intensity and scale length of the disc. After fitting the disc profile, we fit the full function (i.e., disc + S\'ersic) to the full intensity profile. Of course, now the disc profile is held fixed by inserting the values of $I_0$ and $R_d$. The S\'ersic function, thus, only fits the extra light of the disc galaxy. The full profile is,

\begin{equation}
I(R) = I_0 \exp(-R/R_d) + I_{eb} \exp[-b((R/R_e)^{1/n_b}-1)],
\end{equation}

where $n_b$ is the bulge S\'ersic index, $R_e$ is it's effective radius and $I_{eb}$ is the intensity at that radius. We, thus, obtain the disc and bulge parameters, $I_o$, $R_d$, $I_{eb}$, $R_e$, $n_b$, of our disc dominated sample. In Fig.~\ref{fig:profiles}\footnote{The galaxy images have been made using Cubehelix color palette, written by James R. A. Davenport.}, decomposition of rest-frame {\it B} and {\it I}-band profiles of two galaxies is shown along with their images in the respective bands. Note, that all the central light, which is in excess to the underlying disc component, has been regarded as bulge light in this work and fitted with the free S\'ersic component to obtain bulge parameters. The bars, if any, are included in this bulge light and not dealt with separately.

For each profile, it is visually ensured that the decomposition is sensible and the total intensity function superimposes the profile. Generating PSFs from the Tiny-Tim software \citep{Krist1995} and also, directly from the images, we performed PSF deconvolution on the obtained profiles and found that only the central most isophotal value differs in the convolved and de-convolved profiles. The difference of the convolved and de-convolved isophotal values, is susceptible to the ellipticity, position-angle, amongst other placement features of the galaxy \citep{Trujilloetal2001,Ciambur2016}, accounting for which is as yet a non-robust exercise. In our study, we fit the functions (exponential and S\'ersic) individually through visual inspection. We find that the central most isophotal value does not affect the fitting and thereby, the resultant parameters, within the quoted error ranges, for any of the studied profiles.

Of the 565 galaxies, 139 galaxies are well fitted by only the exponential disc function, i.e, do not have any substantial bulge component, in rest-frame {\it B}-band. We have studied the evolution of these pure disc galaxies in \citet{SachdevaandSaha2016}. Our final sample, thus, consists of 426 galaxies with substantial photometric bulges (bulge-to-total ratio $B/T>0.1$) at all redshifts, explored in this paper (361 for $0.4<z<1.0$ and 65 for $0.02<z<0.05$).

The total-light of the galaxy is obtained by computing the area under the full profile till Petrosian radius (Petrosian radius for the full sample was computed in \citet{Sachdevaetal2015}) and bulge light is obtained by integrating the derived bulge profile till the same radius. Then using redshifts, latest cosmological parameters, magnitude zero-points and K-corrections \citep{GrahamandDriver2005,Sachdeva2013}, we obtain intrinsic bulge parameters such as absolute magnitude ($M_b$), effective radius ($R_{e}$), S\'ersic-index ($n_b$), surface brightness at effective radius ($SB_{eb}$) and average surface brightness ($<SB_{eb}>$), in rest-frame {\it B} and {\it I}-band. We obtained total stellar masses of all the galaxies in our intermediate redshift ($0.4 < z < 1.0$) sample from the grism spectra data of 3D-HST survey \citep{Brammeretal2012,Momchevaetal2016} and for galaxies in our local sample from NASA-Sloan Atlas catalog \citep{Blantonetal2011}. In the following, we discuss the classification of these photometric bulges.

\begin{figure*}
\mbox{\includegraphics[width=55mm]{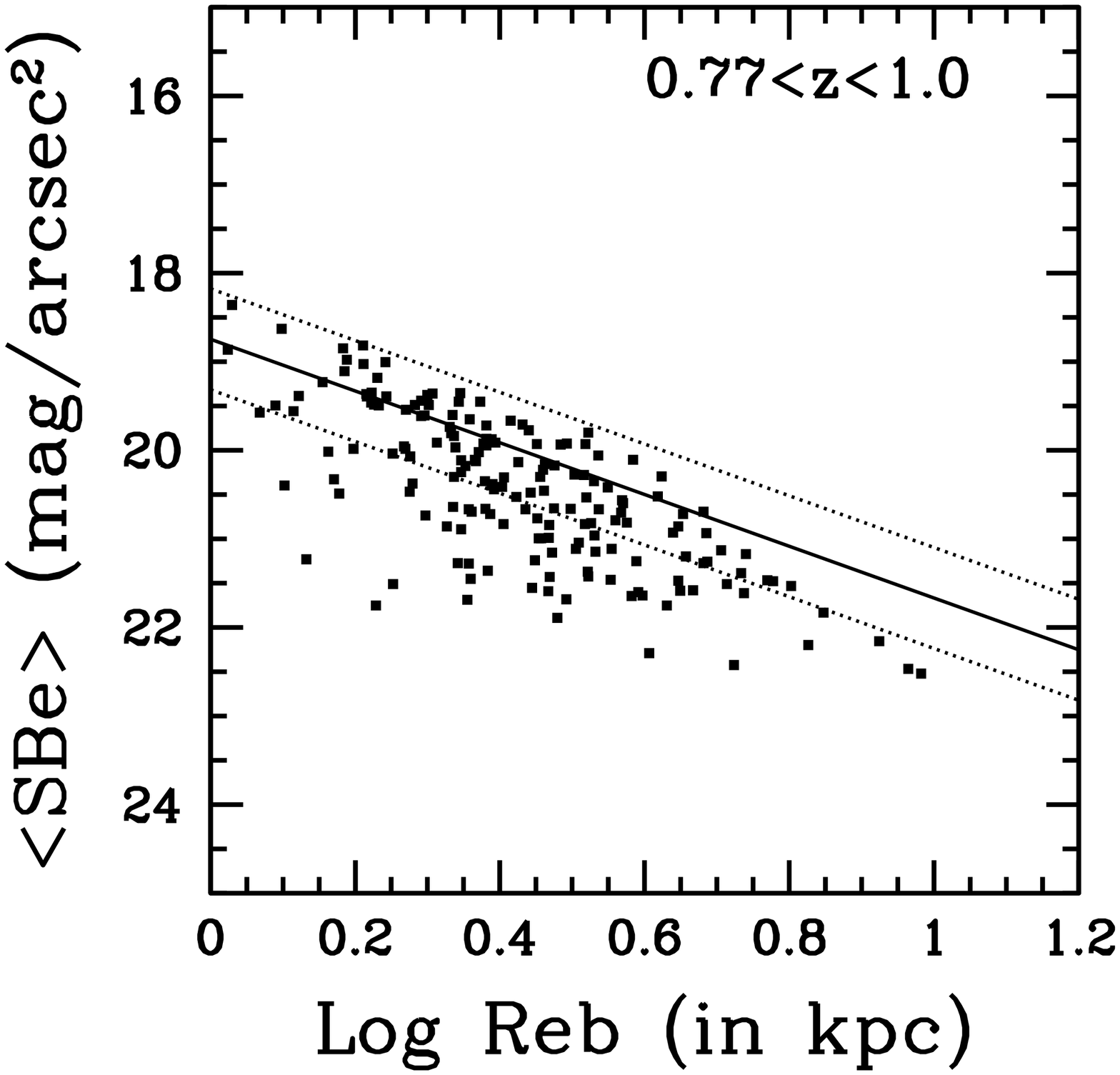}}
\mbox{\includegraphics[width=55mm]{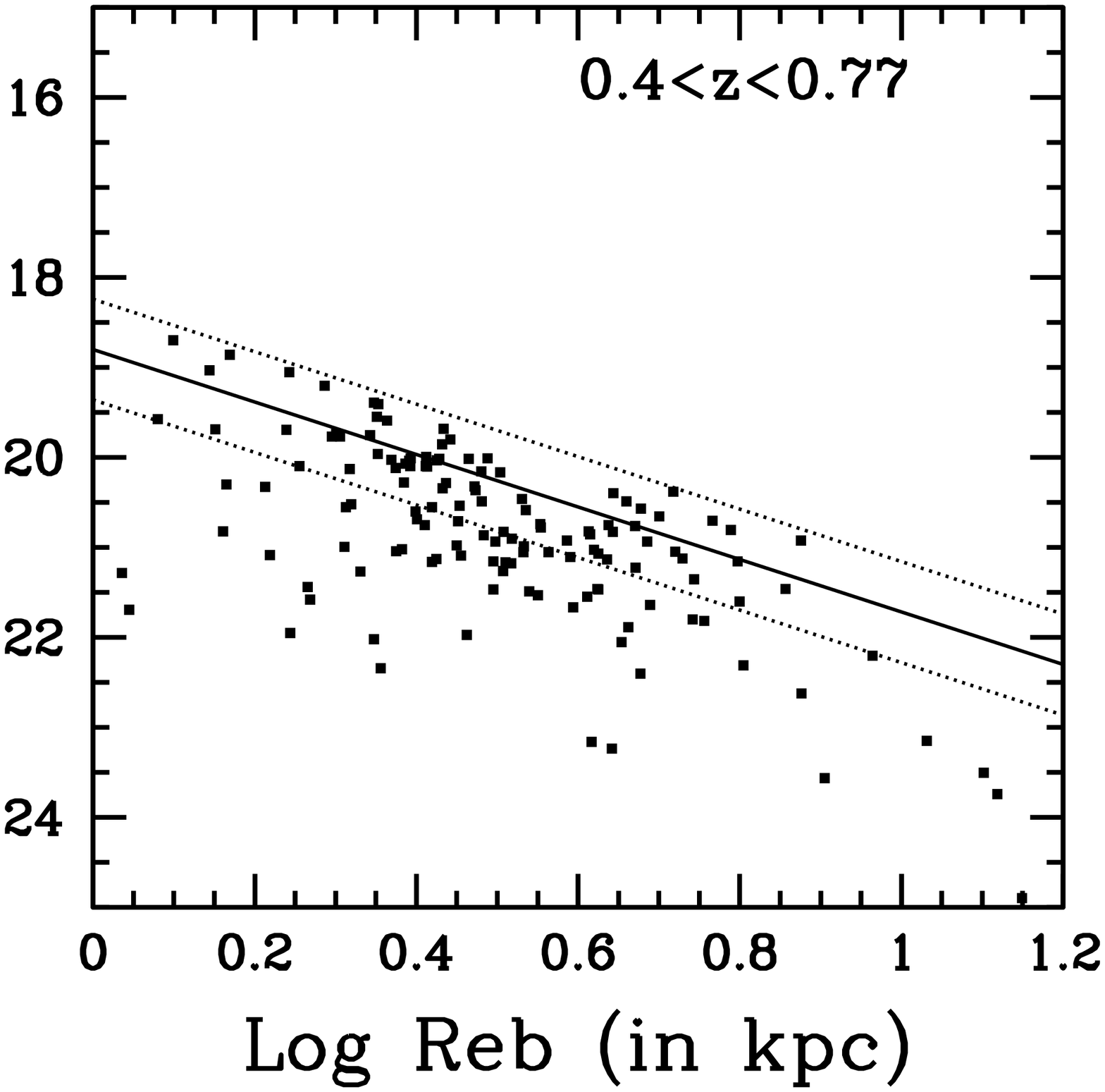}}
\mbox{\includegraphics[width=55mm]{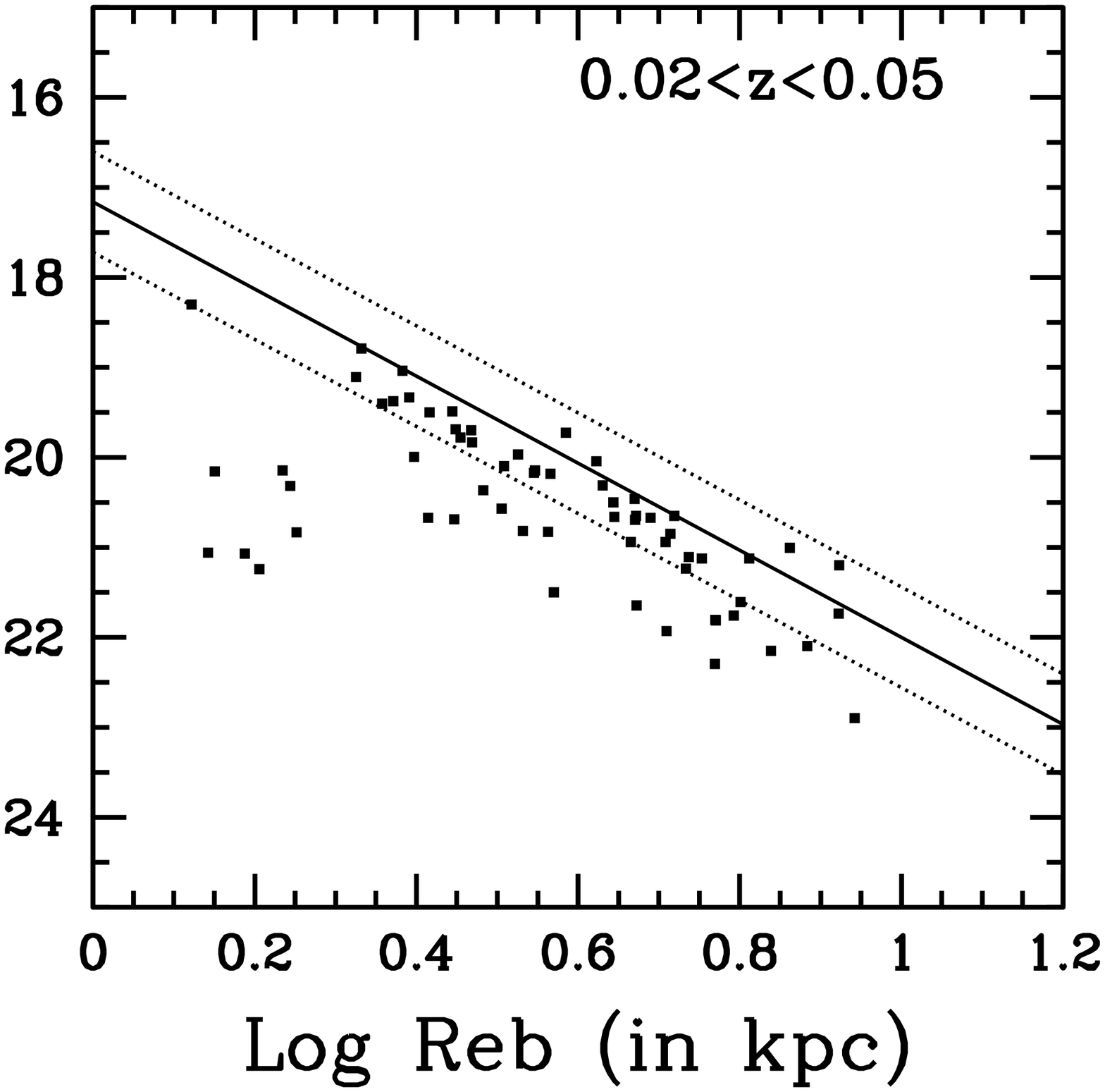}}
\caption{The solid line, in each of these three plots, depicts the Kormendy relation obtained for elliptical galaxies, in rest-frame {\it B}-band, in that redshift range. The dashed lines on the two sides of the solid line, represent the $\pm3\sigma$ boundaries for a fixed slope. The dots refer to our sample of discs with bulges which have been placed according to their rest-frame {\it B}-band bulge parameters, i.e, average surface brightness and effective radius. Sources falling within the $3\sigma$ boundaries are identified as disc galaxies with classical bulges and those below are identified as discs with pseudo bulges.}
\label{fig:Kormendydivision}
\end{figure*}

\begin{table}
\begin{minipage}{100mm}
\caption{Sample size for disc galaxies with pseudo\\and classical bulges}
\begin{tabular}{@{}lllll@{}}
\hline
Redshift & Rest & Discs with & Discs with & Discs with\\
 & frame & ($B/T>0.1$) & pseudo bulges & classical bulges\\
\hline
0.77-1.0 & {\it B}-band & 175 & 64 & 111\\
0.77-1.0 & {\it I}-band & 141 & 48 & 93\\
0.4-0.77 & {\it B}-band & 140 & 62 & 78\\
0.4-0.77 & {\it I}-band & 109 & 44 & 65\\
0.02-0.05 & {\it B}-band & 64 & 27 & 37\\
0.02-0.05 & {\it I}-band & 59 & 24 & 35\\
\hline
\label{sample-size}
\end{tabular}
\end{minipage}
\end{table} 

\subsection{Bulge classification and statistics}
\label{sec:classification}
We employ the Kormendy relation for elliptical galaxies \citep{Kormendyetal2009} to arrive at a first-hand classification of the bulges as per their parameters computed above - bulges which follow the elliptical relation are identified as classical and outliers as pseudo bulges \citep{Gadotti2009}. The relation for ellipticals was computed for galaxies with global S\'ersic index $n_g>3.5$ in \citet{Sachdeva2013}. In Fig.~\ref{fig:Kormendydivision}, we over plot all the bulges in our sample and compare them with the Kormendy relation obtained at that redshift range - bulges lying within $3$~sigma boundaries are considered classical and those lying below $3$~sigma line are selected as pseudo bulges \citep[see][]{Gadotti2009}. According to this criteria, we have 153 pseudo bulges (64 at $0.77<z<1.0$, 62 at $0.4<z<0.77$ and 27 at $0.02<z<0.05$) and 226 classical bulges (111 at $0.77<z<1.0$, 78 at $0.4<z<0.77$ and 37 at $0.02<z<0.05$); 139 no bulge, i.e, pure disc galaxies and the rest which could not be classified into any of these categories. 

Out of the total 153 pseudo bulge galaxies, we have rest-frame {\it I}-band images (and corresponding fitted parameters) for 116 of them (48 at $0.77<z<1.0$, 44 at $0.4<z<0.77$ and 24 at $0.02<z<0.05$) and out of the 226 classical bulge galaxies, we have rest-frame {\it I}-band images (and corresponding fitted parameters) for 193 of them (93 at $0.77<z<1.0$, 65 at $0.4<z<0.77$ and 35 at $0.02<z<0.05$). The number of galaxies in {\it I}-band is lesser than the number of galaxies in the {\it B}-band due to the following reasons: some were lost in the optimal matching of the optical catalog with the 3D-HST catalog to obtain corresponding infrared images; for some the software was not able to converge to provide well defined isophotes, which as stated earlier can happen due to the degeneracy of equally bright points near the centre; for some the profiles were not well defined enough that functions could be fitted, i.e., the functions were not able to coverge. Note that the presence of extra number of sources in {\it B}-band does not affect the results presented in this work. {\it B}-band sample was found to generate the same statistical values, even after removing them. Since, it increases the statistical significance of the results, they have been kept. In one-to-one source comparison, i.e, in comparison plots, these sources are automatically not involved.

The final sample size statistics of discs with bulges are summarized in Table.~\ref{sample-size}. As per rest-frame {\it B}-band, our sample consists of $\sim27$\% pseudo bulges and $\sim40$\% classical bulges. Note that these numbers differ (still within quoted error bars) from the bulge statistics derived from massive galaxies within the local 11 Mpc volume \citep{FisherDrory2011}. We caution the readers that our derived statistics are indicative only as the classification is not based on rigorous kinematic criteria, which can effectively be applied only to nearby galaxies.


\begin{figure*}
{\includegraphics[width=90mm]{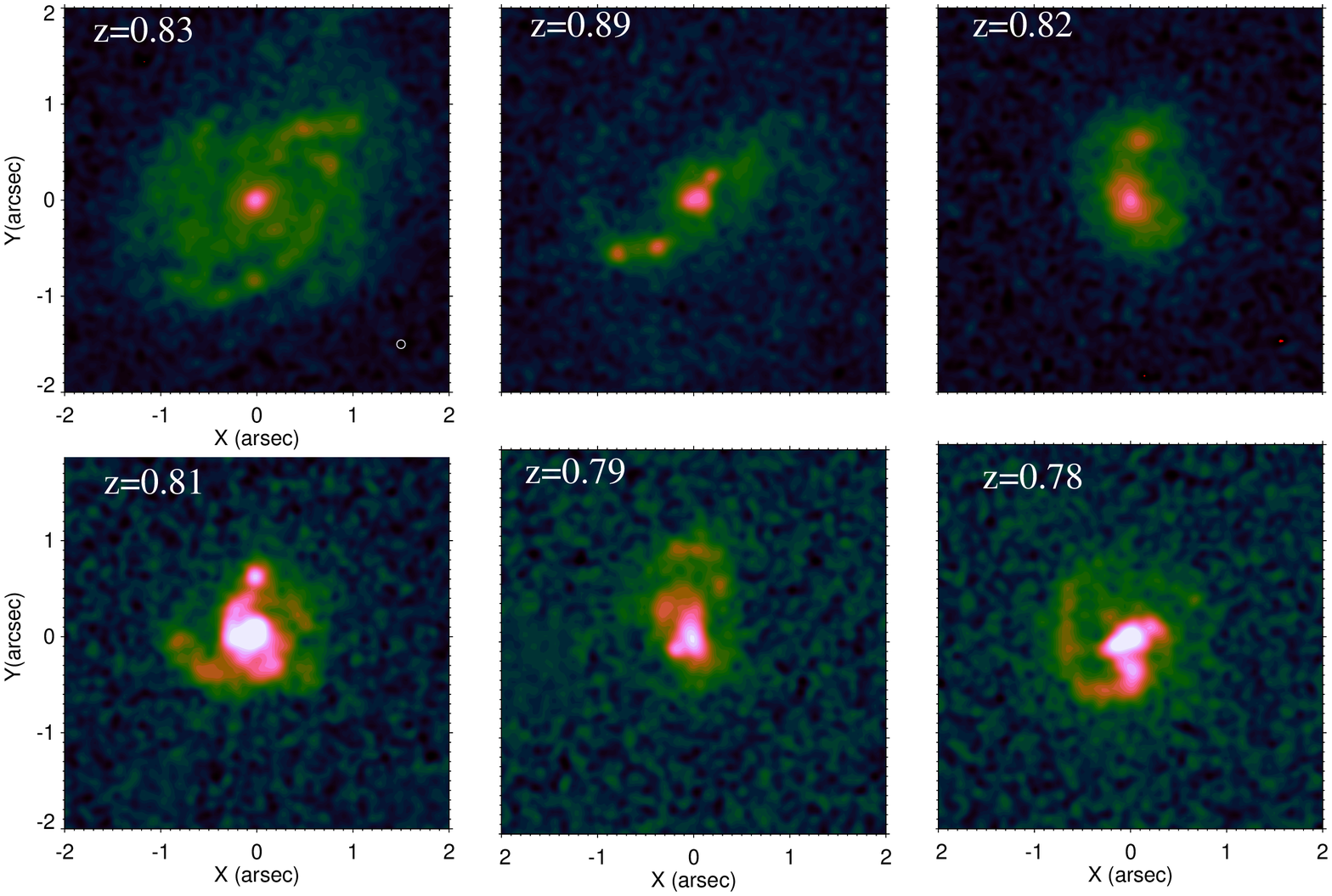}
\includegraphics[width=90mm]{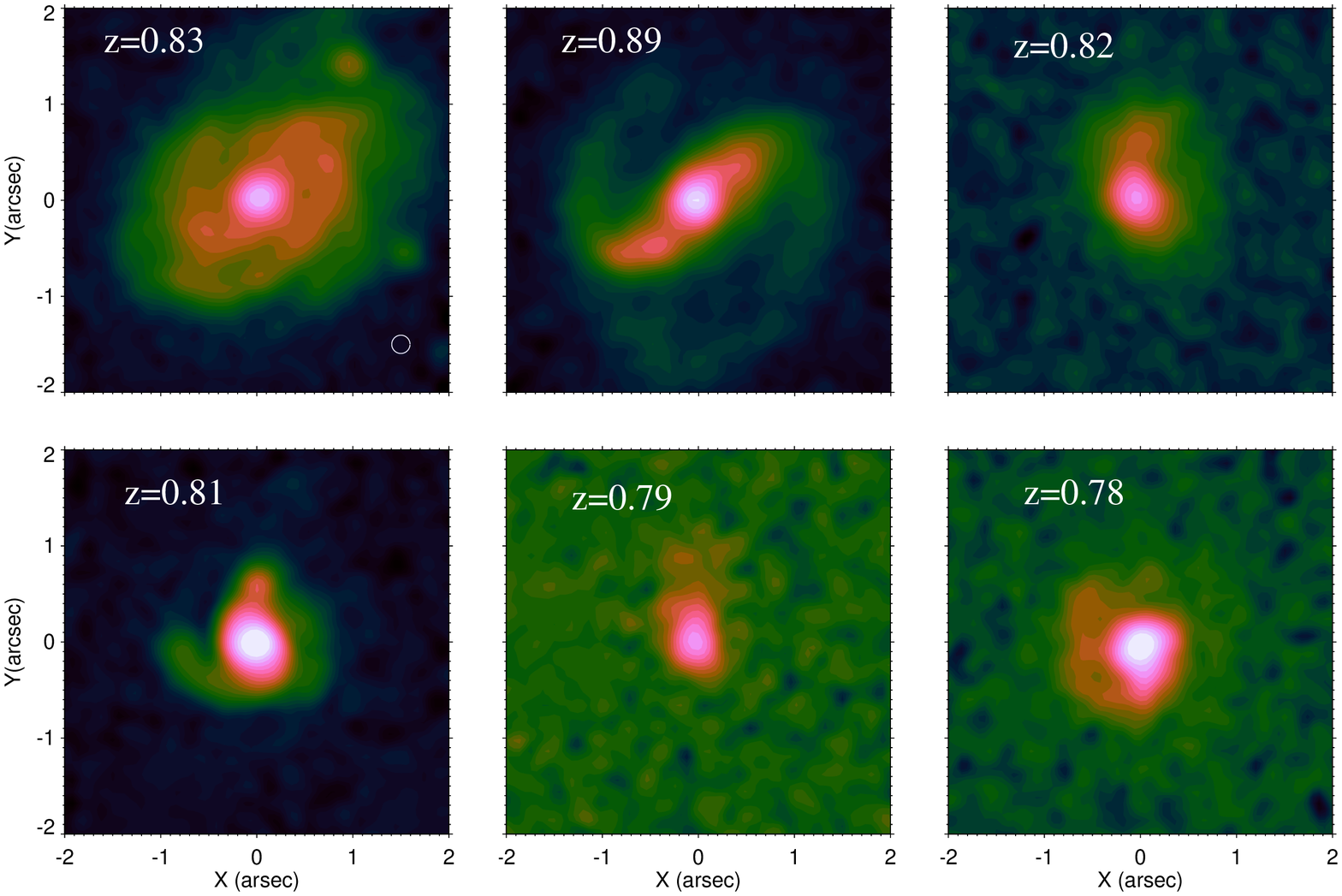}}
{\includegraphics[width=90mm]{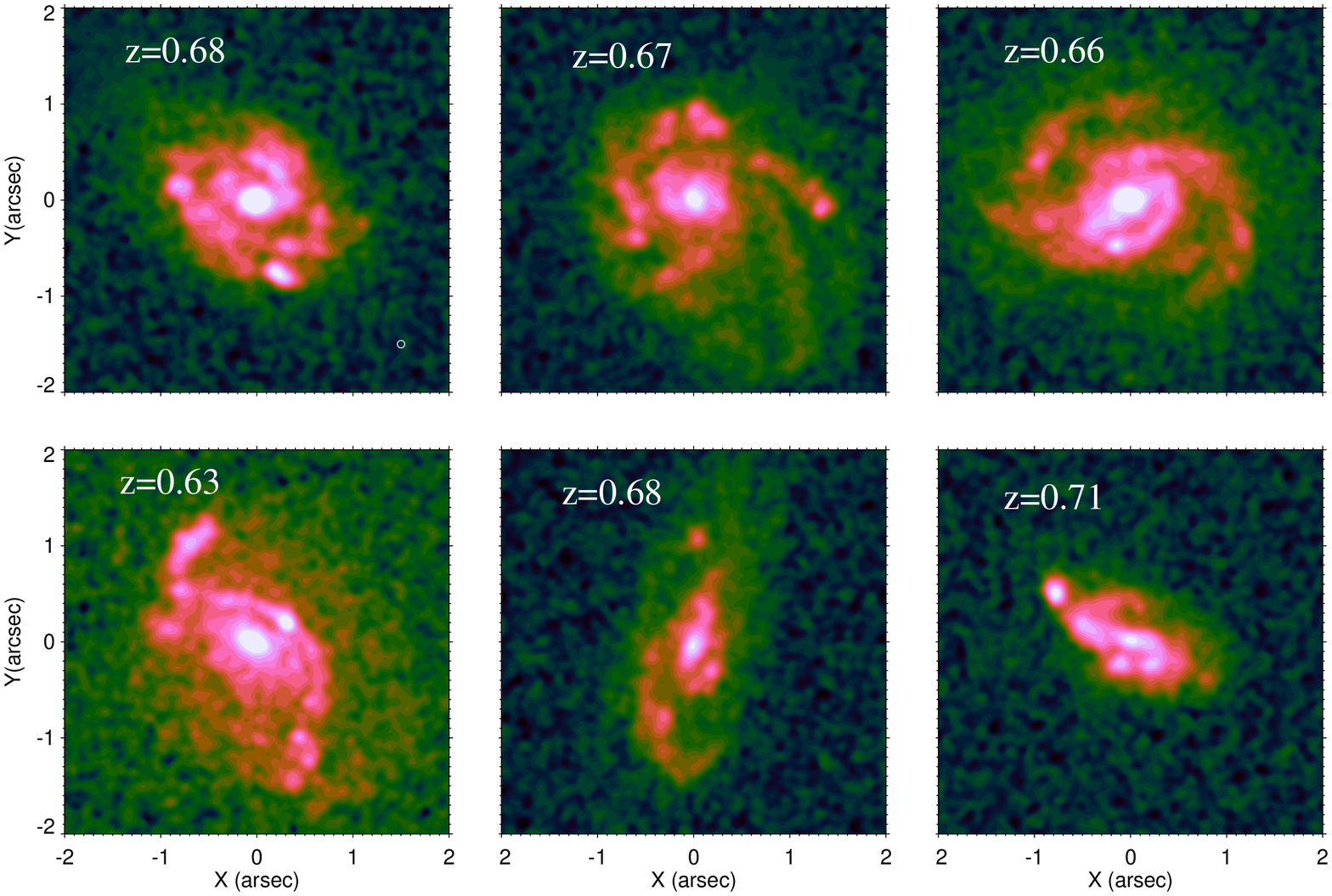}
\includegraphics[width=90mm]{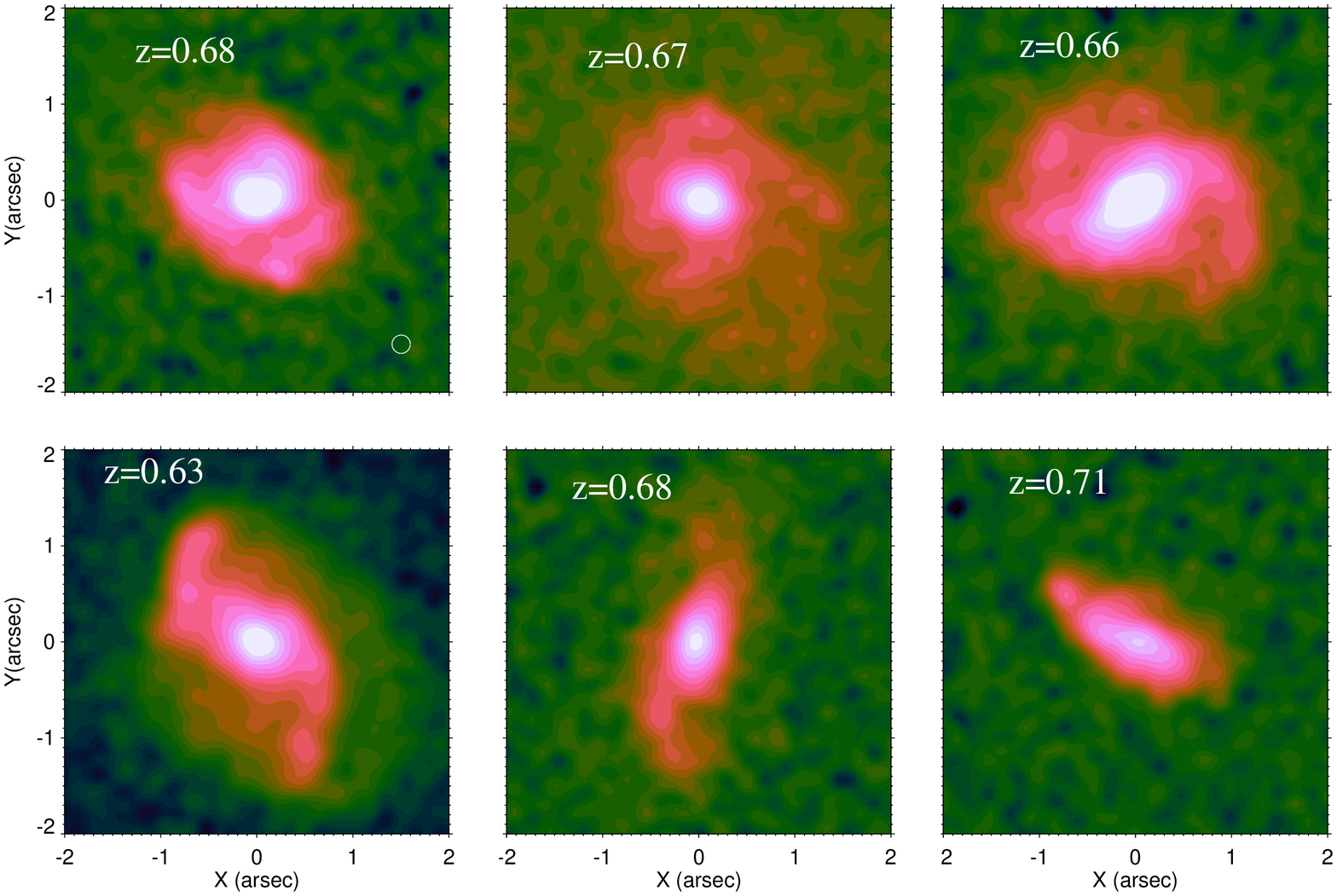}}
\caption{Morphology of {\bf disc galaxies with classical bulges} in rest-frame {\it B}-band (left panel) and their corresponding images in rest-frame {\it I}-band (right panel), at intermediate redshifts. Galaxies in the upper panel have redshifts in the range 0.77-1.0 and those in the lower panel have redshifts in the range 0.4-0.77. The white circle in the first image of each panel depicts the PSF-FWHM (circle's diameter) for that panel of images. The color scale is same for all the images.}
\label{fig:clump-cl}
\end{figure*}

\begin{figure*}
{\includegraphics[width=90mm]{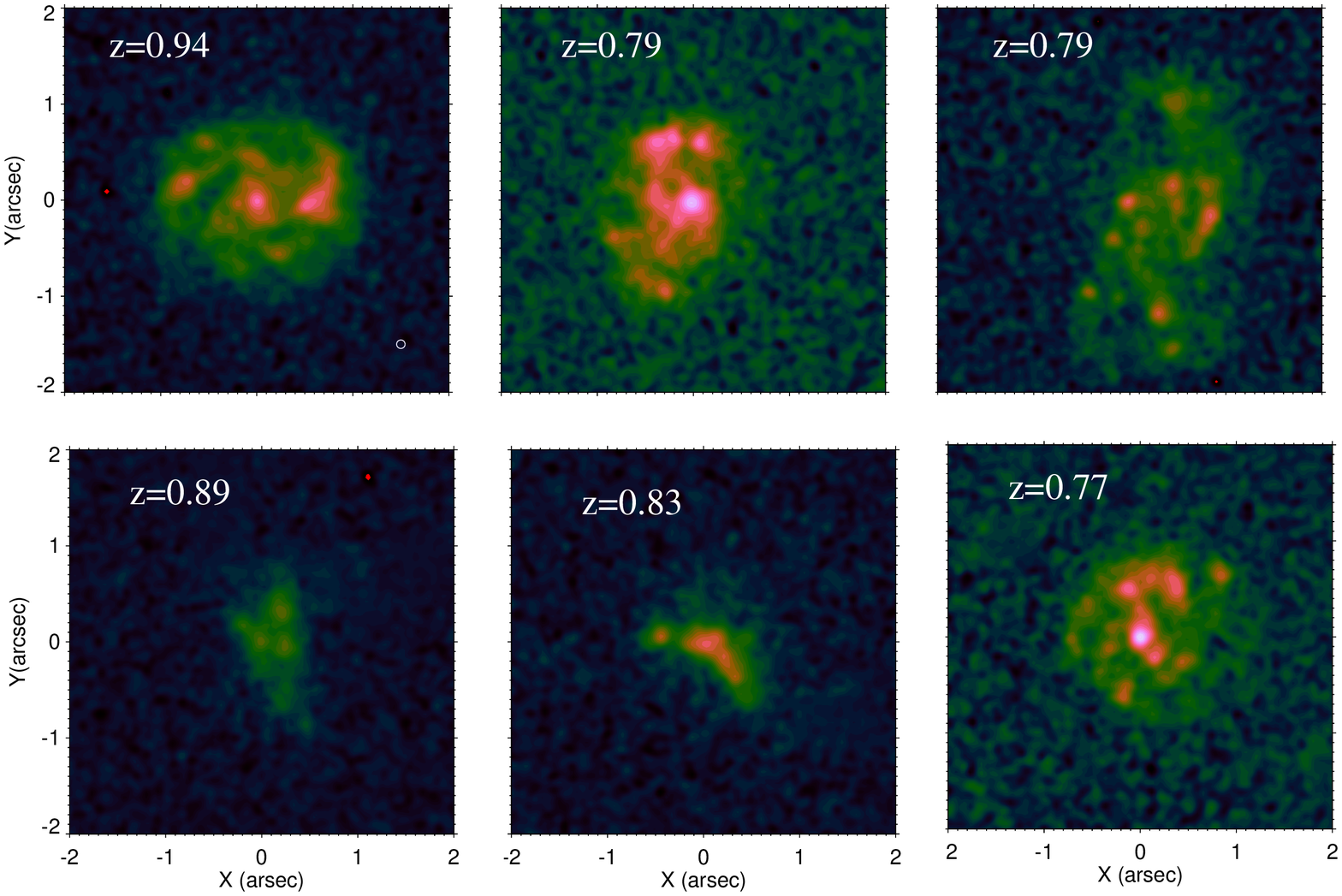}
\includegraphics[width=90mm]{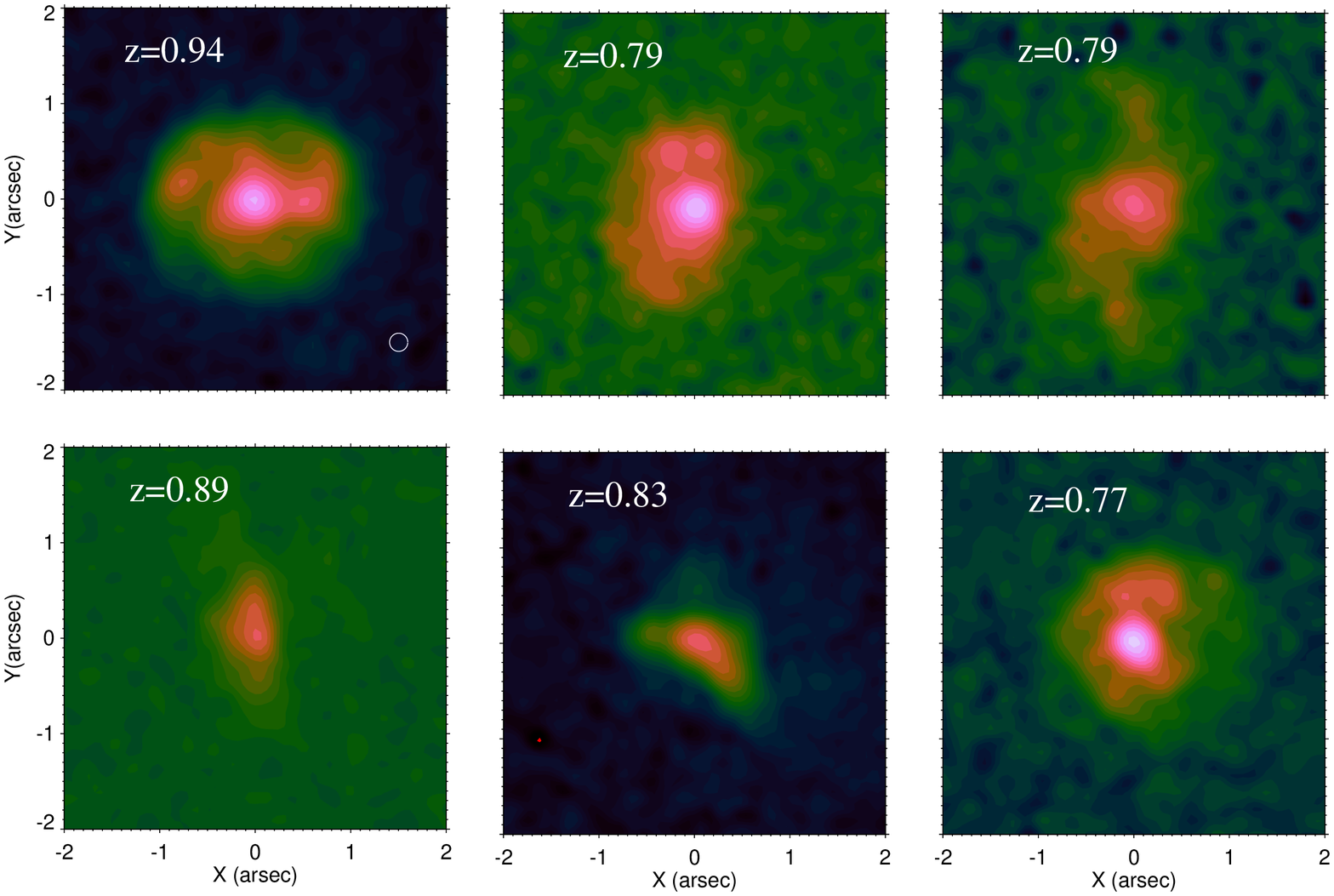}}
{\includegraphics[width=90mm]{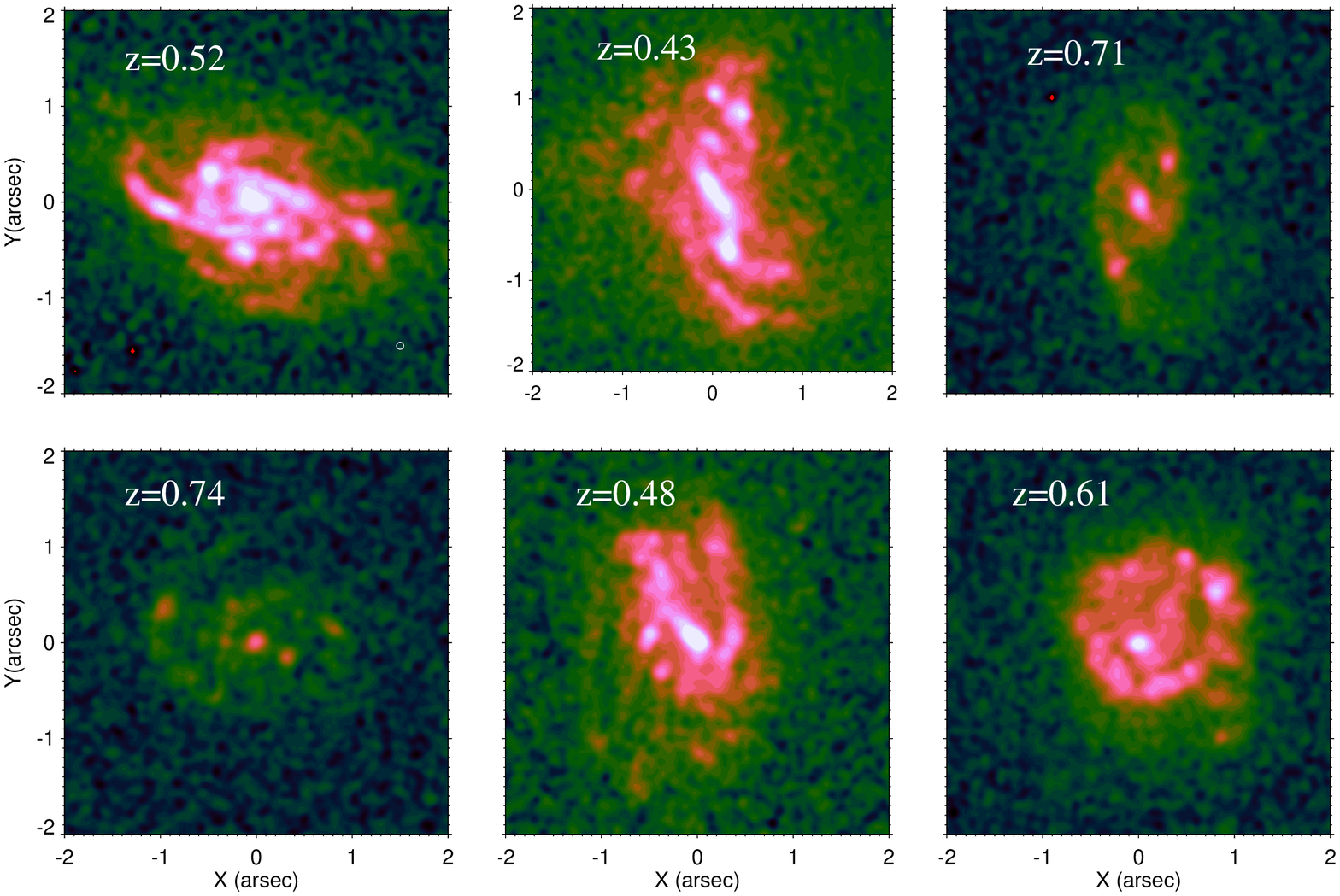}
\includegraphics[width=90mm]{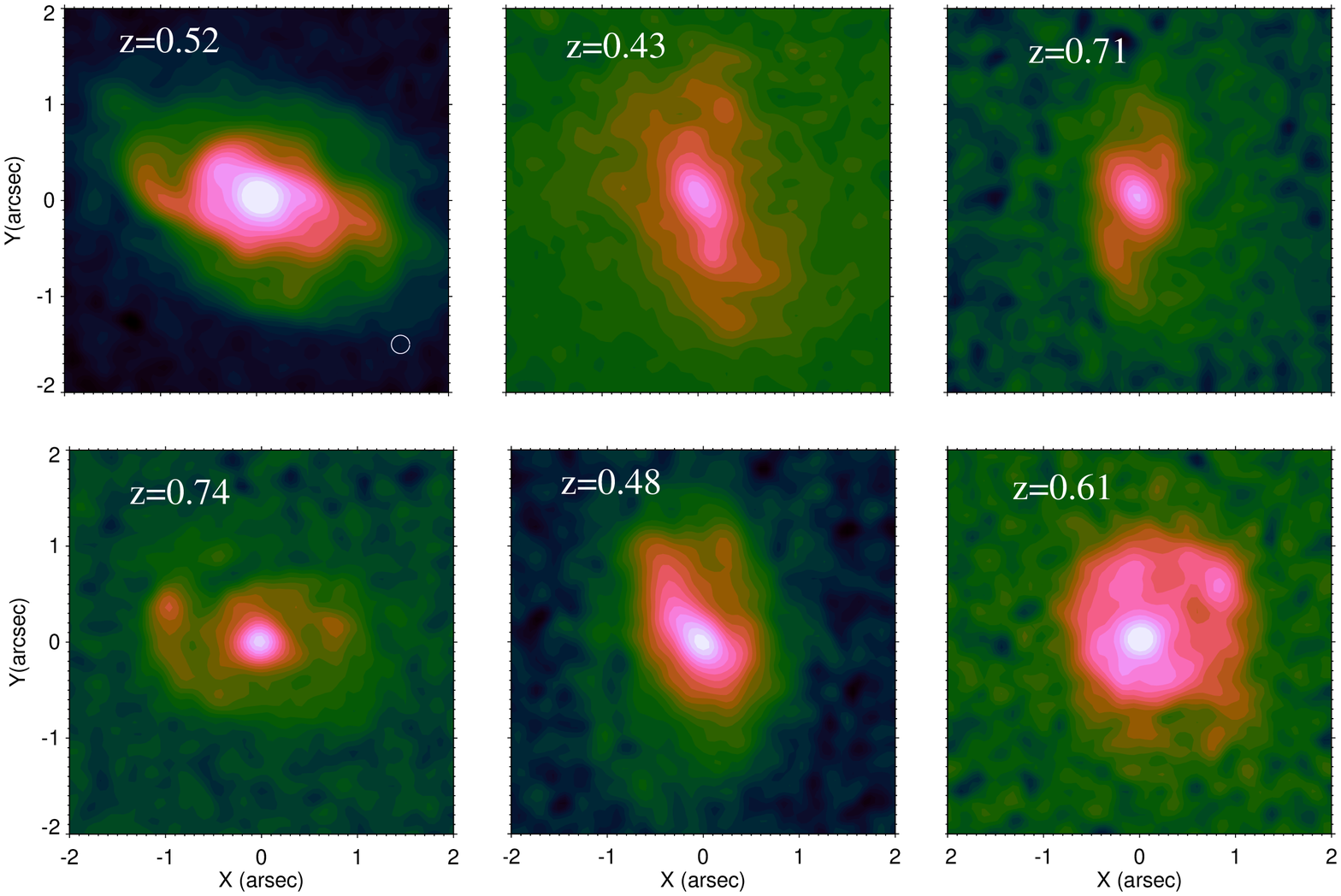}}
\caption{Same as Fig.~\ref{fig:clump-cl} but for {\bf disc galaxies with pseudo bulges}.}
\label{fig:clump-pb}
\end{figure*}

\section{Host galaxy morphology and bulge growth}
\label{sec:bulgegrowth}
There is a growing number of observational evidences that disc galaxies have settled into rotational equilibrium before redshift $z\sim1$ \citep{Kassinetal2007, Buitragoetal2014,HuertasCompanyetal2016,MargalefBentaboletal2016,Simonsetal2016,Tadakietal2017}; but the epoch at which such a transition happened remain obscured - making it harder to pin-point exactly when bulges started forming inside them. Galaxies in the high redshift ($z>1$) are clumpy and these clumps might have played an important role in developing the central bulge \citep{Bournaud2016,Tadakietal2017}. Whether they produce pseudo bulge or classical bulge remains under debate. We visually examine morphologies of all the galaxies in our intermediate redshift (0.4 - 1.0) sample to find special signatures, if any, that lead to the growth of either pseudo or classical bulges. In Fig.~\ref{fig:clump-cl} and Fig.~\ref{fig:clump-pb}, we show the surface brightness maps of some of the galaxies with classical and pseudo bulges, respectively, in rest-frame {\it B}-band and their corresponding images in rest-frame {\it I}-band. For rest-frame {\it B}-band, we find that out of the total disc dominated sample with bulges, one-third of the galaxies have clumps in their host disc morphology. Note that our definition of clump is based on visual identification only. Following which, we report that both pseudo and classical bulge host discs appear to be with similar degree of clumpiness. Statistically, for classical bulge host discs, $32 - 36\%$ (35/111 at $z=0.77-1.0$, 28/78 at $z=0.4-0.77$) are with clumps; and for pseudo bulge host discs, $31 - 32\%$ (20/64 at $z=0.77-1.0$, 20/62 at $z=0.4-0.77$) are with clumps. Thus, the fraction is similar for the two bulge types and does not evolve over the two redshift ranges. For $z>1$, it is understood that clumps form in massive, gaseous, star-forming discs and are themselves sites of intense star-formation \citep{Elmegreenetal2008,Bournaud2016}. We argue that whether these high redshift optical clumps migrate or wither away, similar ones get formed at the same rate at least till the intermediate redshift-range probed in this work. However, we add the caveat that our analysis is only based on visual identification of the clumps.

For the same set of galaxies, examining the surface brightness maps in rest-frame {\it I}-band, we find that the distribution becomes smooth and the clumps recognized in rest-frame {\it B}-band become imperceptible. This could have been caused by the fact that clumps mainly emit in the optical and also the degradation of the PSF from optical to infrared. Nevertheless, insight into whether these clumps have been the major cause of bulge growth or other mechanisms, i.e., accretion and internal evolution have also played their hand, shall be obtained by investigating the growth and properties of bulges as a function of redshift in both optical and infrared. Table.~\ref{med-ps} and Table.~\ref{med-cl} tabulates the median and associated median absolute deviation (MAD) values of the properties of pseudo and classical bulges respectively, in both rest-frame {\it B} and {\it I}-band. We discuss our primary findings below.

\subsection{Bulge growth in optical, infrared and in-terms of stellar mass}

Rest-frame {\it B}-band tracks the growth w.r.t. younger stellar population. We find that both pseudo and classical bulges have steadily grown in terms of size, luminosity, S\'ersic-index, bulge/total ratio (hereafter, B/T), as well as stellar mass, over the three redshift ranges probed. The pseudo bulges in our sample are larger in size than classical ones at higher redshift ranges. But the classical bulges have witnessed comparatively higher growth in size. In terms of luminosity, classical bulges are twice ($\sim0.8 mag$) as luminous as their pseudo counterparts at all redshifts. Both bulges almost double ($\sim1.7$ times, $\sim0.6 mag$) their luminosity with time, see Table.~\ref{med-ps} and Table.~\ref{med-cl}. The increase in luminosity of the bulges gets countered by the expansion in their sizes in such a manner that their effective surface-brightness witnesses insignificant evolution from $z=1$ to $z=0$. Classical bulges, at all redshifts, remain $\sim1 mag$ ($2.5$ times) brighter than their pseudo counterparts. Although their effective surface brightness remains nearly unchanged, both types of bulges have witnessed considerable increase in the B/T values over the last $\sim8 Gyrs$.

\begin{figure*}
\mbox{\includegraphics[width=55mm]{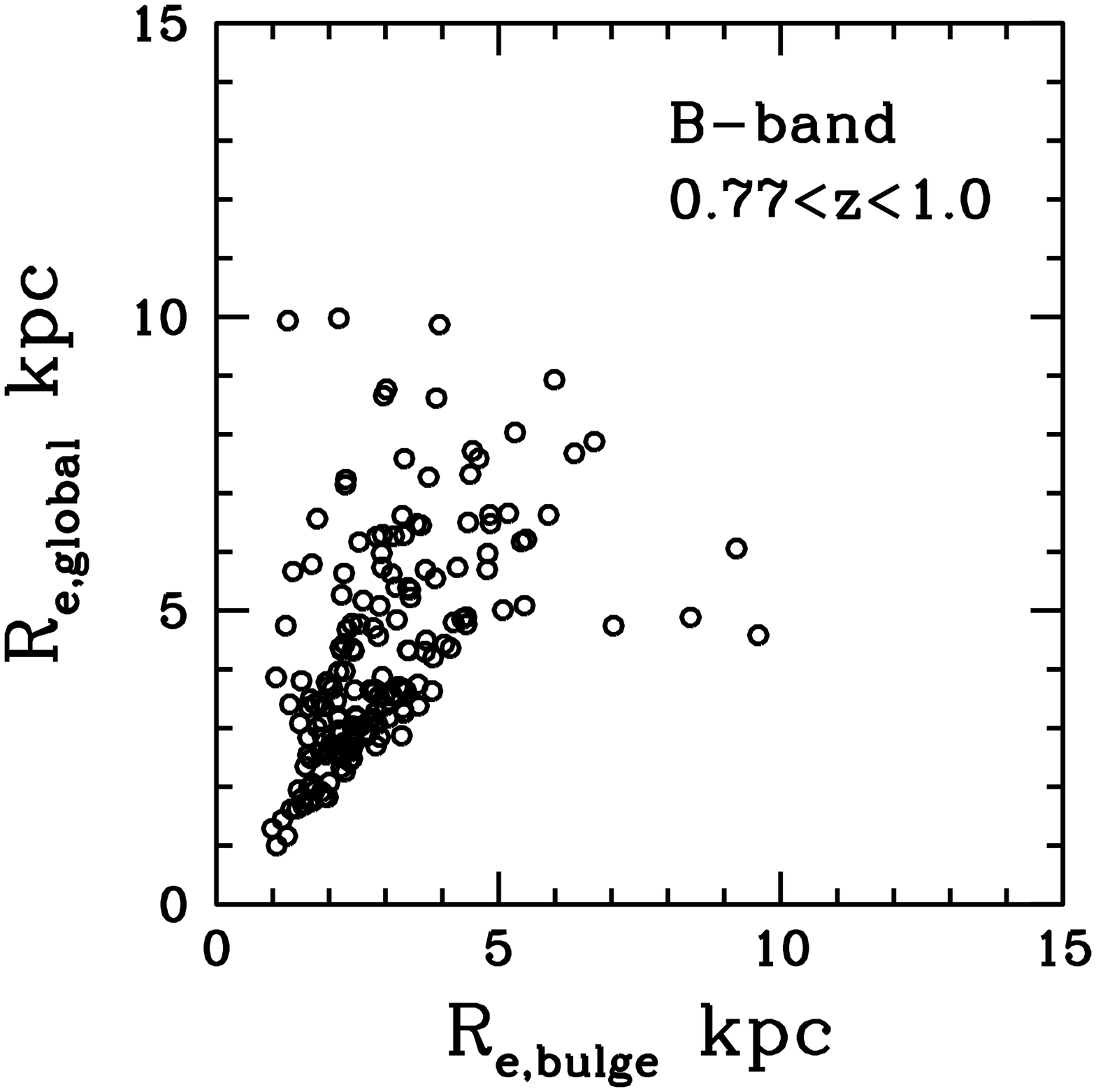}
      \includegraphics[width=55mm]{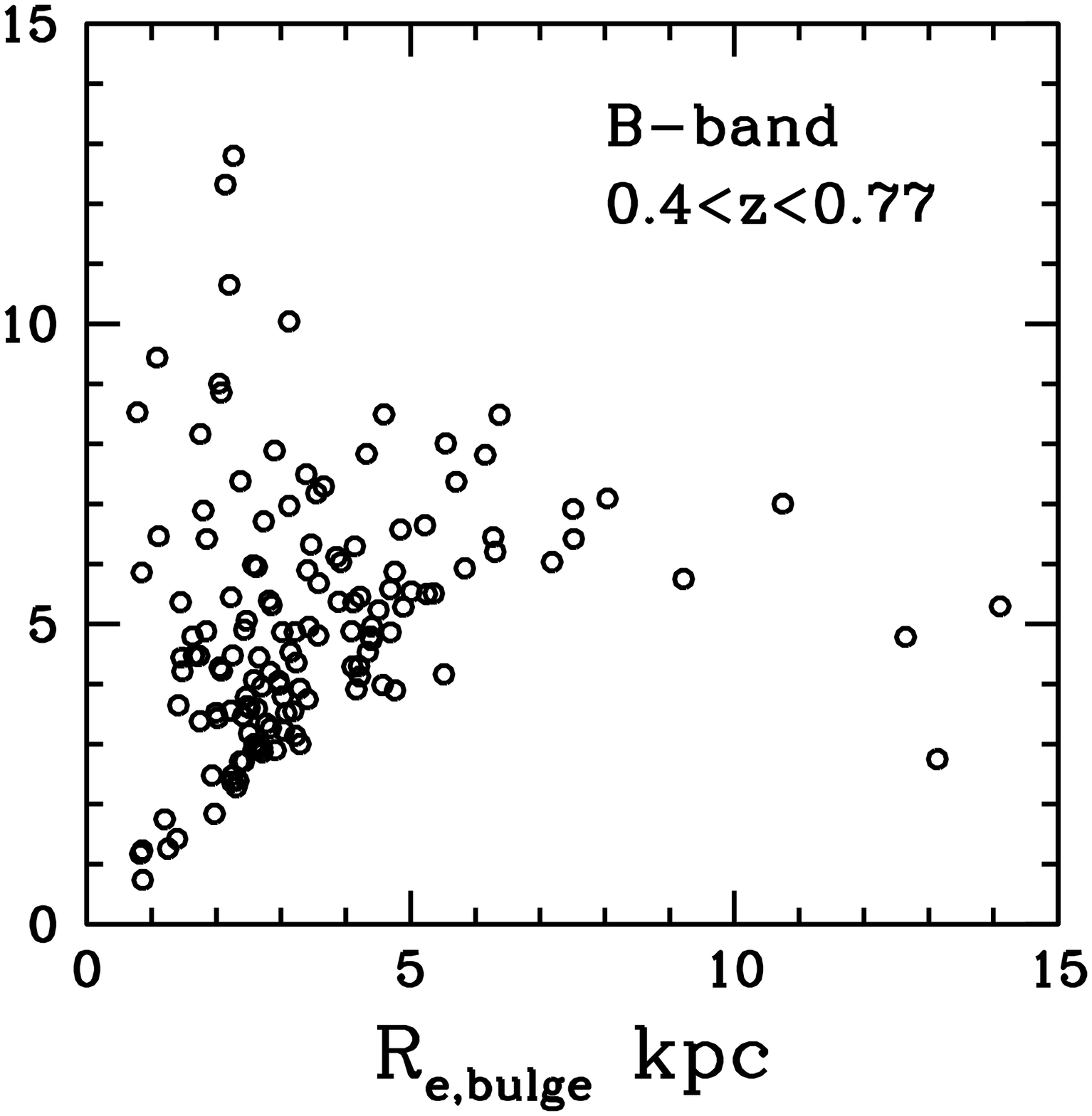}
      \includegraphics[width=55mm]{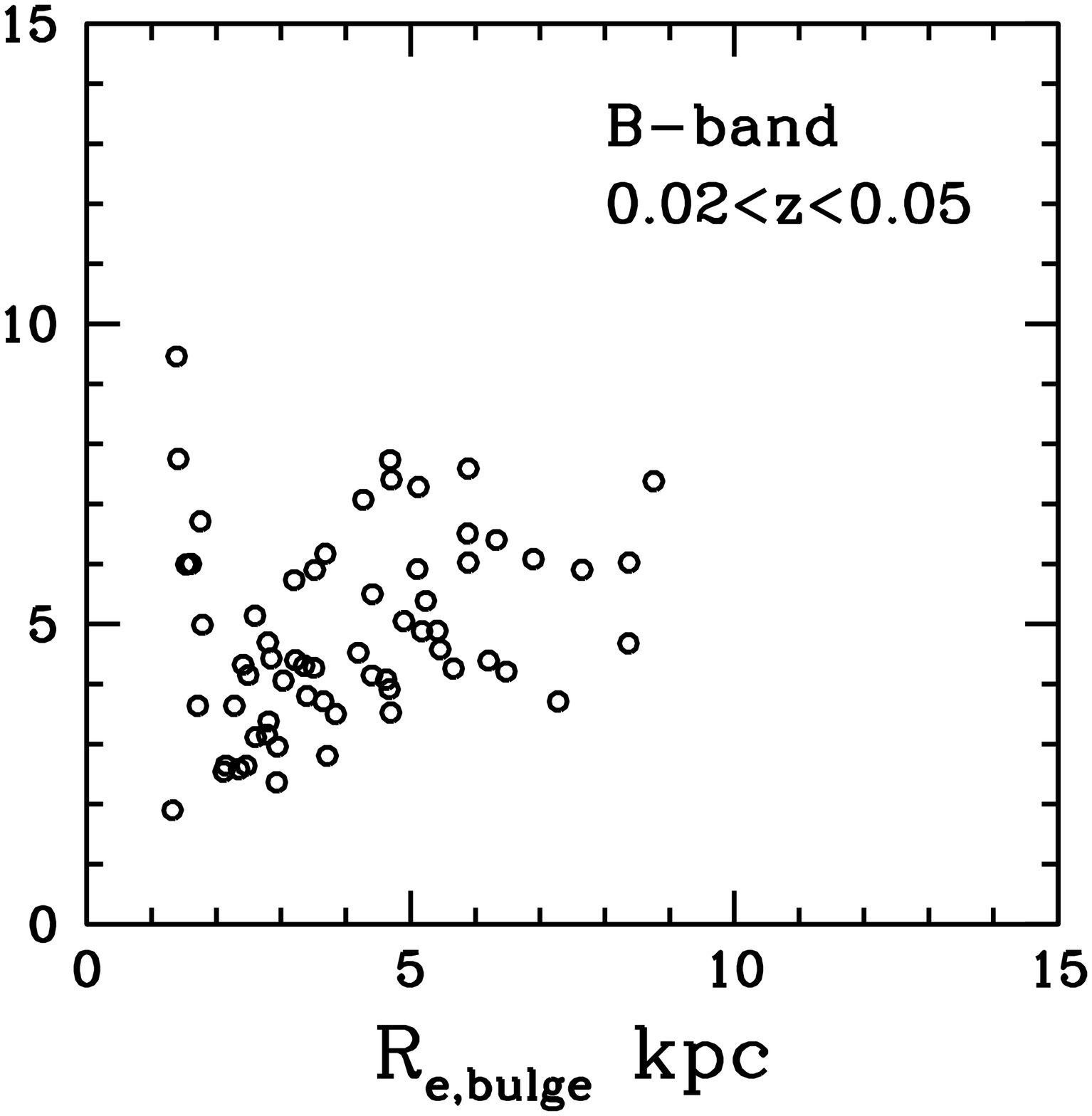}}
\caption{ The half light radius of the bulge is plotted against the global half light radius of the galaxy (computed by fitting single S\'ersic component to galaxy's 2D image using {\it Galfit}), in rest-frame {\it B}-band, for the three redshift ranges. As we move to the local redshifts, the half light radius of the bulge becomes highly comparable to the half light radius of the full galaxy.}
\label{fig:recomp-plots}
\end{figure*}

In Fig.~\ref{fig:recomp-plots} the half light radius of the bulge is plotted against the half light radius of the full galaxy. The latter was computed in Sachdeva (2013) by fitting the 2D rest-frame {\it B}-band image of the galaxy with single S\'ersic component using {\it Galfit}. The figure depicts that as we move from high redshifts to the local epoch, the two radii of the galaxy become highly comparable. In fact, for some of the galaxies, half light radius of the bulge is more than the half light radius of the galaxy. This is because the S\'ersic function correlates with the light concentration and the galaxies get increasingly concentrated as we move to lower redshifts, i.e., increasing fraction of total light of the galaxy is seen to be coming from the bulge component. The results are in agreement with those obtained by \citet{Tascaetal2014}, where they concluded that {\it B}-band emissivity has massively shifted from discs to bulges, since $z\sim0.8$. Separately for the two bulge types we report that, B/T increases from $\sim 10\%$ to $\sim 26\%$ for pseudo bulge discs, whereas for classical bulge discs, it increases from $\sim 21\%$ to $\sim 52\%$. Thus, an average optical bulge that accounts for about half the total light in present day disc galaxies in case of classical bulges, and about one-fourth the total light in case of pseudo bulges, was a not so significant member of the host galaxy at $z\sim1$.

\bigskip

Rest-frame {\it I}-band tracks the growth w.r.t., older stellar population and dust. The results are very insightful. In the case of pseudo bulges, their luminosity does not evolve, additionally, their size contracts by $\sim1 kpc$ reaching the range $2.2 - 2.3 kpc$ at $z=0$, see Table.~\ref{med-ps} and Table.~\ref{med-cl}. Thus, the evolution of pseudo bulges in the infrared is counter to their evolution in the optical, where they were found to expand and become more luminous from $z=1$ to $z=0$. This suggests that older stellar population becomes a more and more insignificant component of the pseudo bulge. The pseudo bulge, perhaps, gets fueled mainly from gas readjustment inside the galaxy which, maintaining the established rotational dynamics, goes through in-situ star formation.  

Classical bulges, on the other hand, follow a different evolutionary trajectory. Their infrared luminosity increases substantially ($2.5$ times, $\sim1 mag$), over the three redshift ranges. Additionally, the contraction in their size is also less intensive, $\sim0.4 kpc$. Both pseudo and classical bulges, reach sizes in the range $2.2-2.3 kpc$ at $z=0$. Note that these present day sizes are same as those reported in \citet{Gadotti2009} for massive bright galaxies in infrared ranges. The classical bulge, in contrast to pseudo, has grown in terms of older stellar population, over the last $\sim 8 Gyrs$, probably from the outside accretion of older satellites. 

\bigskip

We obtained the stellar masses for all the galaxies in the intermediate redshift sample (0.4-1.0) from 3D-HST catalog and for our local sample (0.02-0.05) from NASA-Sloan Atlas catalog. To obtain bulge mass, we multiply the total stellar mass of the galaxy with B/T ratio, where the latter is that obtained for rest-frame {\it I}-band as the {\it I}-band (Near IR) surface brightness profile more closely follows the underlying surface density profile in stellar mass being relatively free of the biases of young stellar population and dust \citep{Peletieretal1999,BellanddeJong2001,Belletal2003}.

We report (see Table.~\ref{med-ps} and Table.~\ref{med-cl}) that the mass of an average pseudo bulge is half of that of classical bulge ($\sim 0.34 \times 10^{10} M_{\odot}$) in the redshift range $0.4 < z <0.77$. As we reach the present epoch, both bulge types witness drastic increase in their stellar mass. While for pseudo bulges, the bulge mass doubles, for classical bulges it grows even higher ($\sim2.6$ times), reaching $\sim 0.88 (e+10) M_{\odot}$. Thus, more than half of the present stellar mass, for both bulge types, has got added during the past $\sim5-6$~Gyrs. This affirms that it is not only in-situ star formation from existing material and ageing of the existing stellar population that makes the bulges more luminous in optical and infrared, in this period.

\begin{table*}
\begin{minipage}{160mm}
\caption{Pseudo bulges in rest-frame {\it B} and {\it I}-band$^*$}
\centering
\begin{tabular}{@{}llllllll@{}}
\hline
Redshift & $R_e$ & $M_b$ & $n_e$ & $SB_e$ & $B/T$ & $Mass$\\
 & $kpc$ & $mag$ & & $mag/arcsec^2$ & & $(\times 10^{10}) M_{\odot}$\\
\hline
Rest-frame {\it B}-band\\
\hline
0.77-1.0 & 2.93($\pm$0.72) & -19.77($\pm$0.43) & 0.93($\pm$0.26) & 21.83($\pm$0.44) & 0.102($\pm$0.032) & 0.06718($\pm$0.05226)\\
0.4-0.77 & 3.14($\pm$1.03) & -19.78($\pm$0.45) & 1.03($\pm$0.24) & 21.98($\pm$0.45) & 0.169($\pm$0.081) & 0.07078($\pm$0.04368)\\
0.02-0.05 & 3.65($\pm$1.94) & -20.38($\pm$0.39) & 1.11($\pm$0.31) & 21.78($\pm$0.77) & 0.255($\pm$0.088) & 0.42549($\pm$0.15806)\\
\hline
Rest-frame {\it I}-band\\
\hline
0.77-1.0 & 3.34($\pm$0.68) & -21.03($\pm$0.55) & 0.86($\pm$0.18) & 21.01($\pm$0.44) & 0.496($\pm$0.204) & 0.35321($\pm$0.25046)\\
0.4-0.77 & 3.33($\pm$0.88) & -20.58($\pm$0.82) & 0.89($\pm$0.21) & 21.08($\pm$0.67) & 0.423($\pm$0.156) & 0.16609($\pm$0.11083)\\
0.02-0.05 & 2.31($\pm$0.77) & -21.10($\pm$0.61) & 1.21($\pm$0.28) & 19.96($\pm$0.43) & 0.178($\pm$0.117) & 0.37712($\pm$0.18255)\\
\hline
\label{med-ps}
\end{tabular}
$^*$The values provided in the table are Median($\pm$Median Absolute Deviation)\\
\end{minipage}
\end{table*}

\begin{table*}
\begin{minipage}{160mm}
\caption{Classical bulges in rest-frame {\it B} and {\it I}-band$^*$}
\centering
\begin{tabular}{@{}llllllll@{}}
\hline
Redshift & $R_e$ & $M_b$ & $n_e$ & $SB_e$ & $B/T$ & $Mass$\\
 & $kpc$ & $mag$ & & $mag/arcsec^2$ & & $(\times 10^{10}) M_{\odot}$\\
\hline
Rest-frame {\it B}-band\\
\hline
0.77-1.0 & 2.41($\pm$0.66) & -20.58($\pm$0.28) & 0.83($\pm$0.15) & 20.56($\pm$0.50) & 0.209($\pm$0.038) & 0.1331($\pm$0.0776)\\
0.4-0.77 & 2.75($\pm$0.79) & -20.64($\pm$0.29) & 0.91($\pm$0.19) & 20.82($\pm$0.49) & 0.372($\pm$0.067) & 0.2191($\pm$0.1300)\\
0.02-0.05 & 3.84($\pm$1.06) & -21.22($\pm$0.13) & 1.08($\pm$0.13) & 20.93($\pm$0.57) & 0.521($\pm$0.106) & 1.3502($\pm$0.3394)\\
\hline
Rest-frame {\it I}-band\\
\hline
0.77-1.0 & 2.60($\pm$0.51) & -21.13($\pm$0.56) & 0.71($\pm$0.14) & 19.93($\pm$0.41) & 0.605($\pm$0.171) & 0.3991($\pm$0.2506)\\
0.4-0.77 & 3.14($\pm$0.82) & -21.43($\pm$0.59) & 0.77($\pm$0.15) & 20.09($\pm$0.42) & 0.632($\pm$0.177) & 0.3370($\pm$0.1677)\\
0.02-0.05 & 2.21($\pm$0.63) & -22.03($\pm$0.53) & 1.16($\pm$0.22) & 19.08($\pm$0.39) & 0.398($\pm$0.123) & 0.8842($\pm$0.5761)\\
\hline
\label{med-cl}
\end{tabular}
$^*$The values provided in the table are Median($\pm$Median Absolute Deviation)\\
\end{minipage}
\end{table*}


\begin{figure*}
\mbox{\includegraphics[width=55mm]{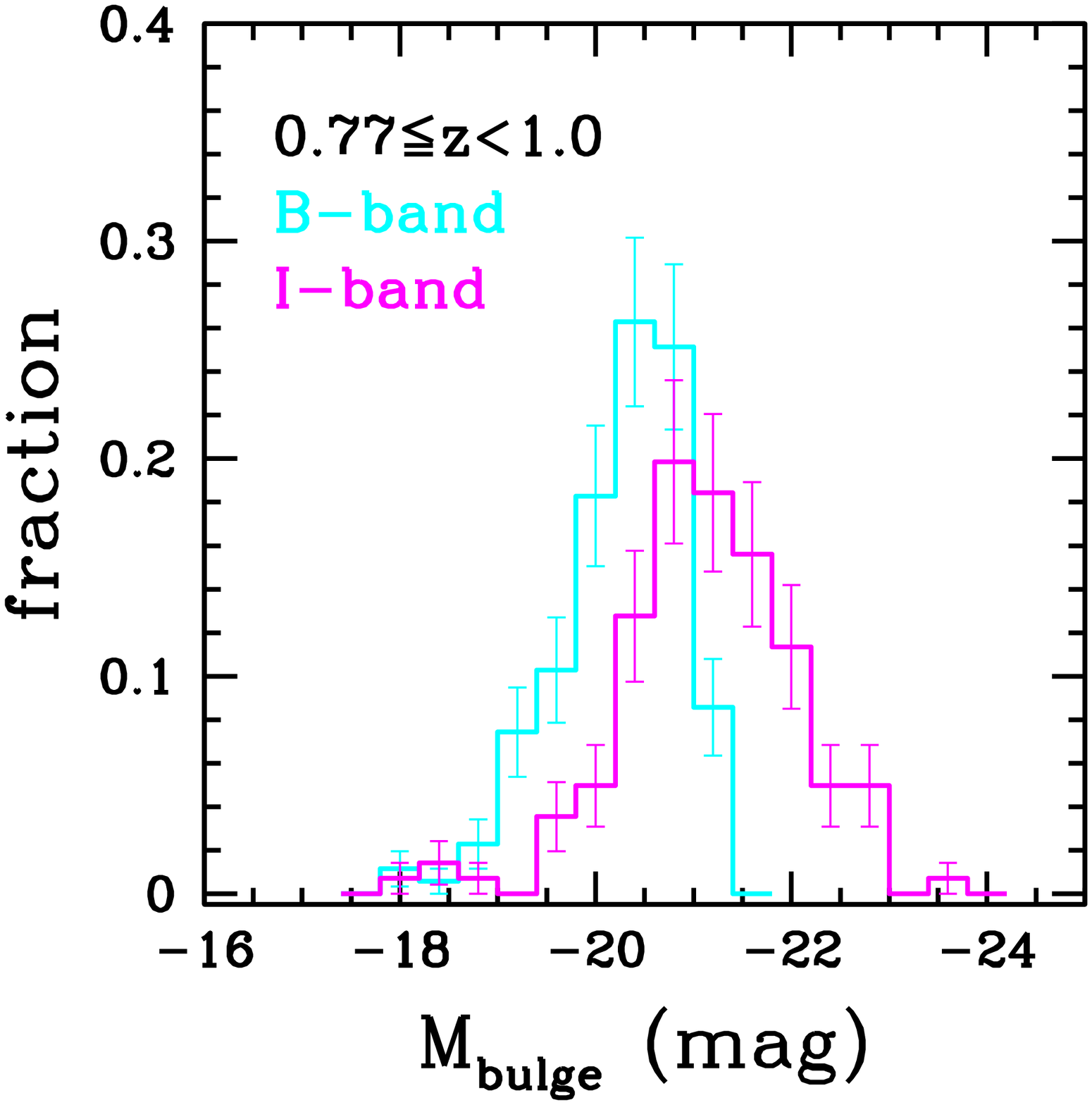}
      \includegraphics[width=55mm]{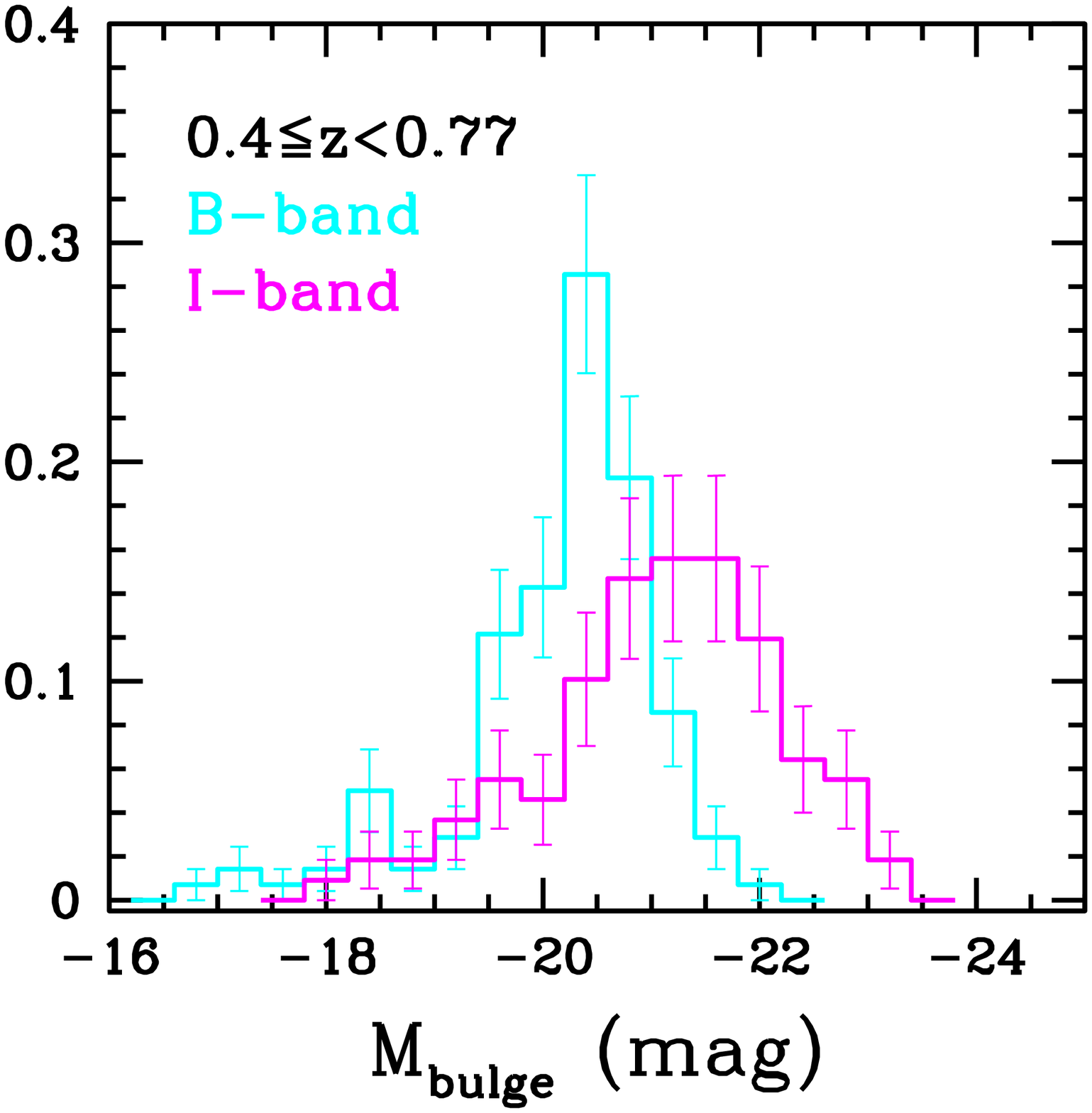} 
      \includegraphics[width=55mm]{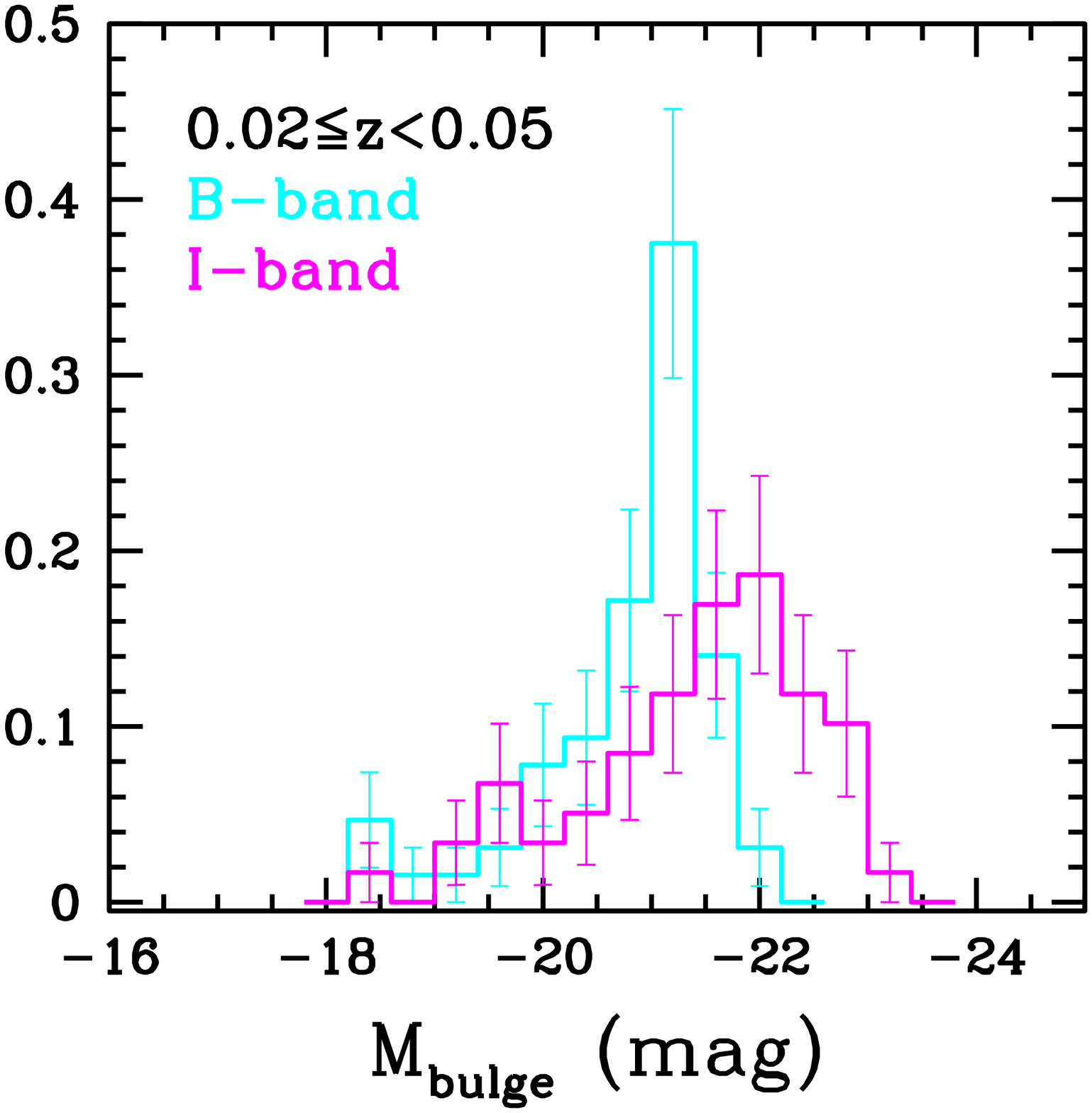}}
\mbox{\includegraphics[width=55mm]{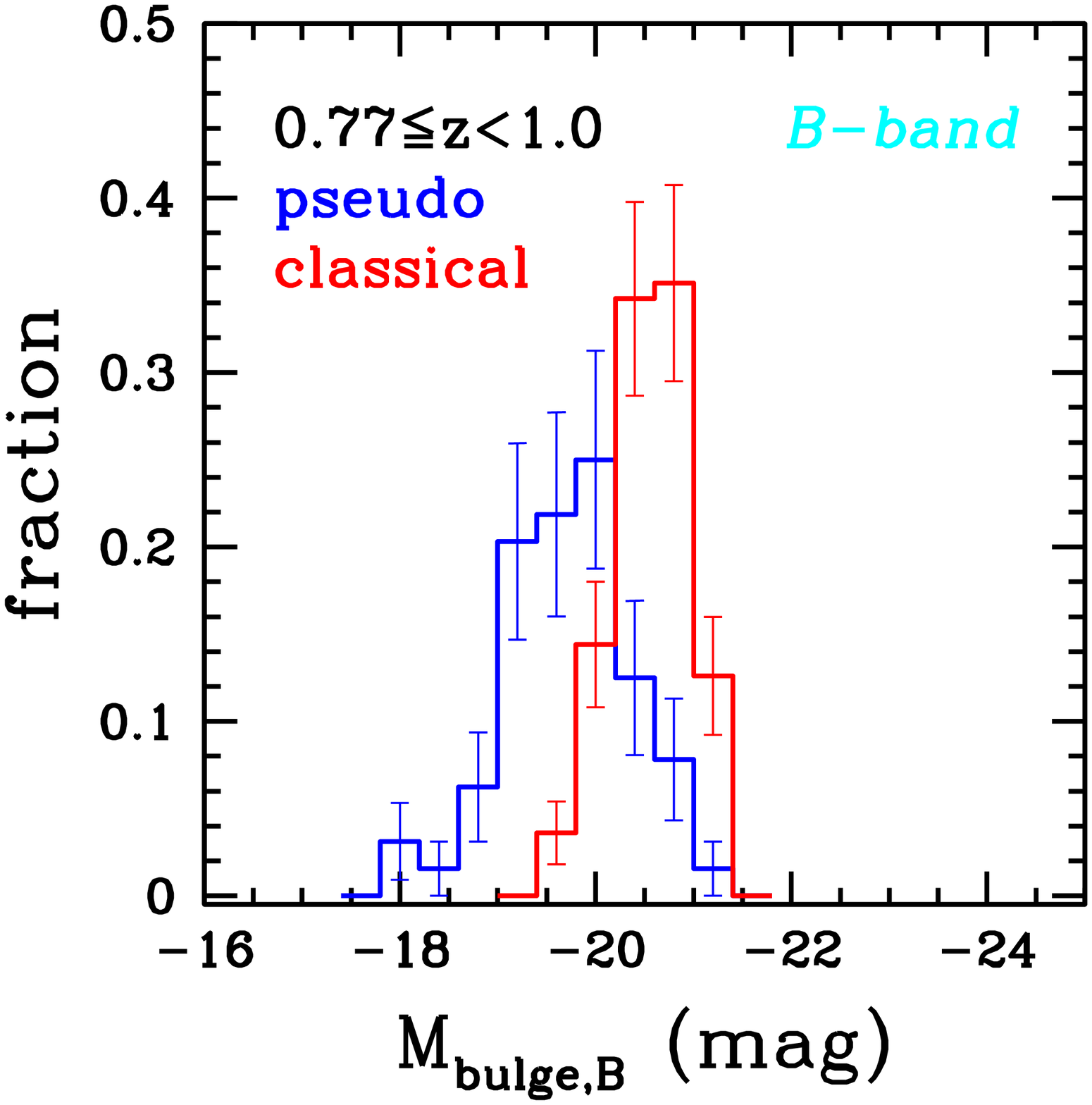}
      \includegraphics[width=55mm]{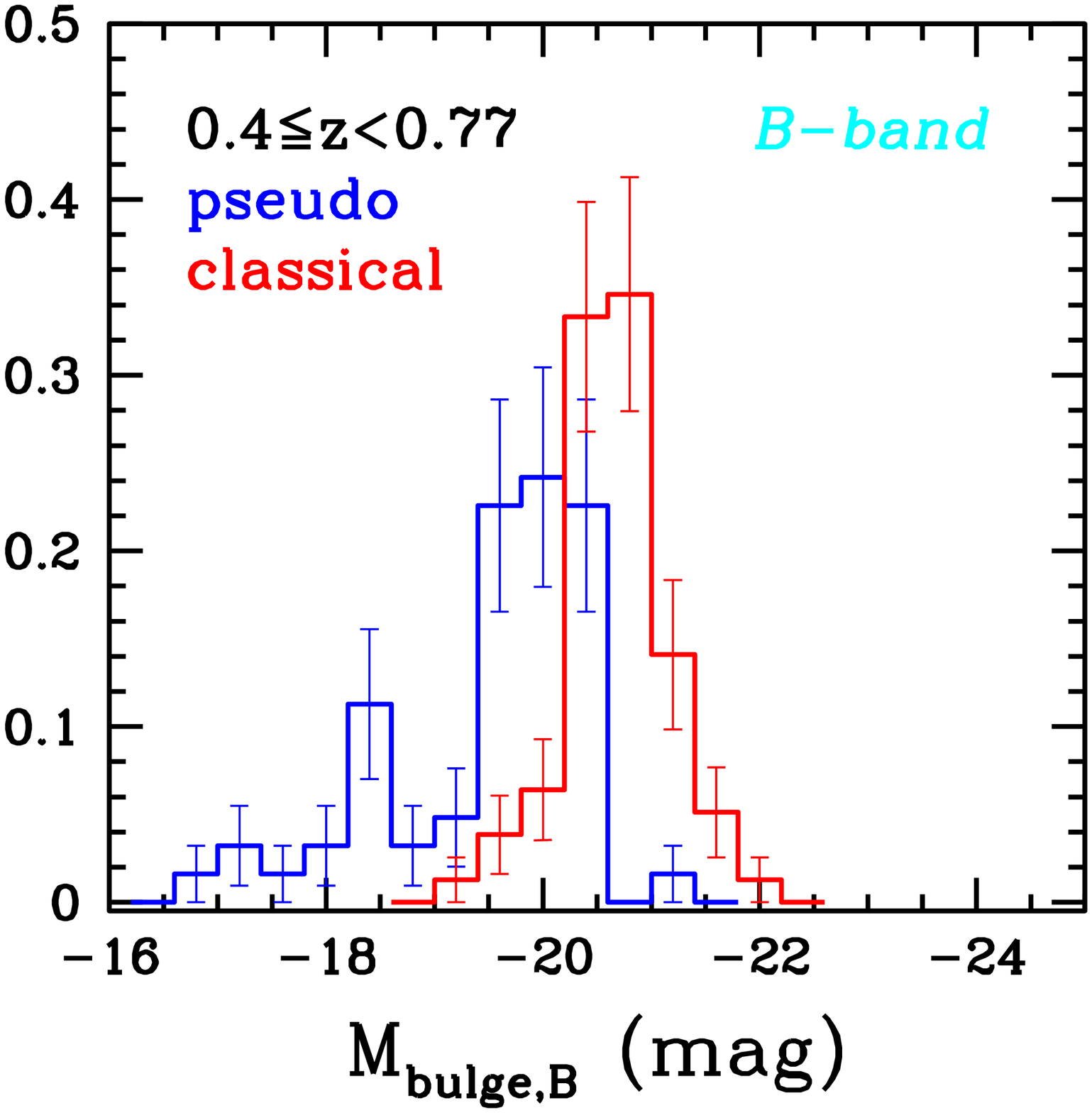}
      \includegraphics[width=55mm]{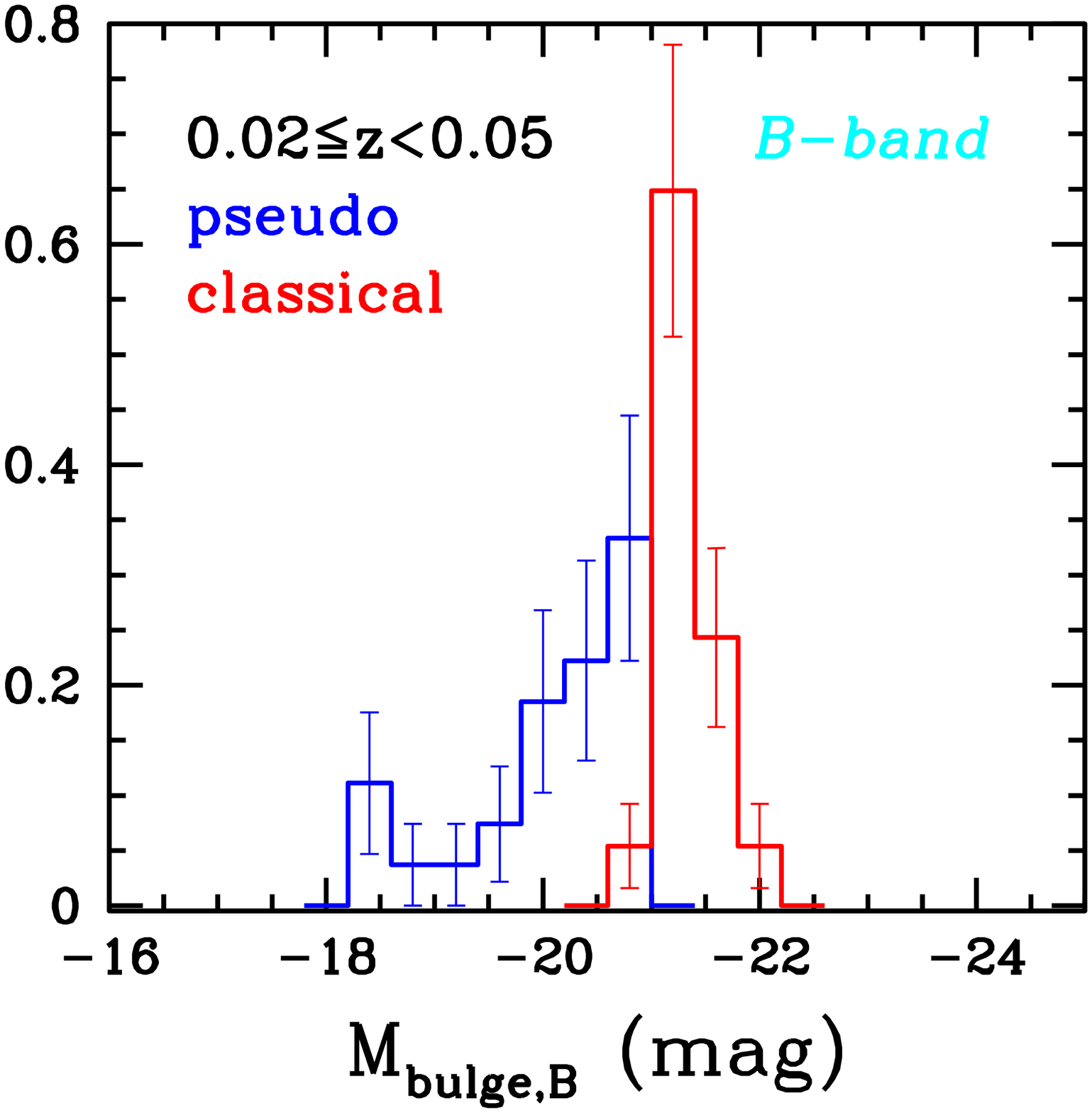}}
\mbox{\includegraphics[width=55mm]{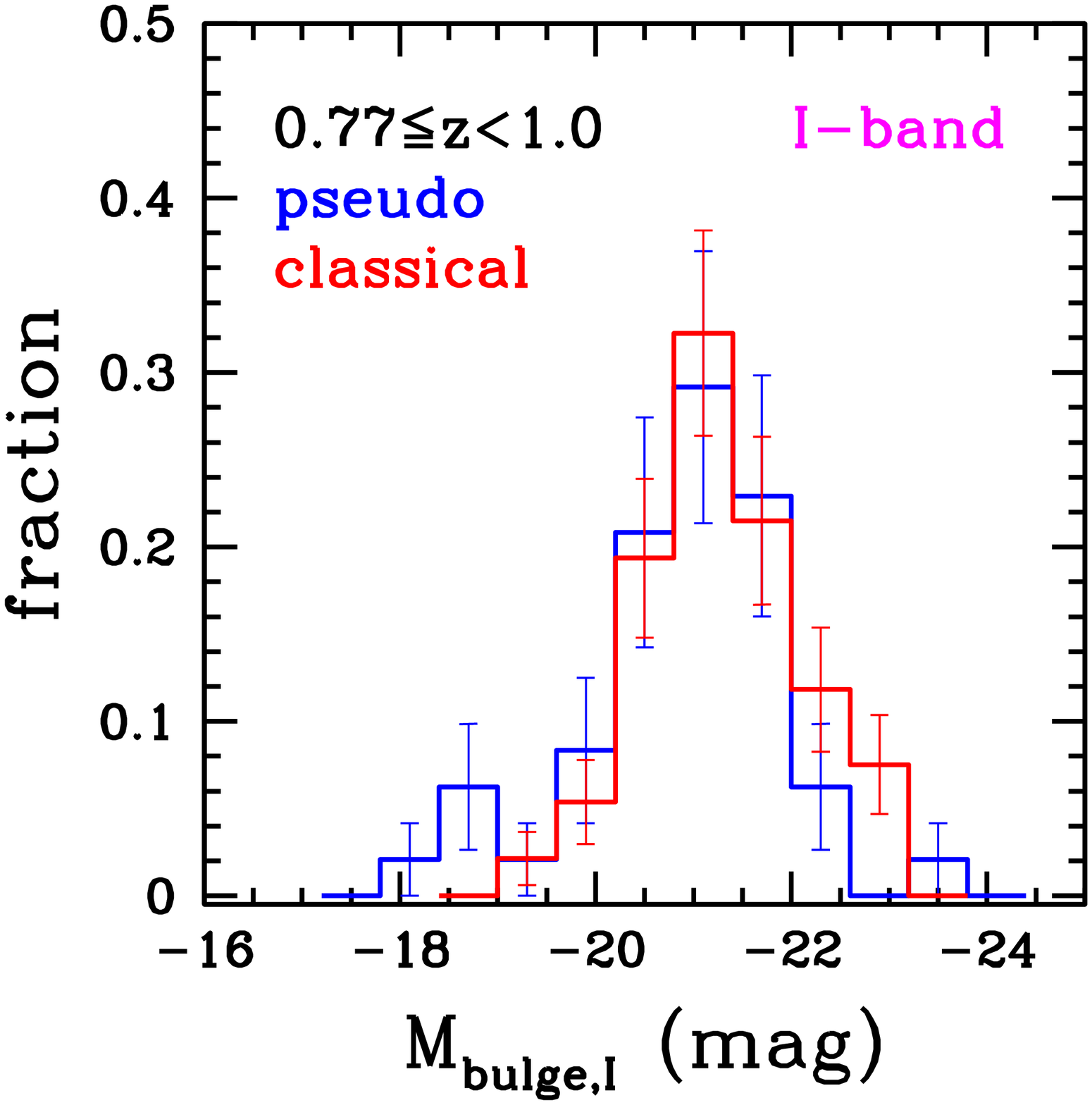}
      \includegraphics[width=55mm]{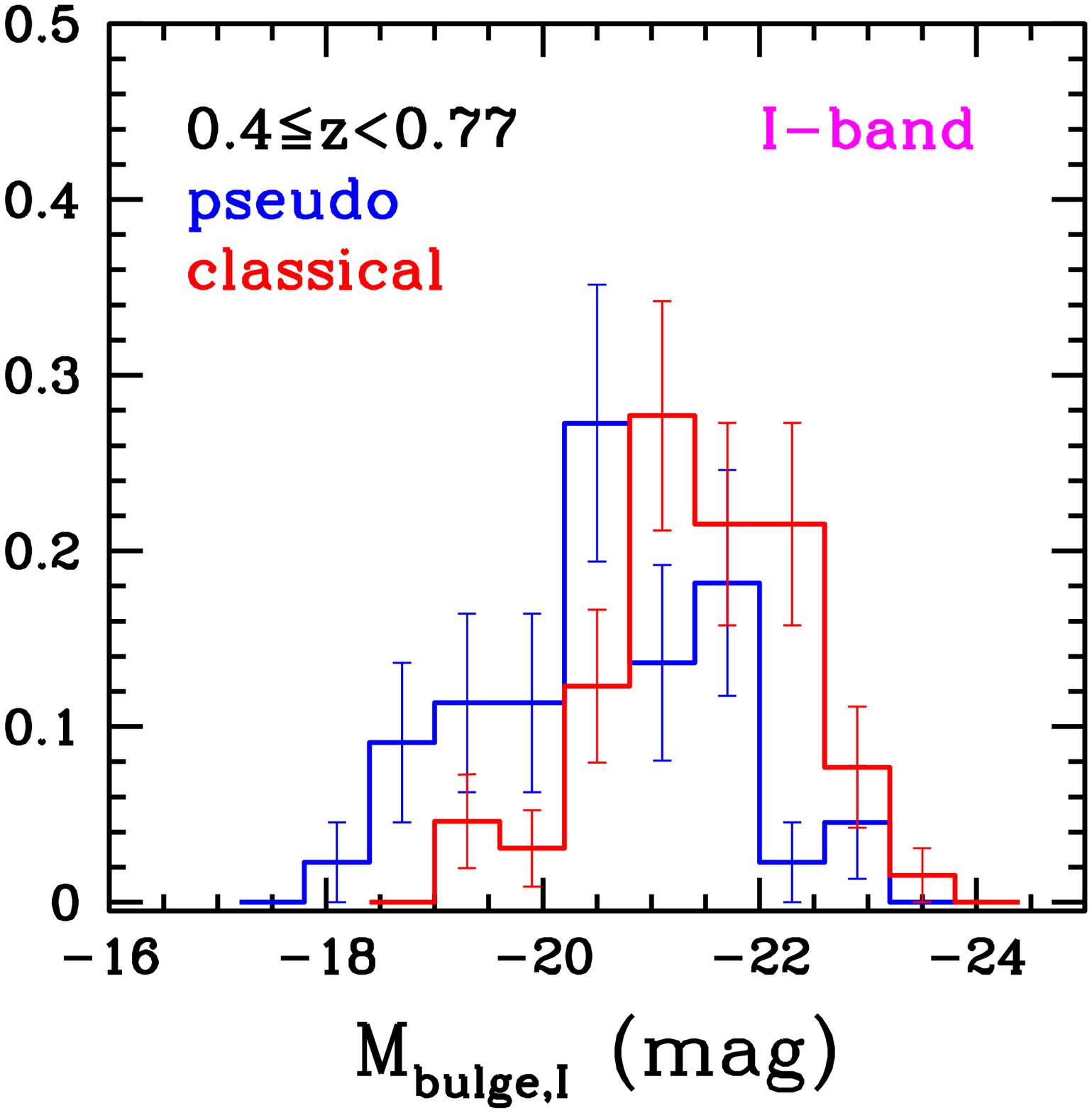}
      \includegraphics[width=55mm]{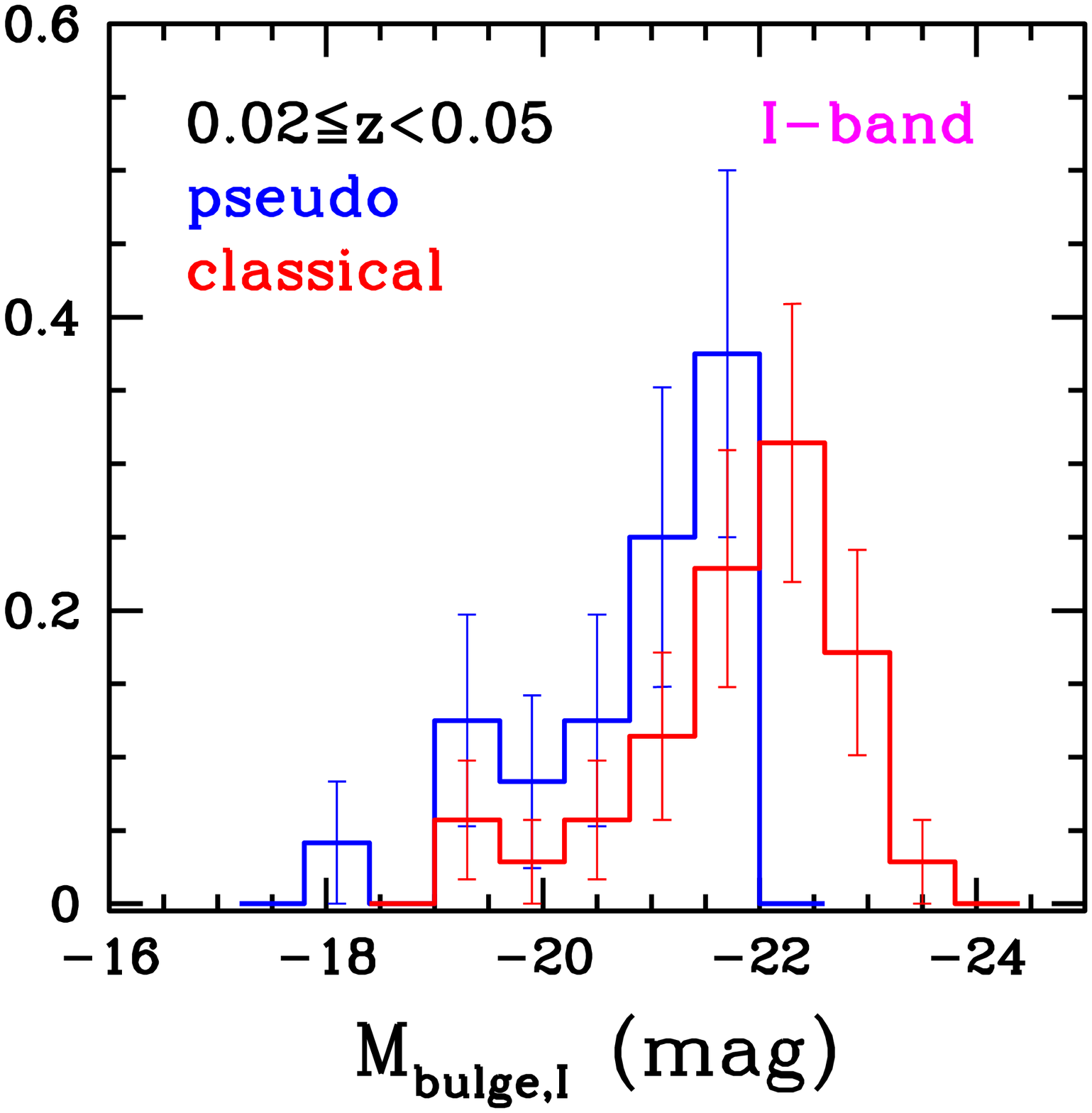}}
\caption{{\bf First row:} The distribution of bulge luminosity in rest-frame {\it B}-band (cyan) is shown in comparison to that in rest-frame {\it I}-band (magenta), for the three redshift ranges. {\bf Second row:} The distribution of bulge luminosity for pseudo bulges (blue) is shown in comparison to that for classical bulges (red), for the three redshift ranges, in rest-frame {\it B}-band. {\bf Third row:} Same as the second row, but, in rest-frame {\it I}-band.}
\label{fig:histmbulge}
\end{figure*}

\begin{figure*}
\mbox{\includegraphics[width=55mm]{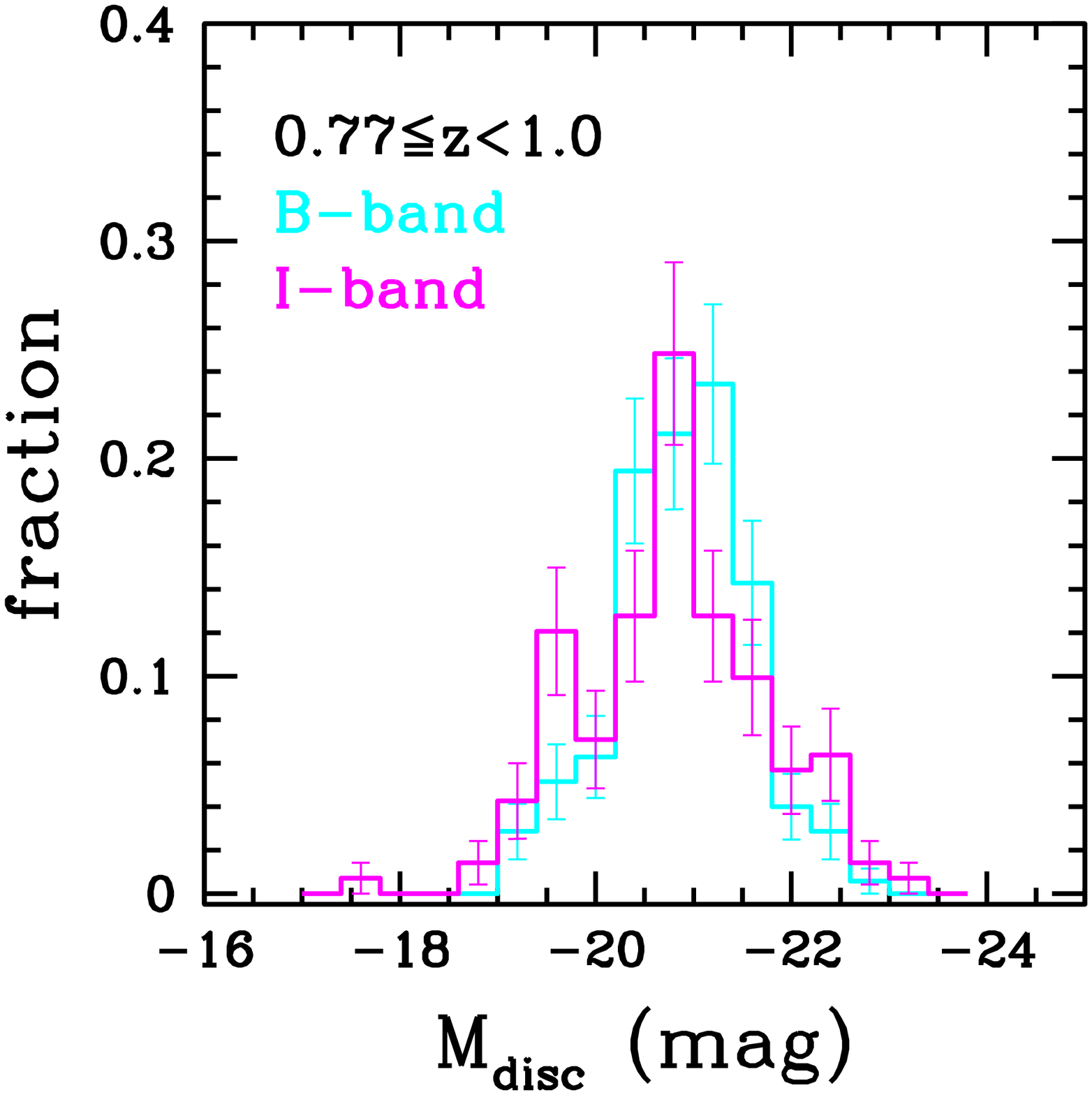}
      \includegraphics[width=55mm]{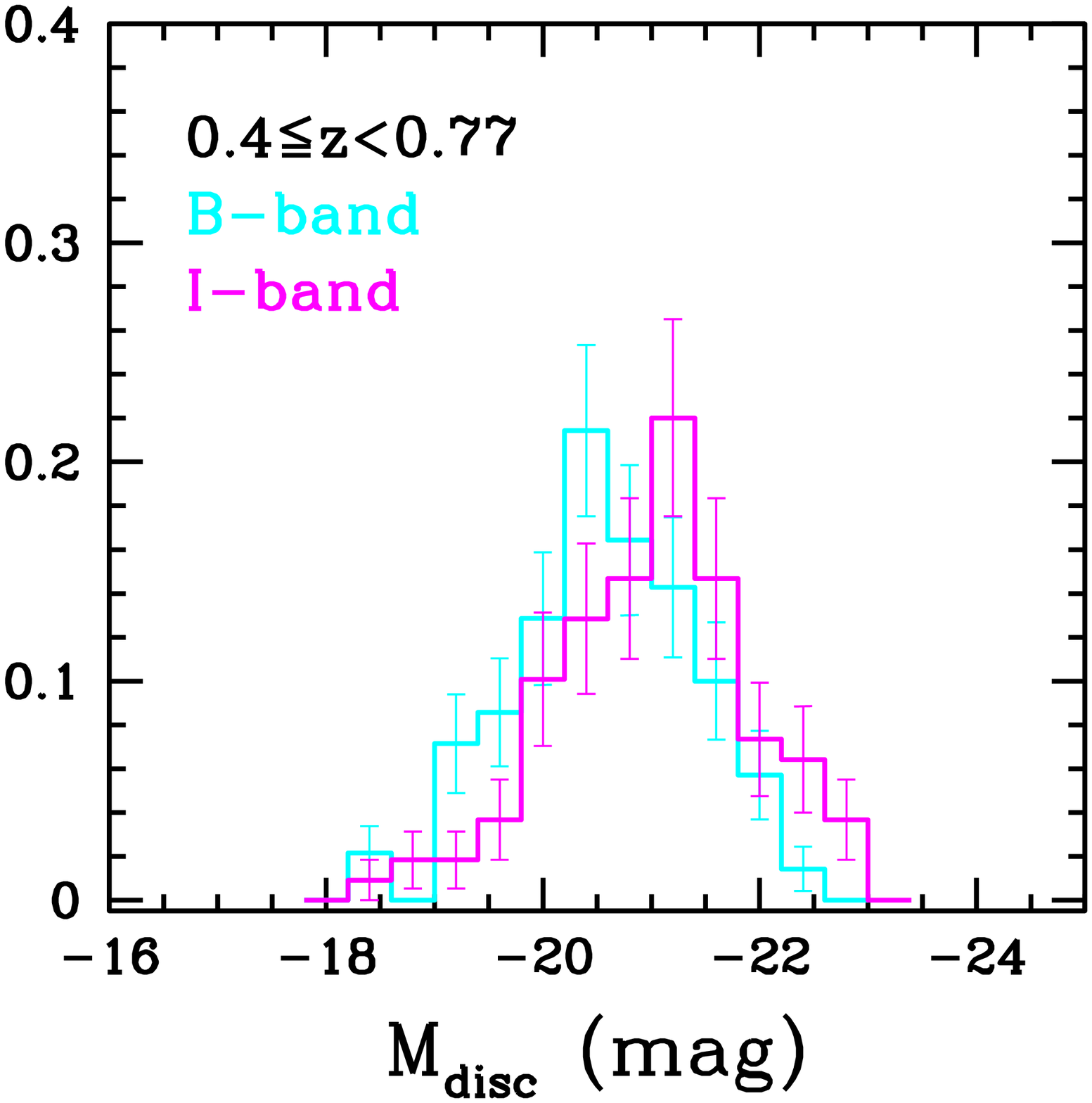}
      \includegraphics[width=55mm]{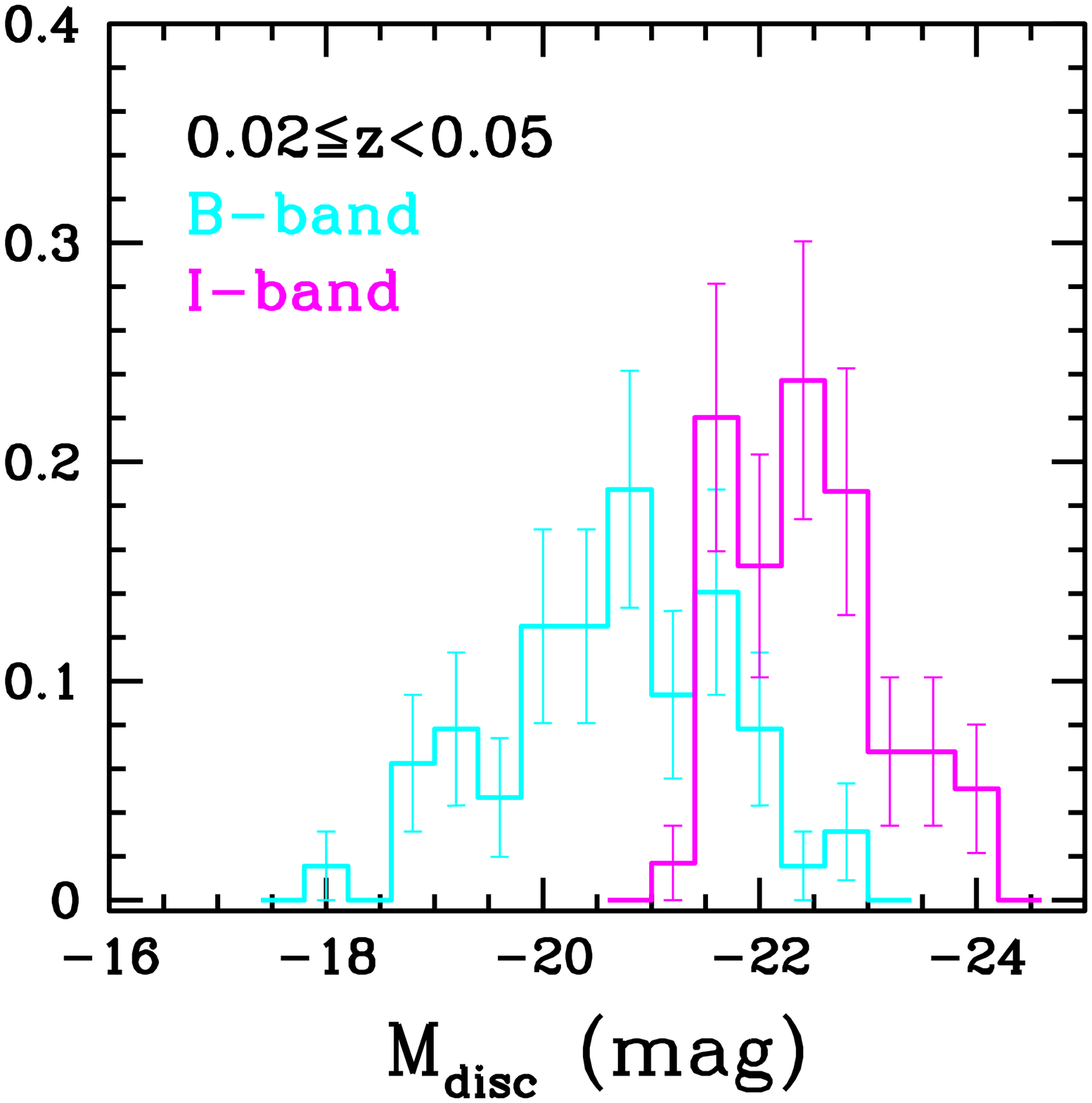}}
\mbox{\includegraphics[width=55mm]{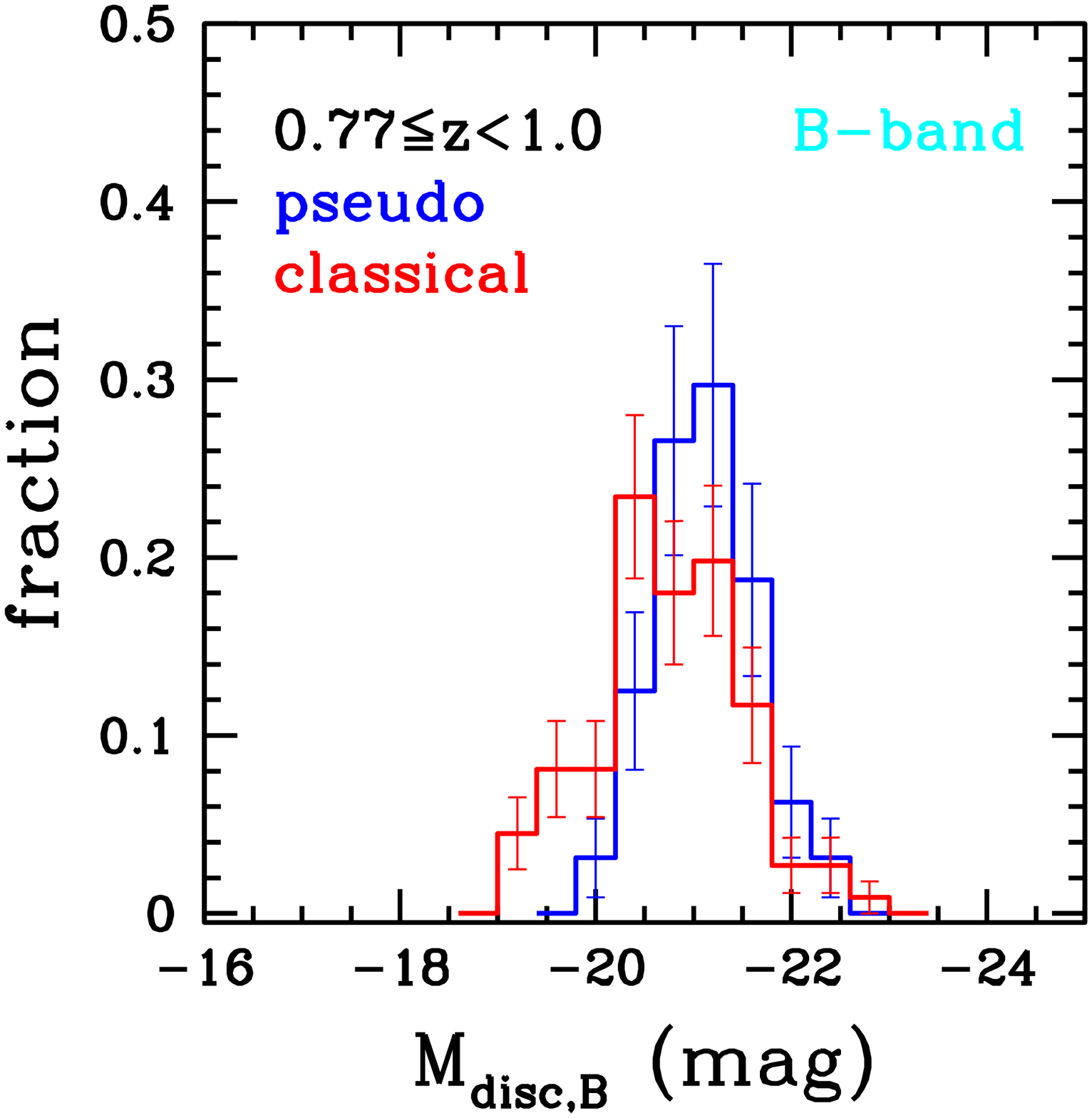}
      \includegraphics[width=55mm]{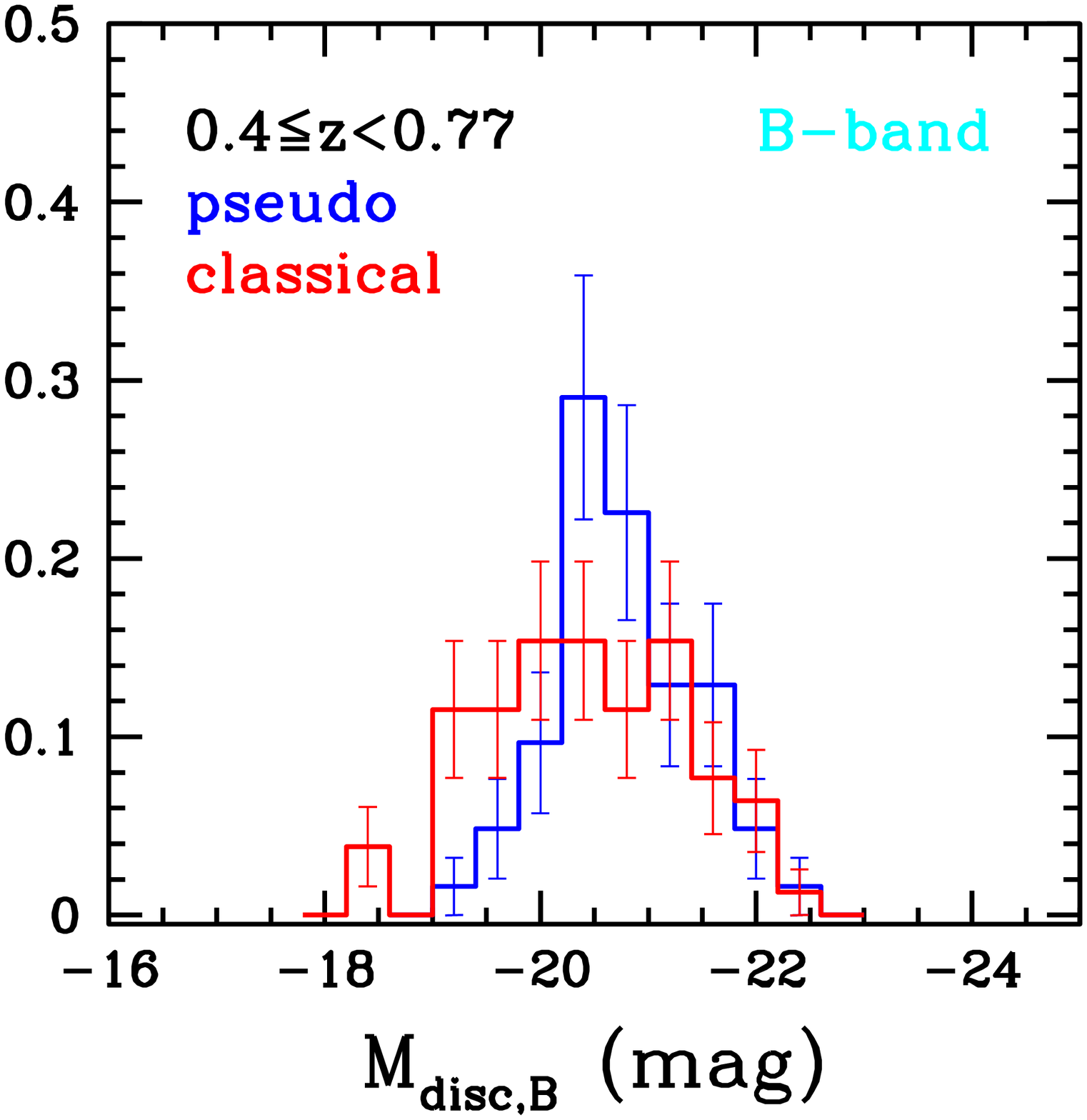}
      \includegraphics[width=55mm]{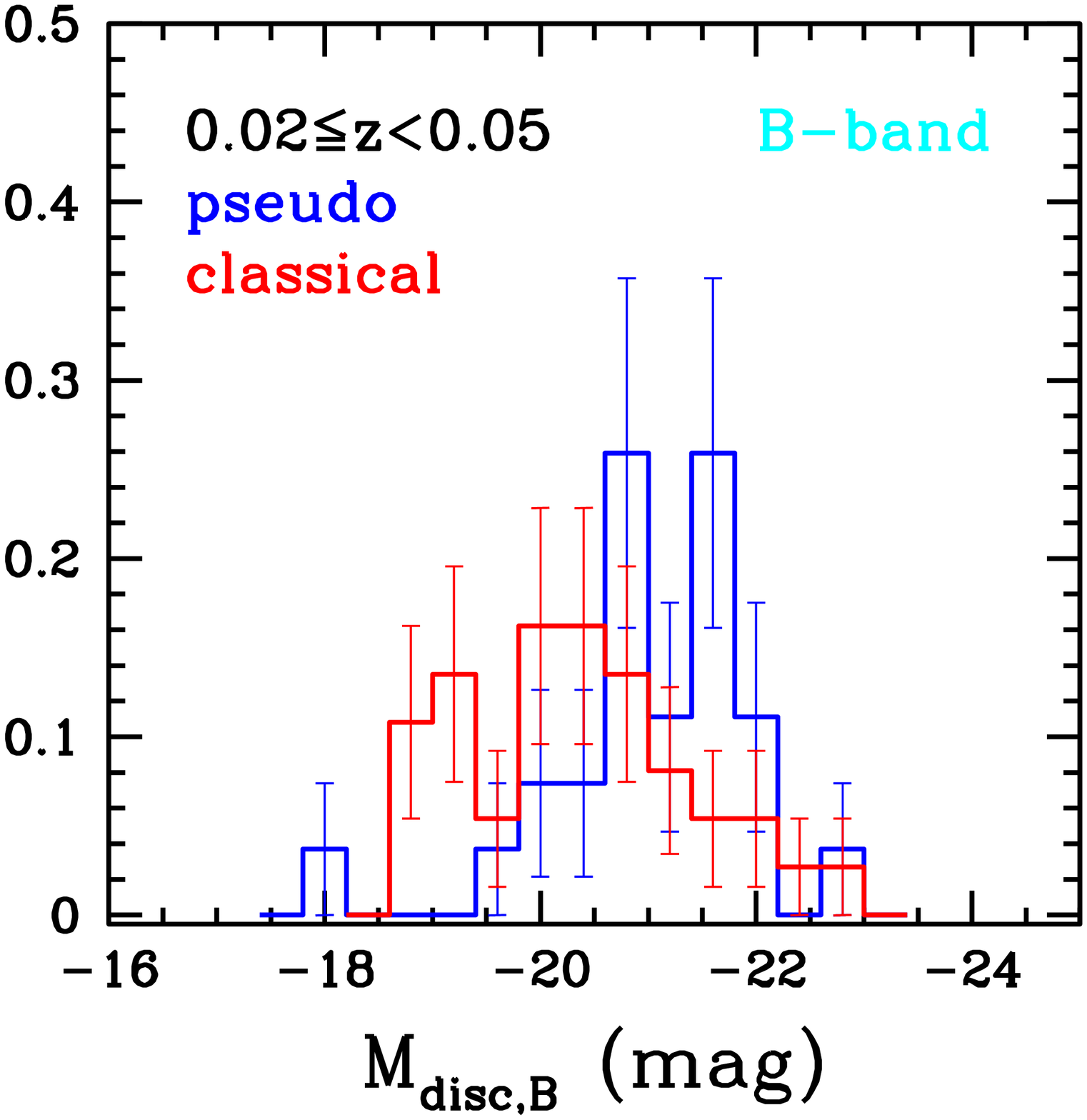}}
\mbox{\includegraphics[width=55mm]{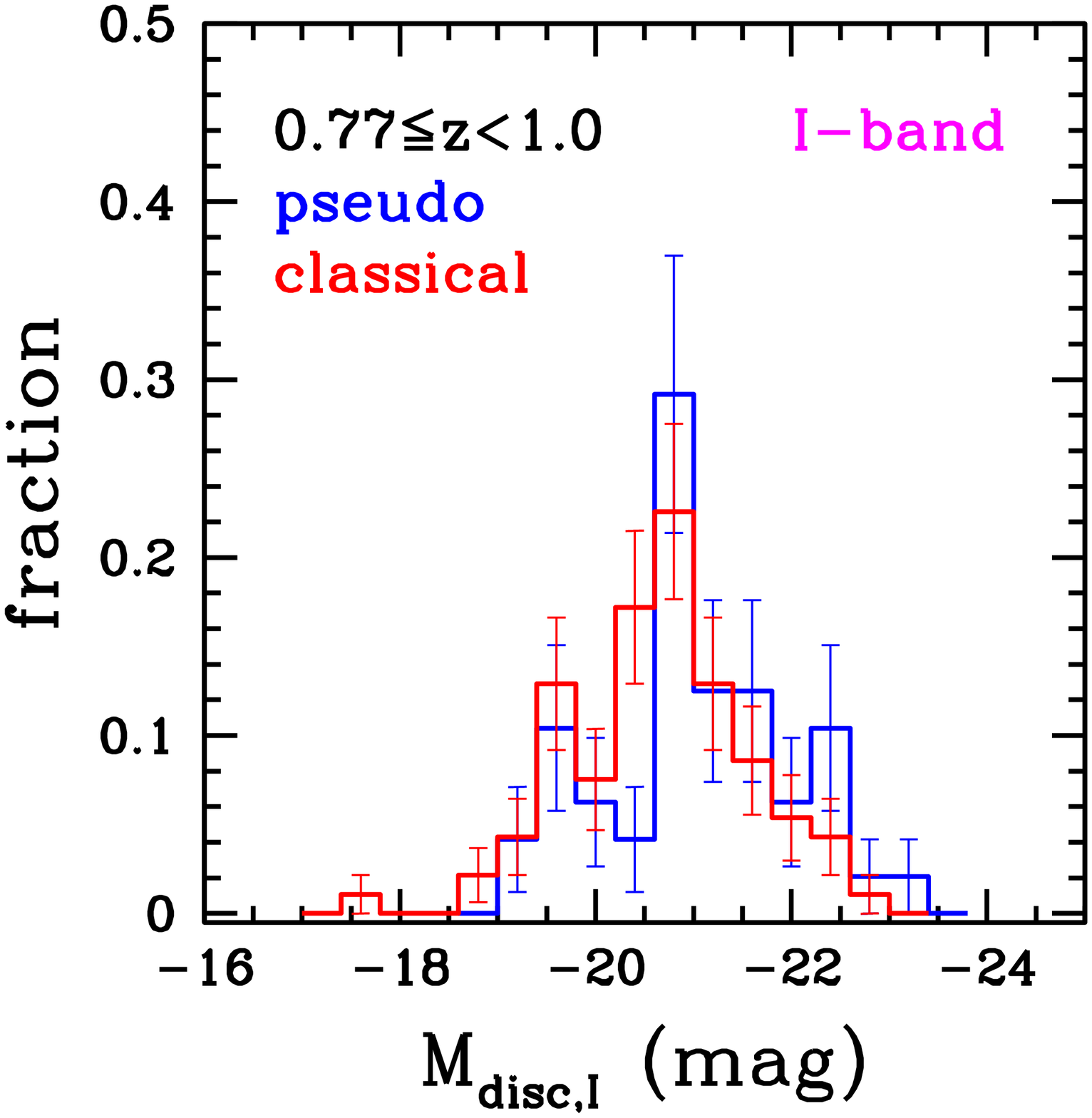}
      \includegraphics[width=55mm]{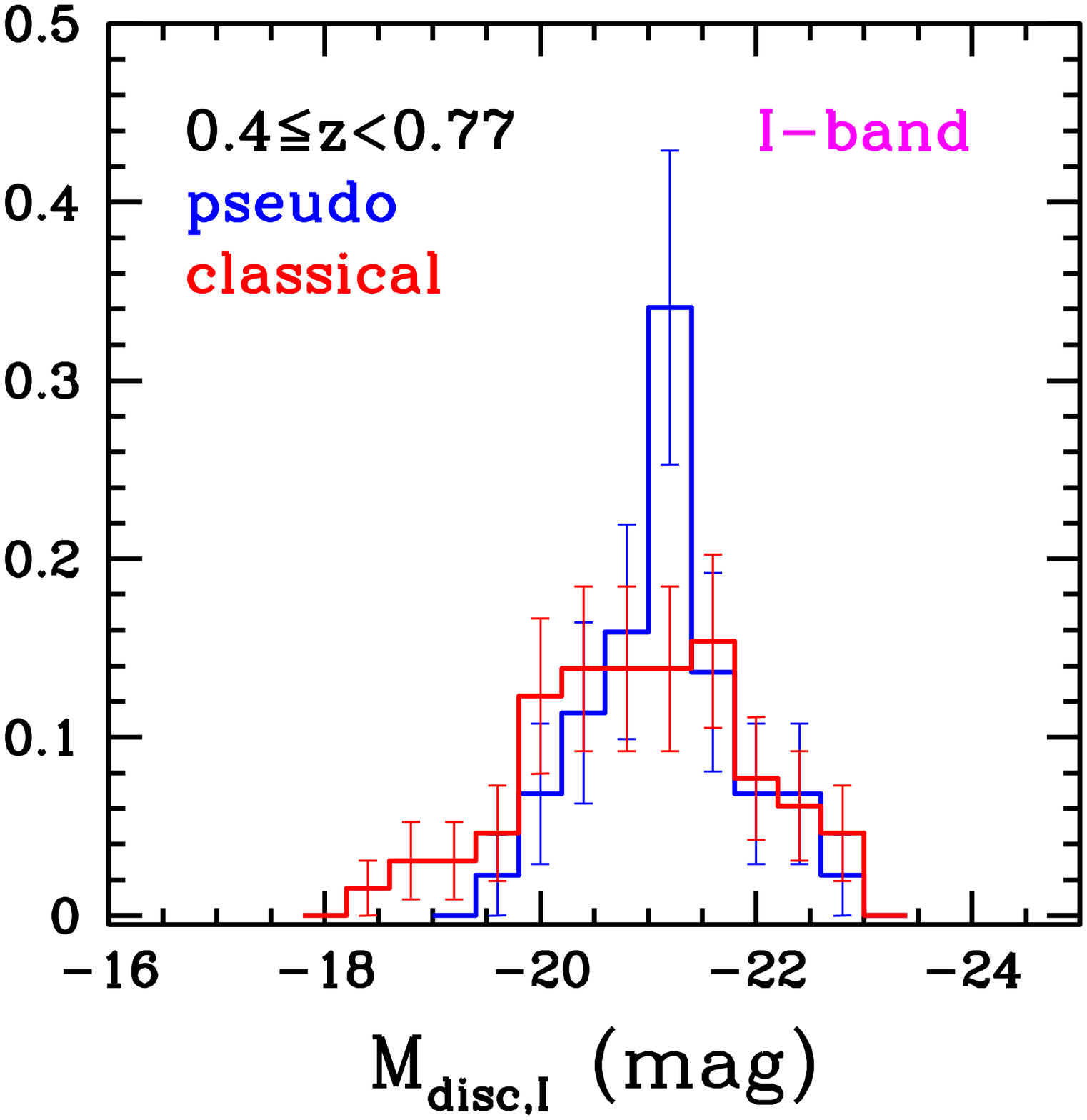}
      \includegraphics[width=55mm]{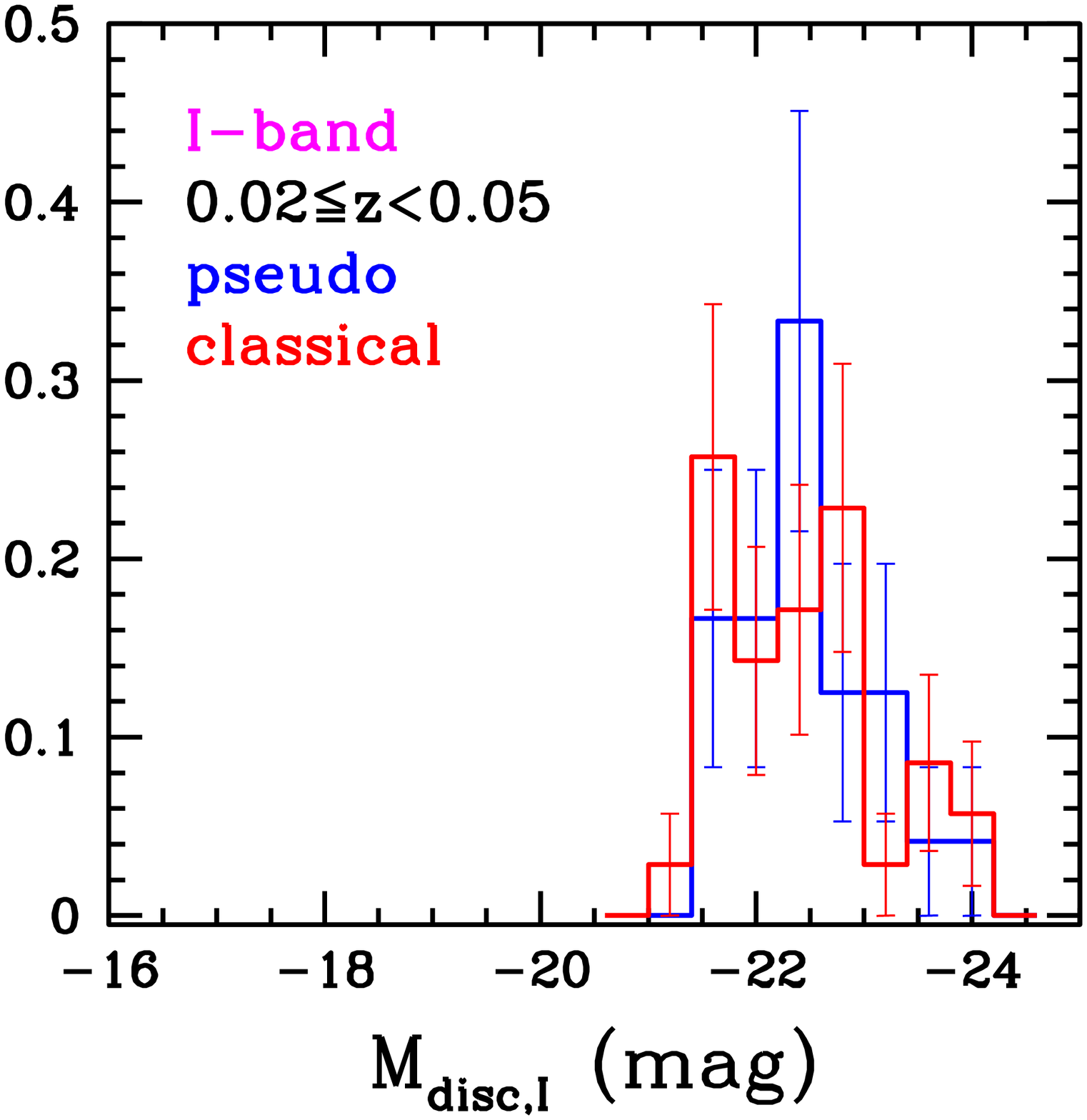}}
\caption{{\bf First row:} The distribution of host-disc luminosity in rest-frame {\it B}-band (cyan) is shown in comparison to that in rest-frame {\it I}-band (magenta), for the three redshift ranges. {\bf Second row:} The distribution of host-disc luminosity for those which host pseudo bulges (blue) is shown in comparison to those which host classical bulges (red), for the three redshift ranges, in rest-frame {\it B}-band. {\bf Third row:} Same as the second row, but, in rest-frame {\it I}-band.}
\label{fig:histmdisc}
\end{figure*}

\begin{figure*}
\mbox{\includegraphics[width=55mm]{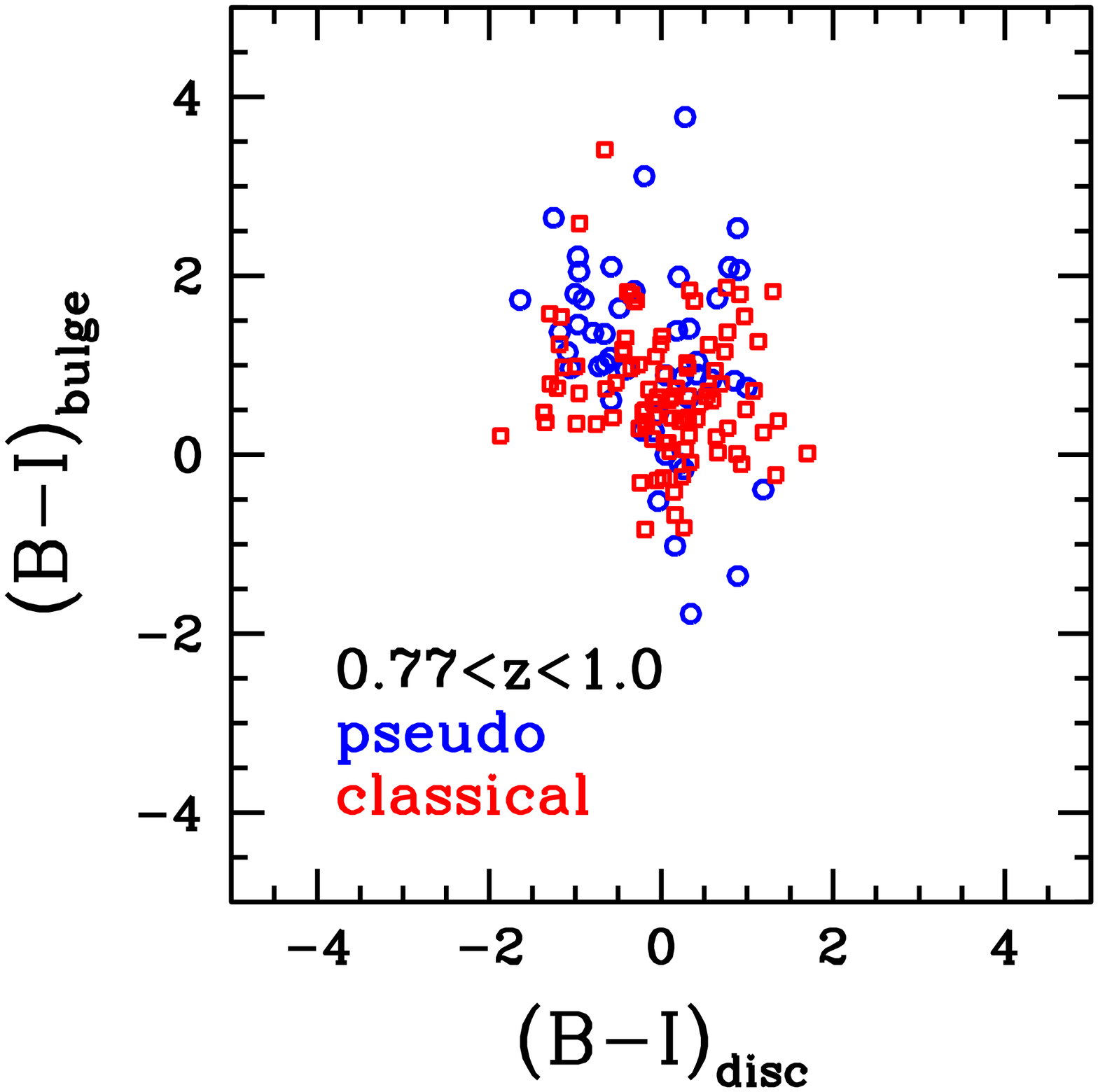}
      \includegraphics[width=55mm]{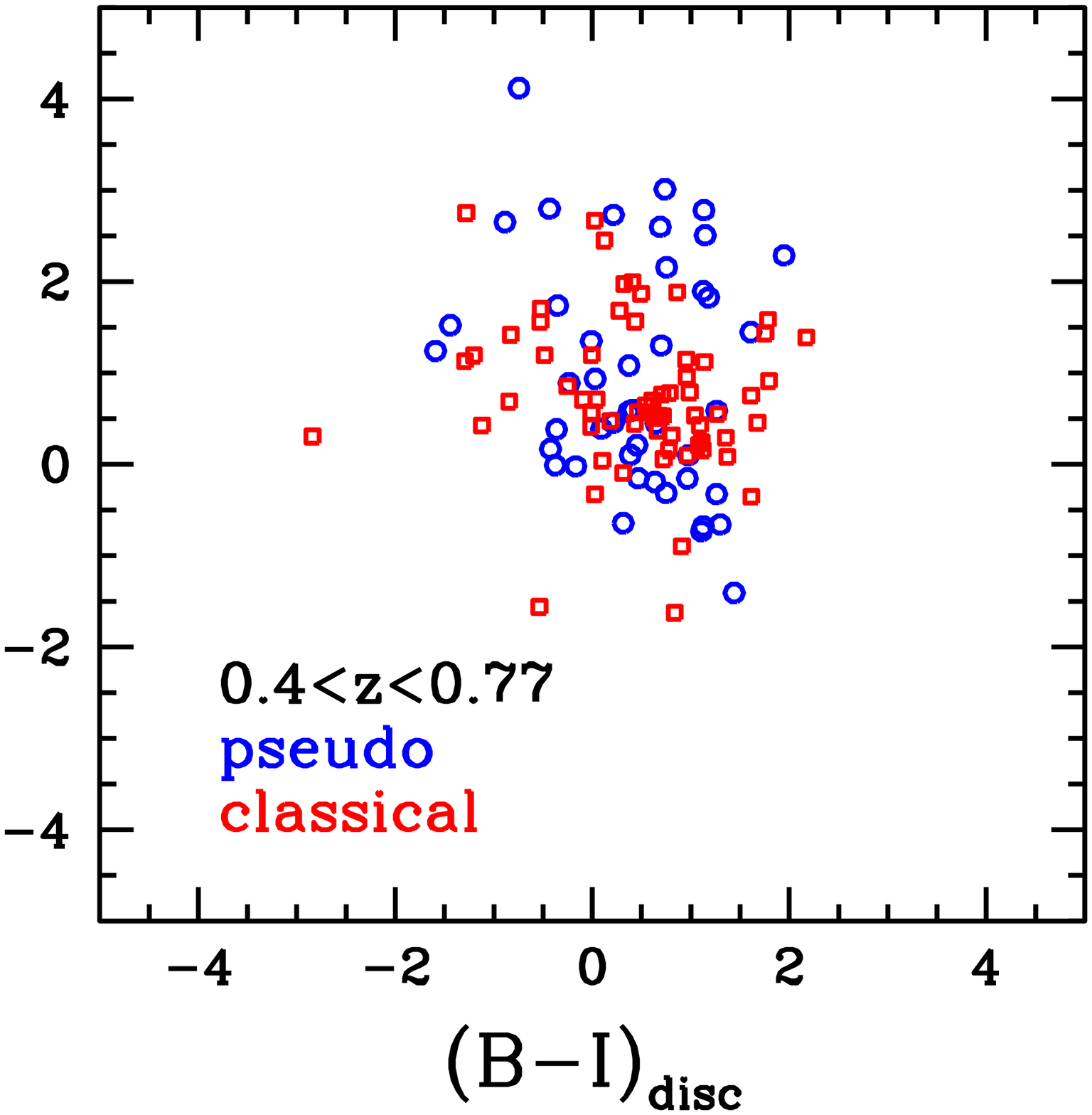}
      \includegraphics[width=55mm]{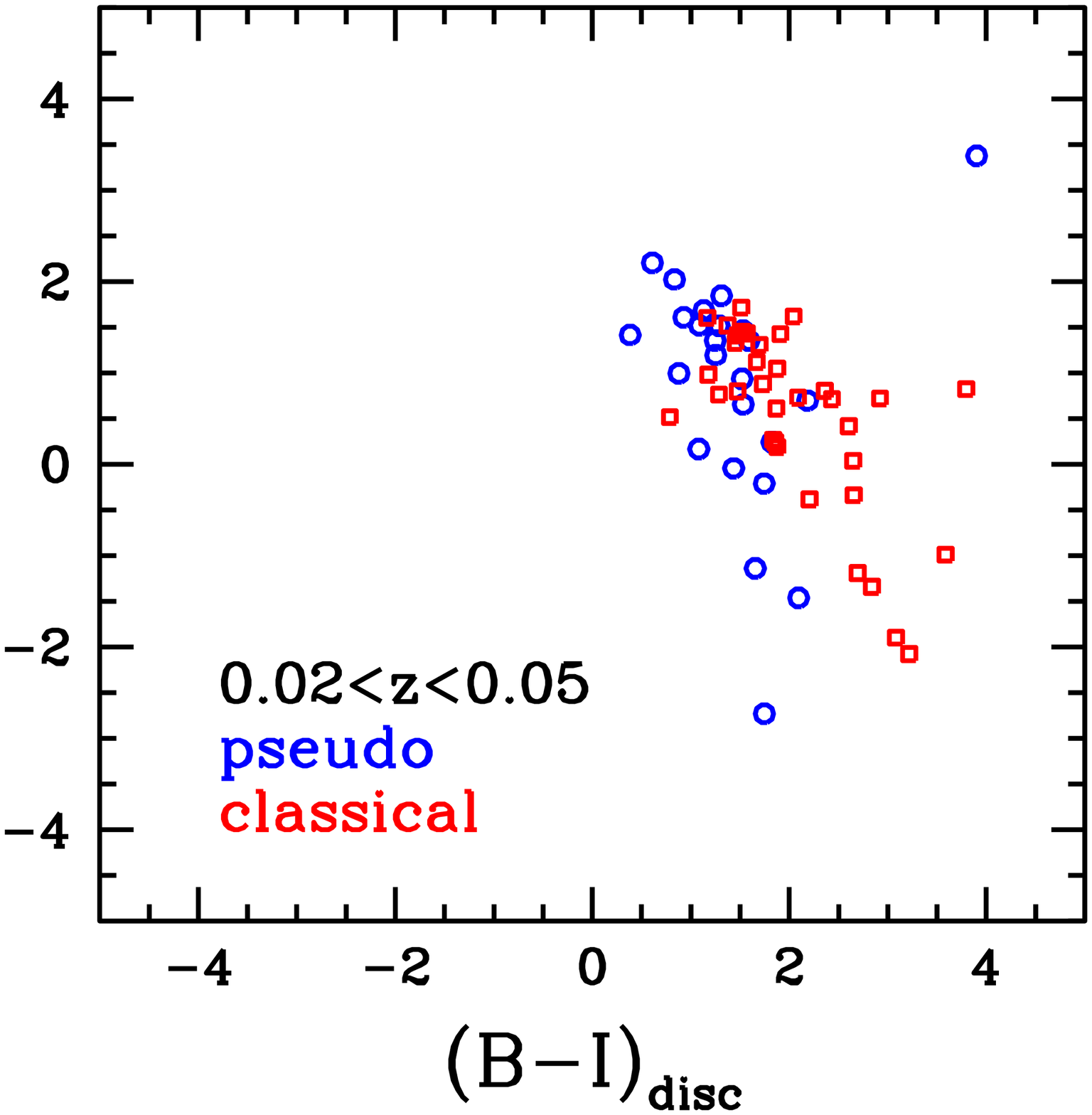}}
\caption{ The colour of the bulge is plotted against the colour of its host disc, for the three redshift ranges. Pseudo bulges have been marked with blue colour and classical bulges have been marked with red colour.}
\label{fig:colour-plots}
\end{figure*}

\section{Absolute magnitude distribution - bulge versus host disc}
\label{sec:luminosity}
In Fig.~\ref{fig:histmbulge} and Fig.~\ref{fig:histmdisc}, we show the distribution of bulge and host-disc absolute magnitude of our pseudo and classical bulge sample in rest-frame {\it B} (i.e, optical)and {\it I}-band (i.e, infrared). First consider the top panel of both the figures where we compare the absolute magnitude distribution of the total bulge population (i.e., combined pseudo and classical) in optical and infrared. We find that bulges are overall more luminous in the infrared, by about $0.5-0.6$~mag, than in optical. This lead is maintained over all redshift ranges, which is expected as old stars make the bulk of the bulge component (especially, for classicals). The behaviour of host-discs is also insightful. At high redshifts (1.0 - 0.77), host-discs are equally luminous in the optical and infrared (see top panels in Fig.~\ref{fig:histmdisc}). However, they become $\sim 1$~mag brighter in the infrared in comparison to their optical magnitude, as we reach the present epoch. This suggests the ageing of existing stars and also loss of star-forming material, i.e, clumps and gas, in the disc, which probably shifts towards the bulge as both bulge types become more luminous in the optical with the passage of time.

The middle panels of Fig.~\ref{fig:histmbulge} and Fig.~\ref{fig:histmdisc}, depicts the absolute magnitude distribution of pseudo and classical bulges and their host discs in the rest-frame {\it B}-band. In this band, pseudo bulges are less luminous than classical bulges, at all redshift ranges, however, this is not the case for their host discs. To begin with, i.e., at $0.77<z<1.0$, host-discs of the two bulge types have similar distribution (see results from KS test below). As we move to lower redshifts, classical bulge host discs faint moderately ($\sim0.5$ $mag$), whereas, pseudo bulge host discs maintain the same distribution. Thus, at local redshifts, pseudo bulge host discs become brighter than classical bulge host discs, indicating, that the loss of star-forming material has been more severe in the latter's case, probably through the migration of clumps to the centre.

In the last panels of Fig.~\ref{fig:histmbulge} and Fig.~\ref{fig:histmdisc}, we show the distribution of absolute magnitude of pseudo and classical bulges and their host discs in rest-frame {\it I}-band. Both bulge types are equally luminous in infrared at the highest redshift range (1.0 - 0.77). Note that, this is in contrast with the optical where classical bulges were always brighter than pseudobulges. Later, i.e, as we reach $z=0$, classical bulges become more luminous than their pseudo counterparts, in the infrared as well. Thus, the bulges had an equal amount of old stellar population at earlier redshifts, it is only with time that classical have become more populous with them. Interestingly, the distribution of absolute magnitude for pseudo and classical bulge host discs overlaps at all redshift ranges. Both witness significant increase in their absolute magnitude ($\sim1$ $mag$), as we move to $z=0$, however, the increase is equal for the two bulge types, i.e., they become equally redder with time. Thus, the evolution of the host-discs, in infrared, is irrespective of the bulge they host.

We perform the Kolmogorov-Smirnov test (KS-test) to check if the sample sets, i.e, pseudo and classical bulges and their host discs, in rest-frame {\it B} and {\it I}-band, actually differ or are part of the same distribution. We find that at $95\%$ confidence level, i.e., $\alpha=0.05$, host discs of pseudo and classical bulges differ considerably from each other in the optical, however, the two samples are quite alike in the infrared, at all redshift ranges. The KS-test results are highly consistent with the more sensitive Anderson Darling Test (AD-test), which we additionally performed to confirm the findings (see Table.~\ref{ks-test}). Here also, host discs are similar in the infrared and differ in the optical to $95\%$ confidence level. This asserts the result found above that, in infrared host-discs evolve irrespective of the bulge they host. 

Comparing the bulges, we find that the samples differ, again at $95\%$ confidence level, both for optical and infrared, at all redshift ranges. As seen from the table, the KS and AD statistic values (KS-stat, AD-stat) are much higher than the critical values (KS-crit, AD-crit) at $\alpha=0.05$, and the probability (p) of the two samples coming from the same distribution is much less than 0.05, i.e., 5\%. However, interestingly, at the highest redshift range (0.77-1.0), the bulge samples are similar in the infrared (see Table.~\ref{ks-test}). This also affirms the result found above that the two bulge types are equally bright in the infrared to begin with, it is only later that classical bulges becomes more populous with older stellar population.

In Fig.~\ref{fig:colour-plots}, we plot the integrated colour of the bulge against the colour of it's host disc, for the three redshift ranges. At the highest redshift range (0.77-1.0), while the host discs appear to be equally blue and red, the bulges are dominantly redder in colour. As we proceed towards the intermediate redshift range (0.4-0.77), while the bulge component maintains it's colour position, the host disc shifts towards the redder end. As we reach the local epoch, the bulge population continues to reside in the same colour domain as it was at high redshifts, nevertheless, the host discs become even redder and the effect is more pronounced for classical bulge host discs. We thereby interpret that the host discs are losing star-forming material with time, which eventually adds to the bulge. Also, the effect being more severe for discs which end up as classical bulges, suggests the migration of star-forming clumps.  

\begin{table}
\begin{minipage}{80mm}
\caption{KS and AD-test for pseudo(ps) and classical(cl)\\bulge and disc luminosities}
\centering
\begin{tabular}{@{}llllll@{}}
\hline
Redshift & Size & KS & p & KS-crit & AD$^*$\\
 & ps,cl & stat & value & ($\alpha$$=$$0.05$) & stat\\
\hline
R.f. {\bf {\it B}-band}\\bulge\\
\hline
0.77-1.0 & 64,111 & 0.6822 & 0.0000 & 0.2134 & 47.725\\
0.4-0.77 & 62,78 & 0.6944 & 0.0000 & 0.2314 & 44.818\\
0.02-0.05 & 27,37 & 0.9459 & 0.0000 & 0.3442 & 30.647\\
\hline
R.f. {\bf {\it B}-band}\\disc\\
\hline
0.77-1.0 & 64,111 & 0.3212 & 0.0003 & 0.2134 & 8.323\\
0.4-0.77 & 62,78 & 0.2812 & 0.0065 & 0.2314 & 4.494\\
0.02-0.05 & 27,37 & 0.4805 & 0.0009 & 0.3442 & 5.133\\
\hline
R.f. {\bf {\it I}-band}\\bulge\\
\hline
0.77-1.0 & 48,93 & 0.1714 & 0.2829 & 0.2417 & 1.672\\
0.4-0.77 & 44,65 & 0.4591 & 0.0000 & 0.2655 & 11.918\\
0.02-0.05 & 24,35 & 0.5143 & 0.0006 & 0.3604 & 8.322\\
\hline
R.f. {\bf {\it I}-band}\\disc\\
\hline
0.77-1.0 & 48,93 & 0.2124 & 0.0997 & 0.2417 & 2.254\\
0.4-0.77 & 44,65 & 0.2175 & 0.1454 & 0.2655 & 1.216\\
0.02-0.05 & 24,35 & 0.2333 & 0.3721 & 0.3604 & 0.259\\
\hline
\label{ks-test}
\end{tabular}
$^*$The value for AD critical for $\alpha$$=$$0.05$ is 3.752.
\end{minipage}
\end{table}


\begin{figure*}
\mbox{\includegraphics[width=55mm]{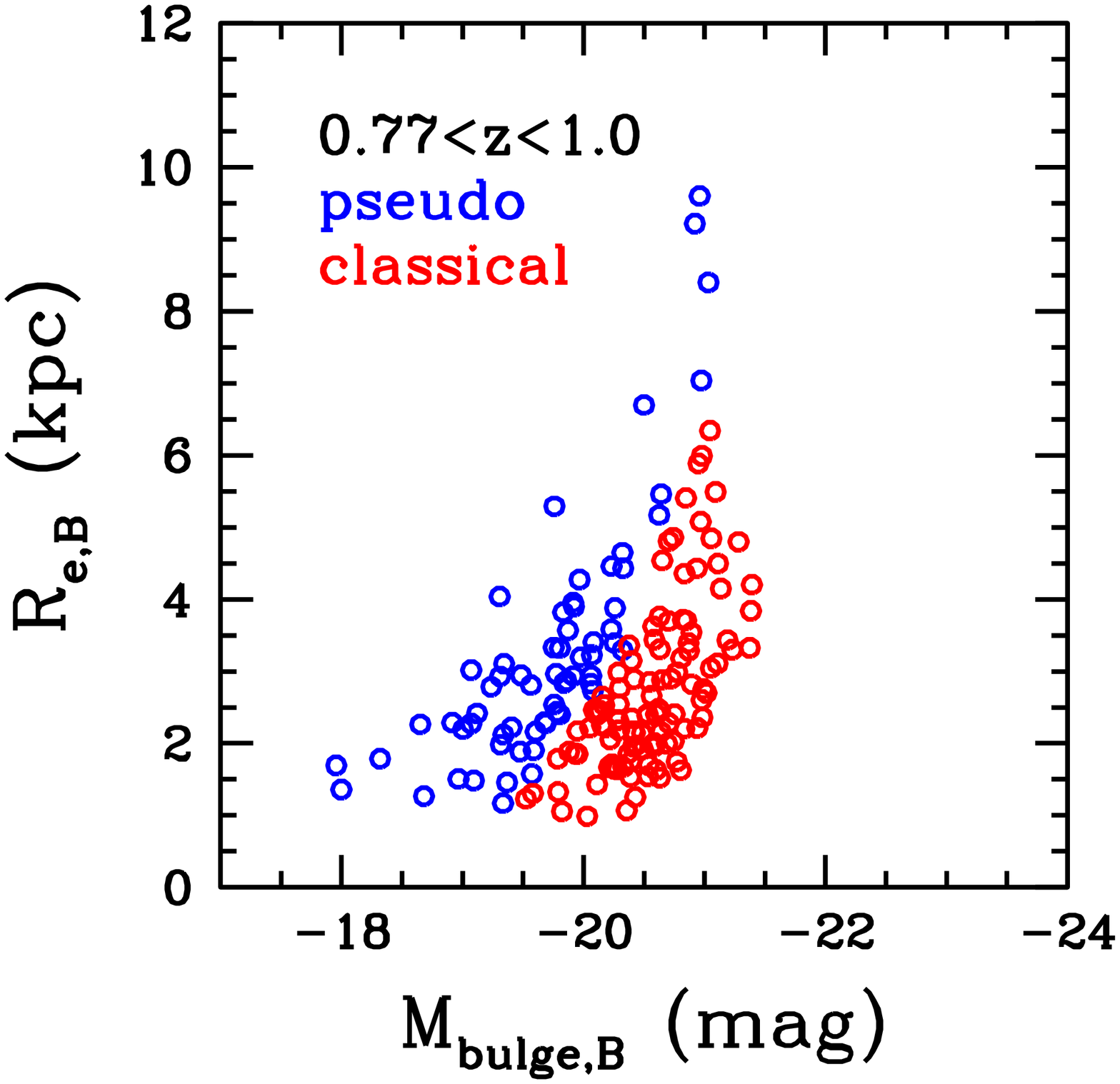}
      \includegraphics[width=55mm]{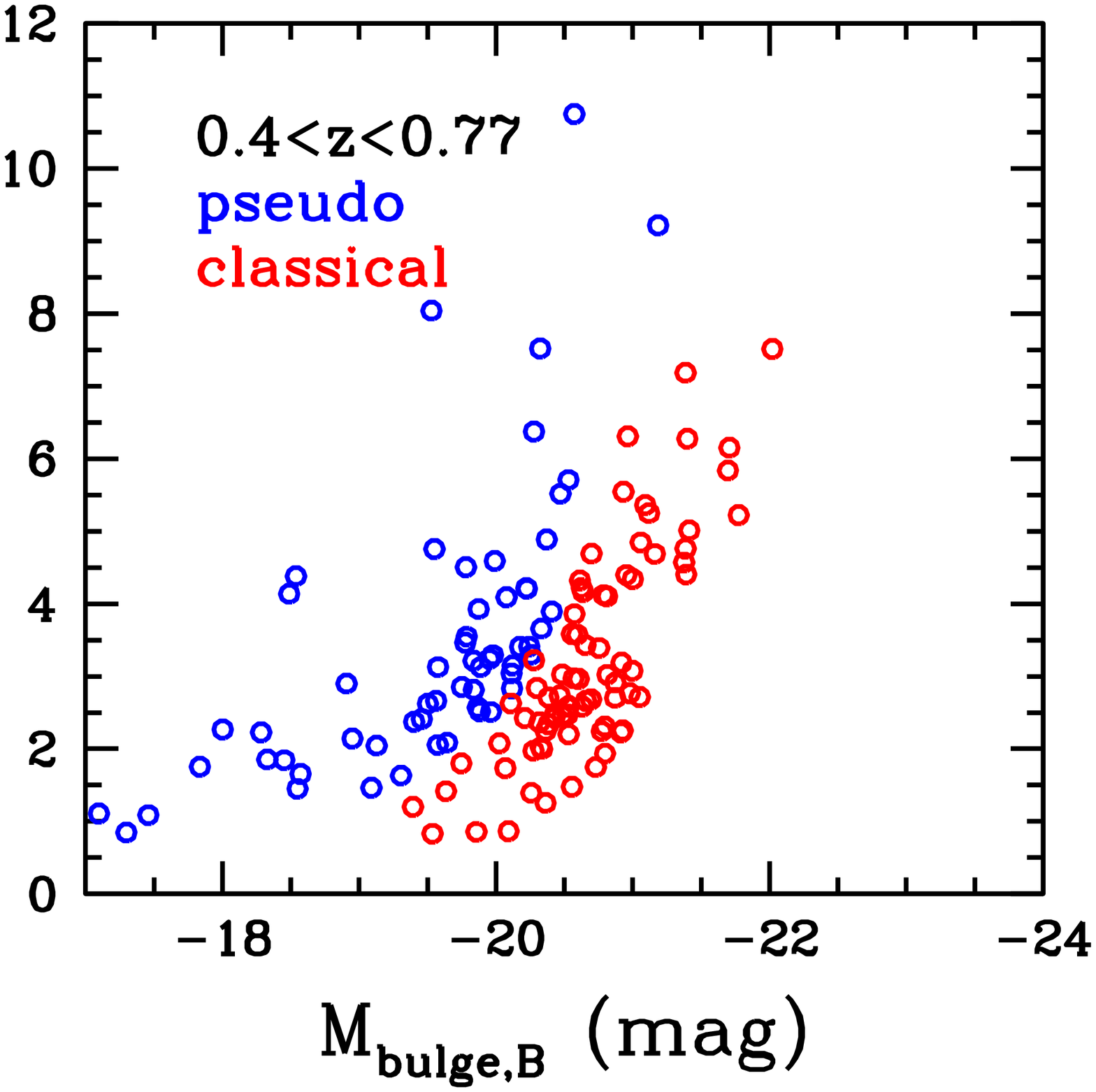}
      \includegraphics[width=55mm]{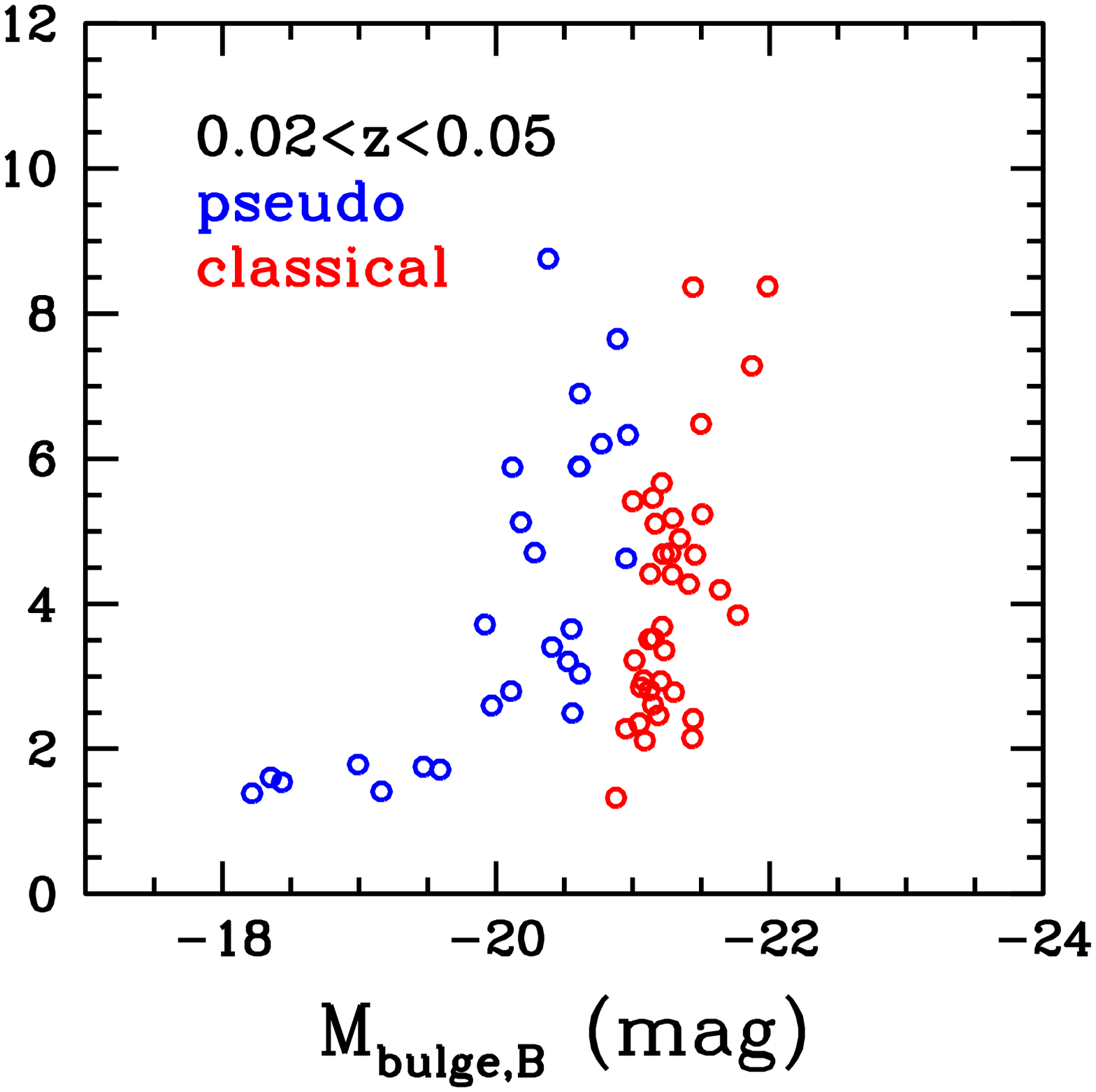}}
\mbox{\includegraphics[width=55mm]{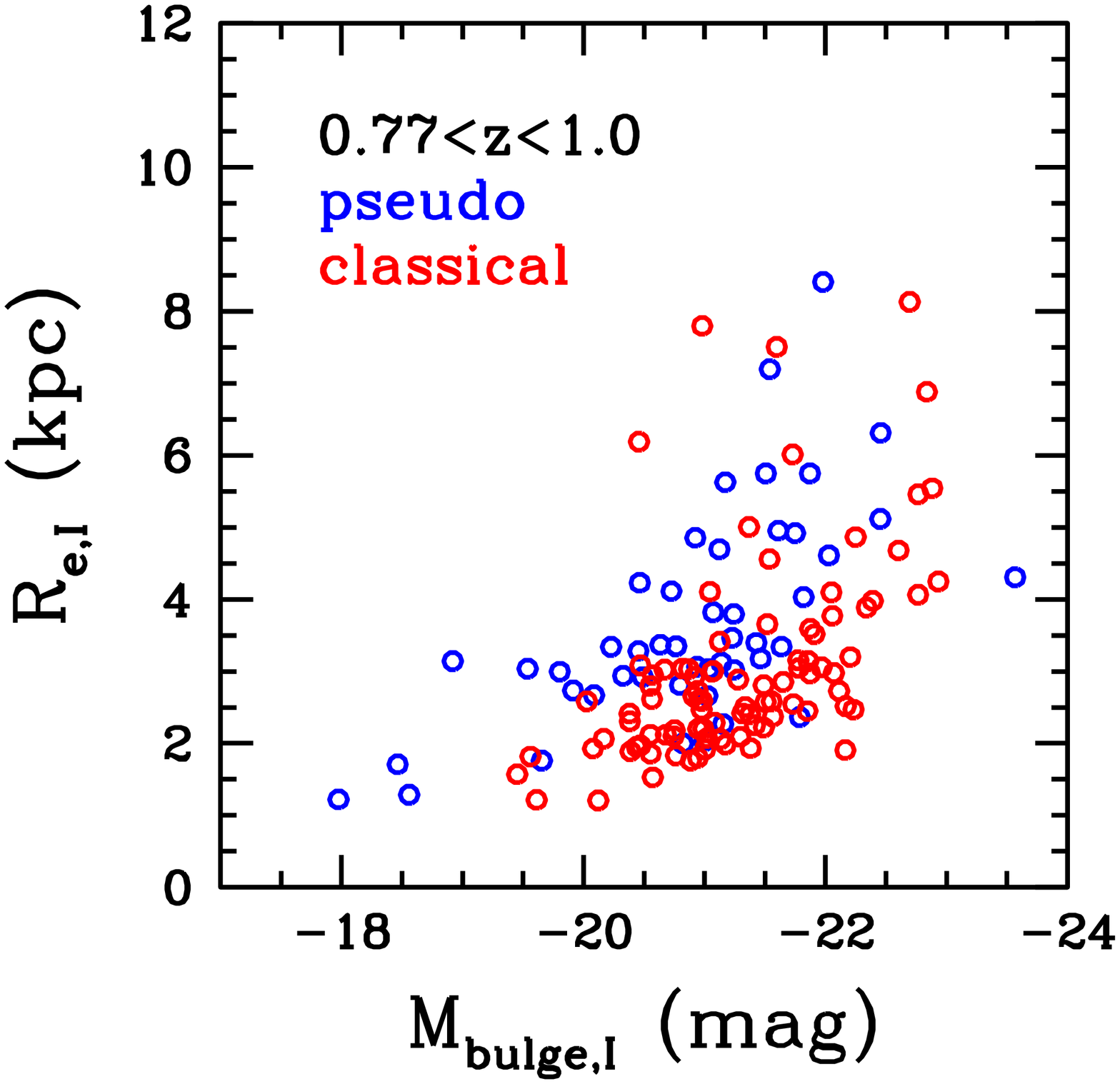}
      \includegraphics[width=55mm]{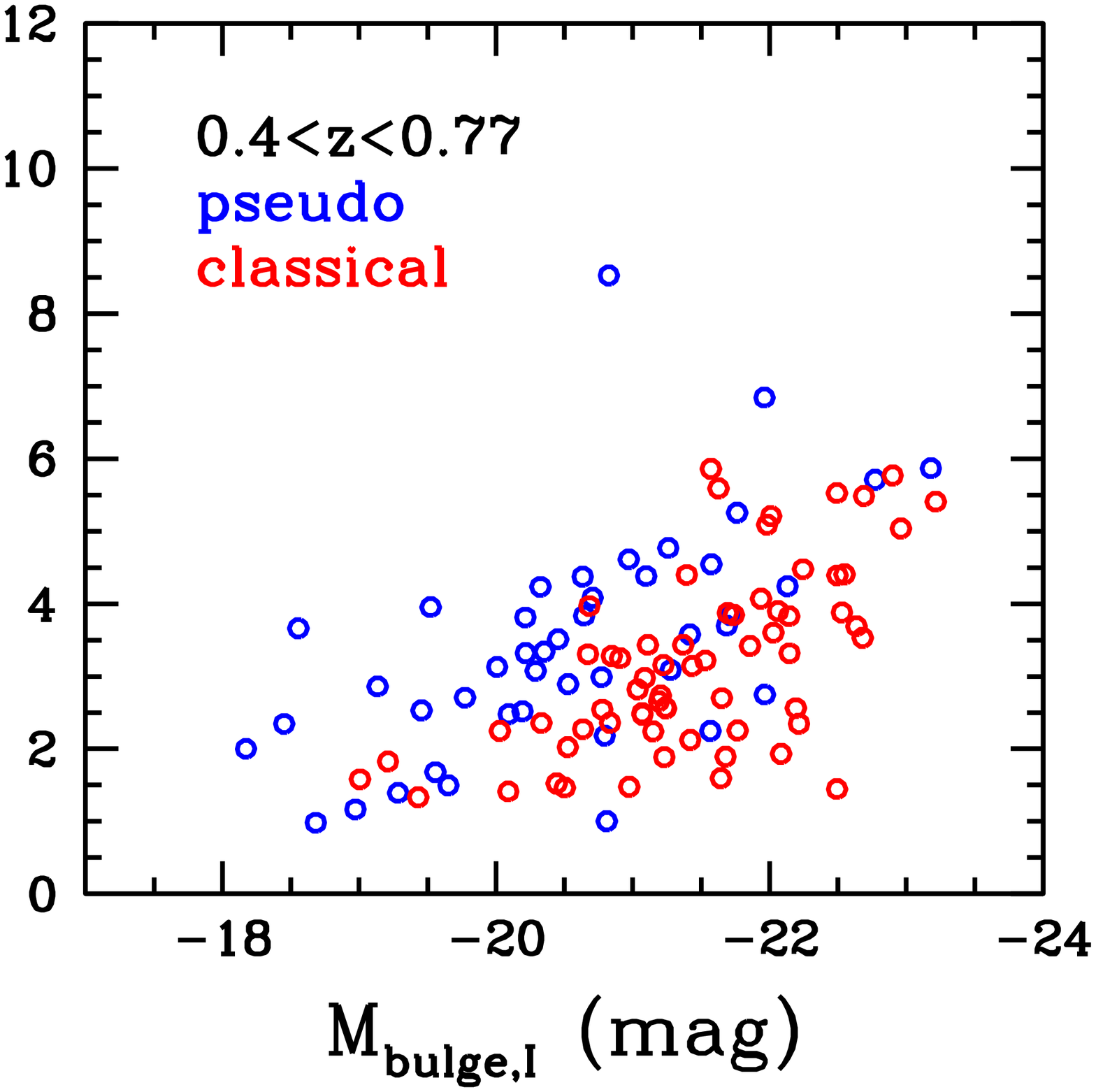}
      \includegraphics[width=55mm]{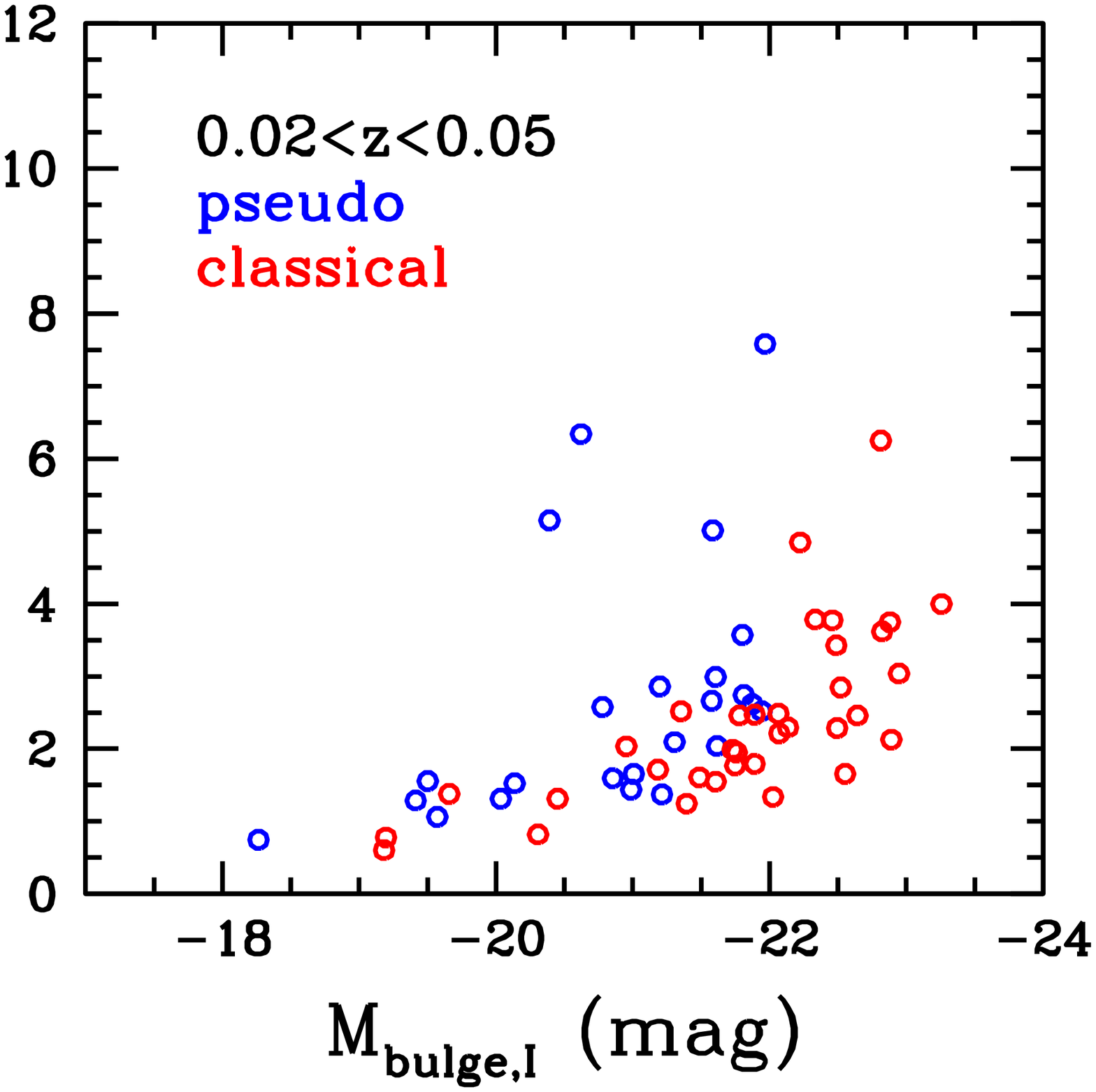}}
\mbox{\includegraphics[width=55mm]{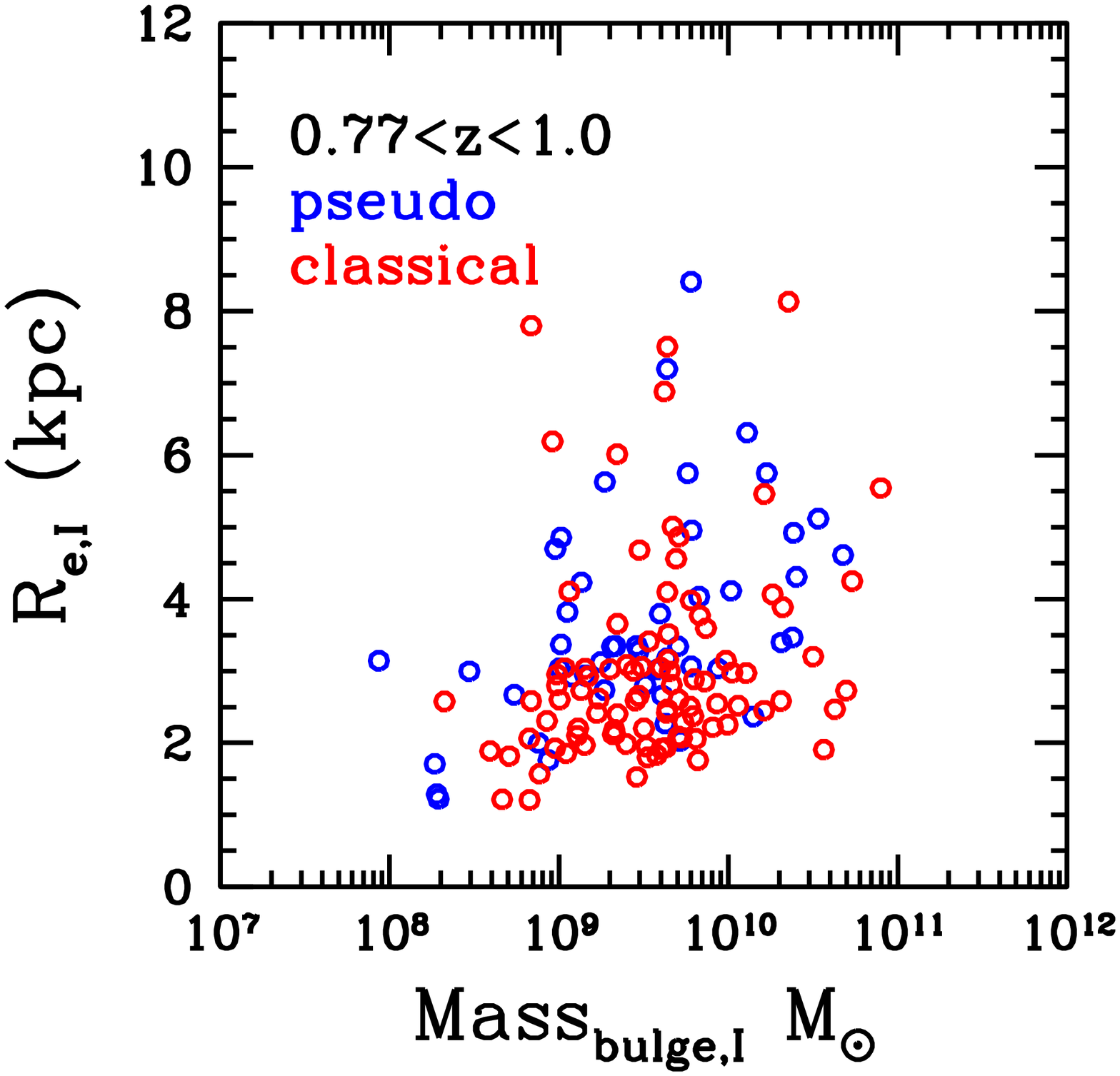}
      \includegraphics[width=55mm]{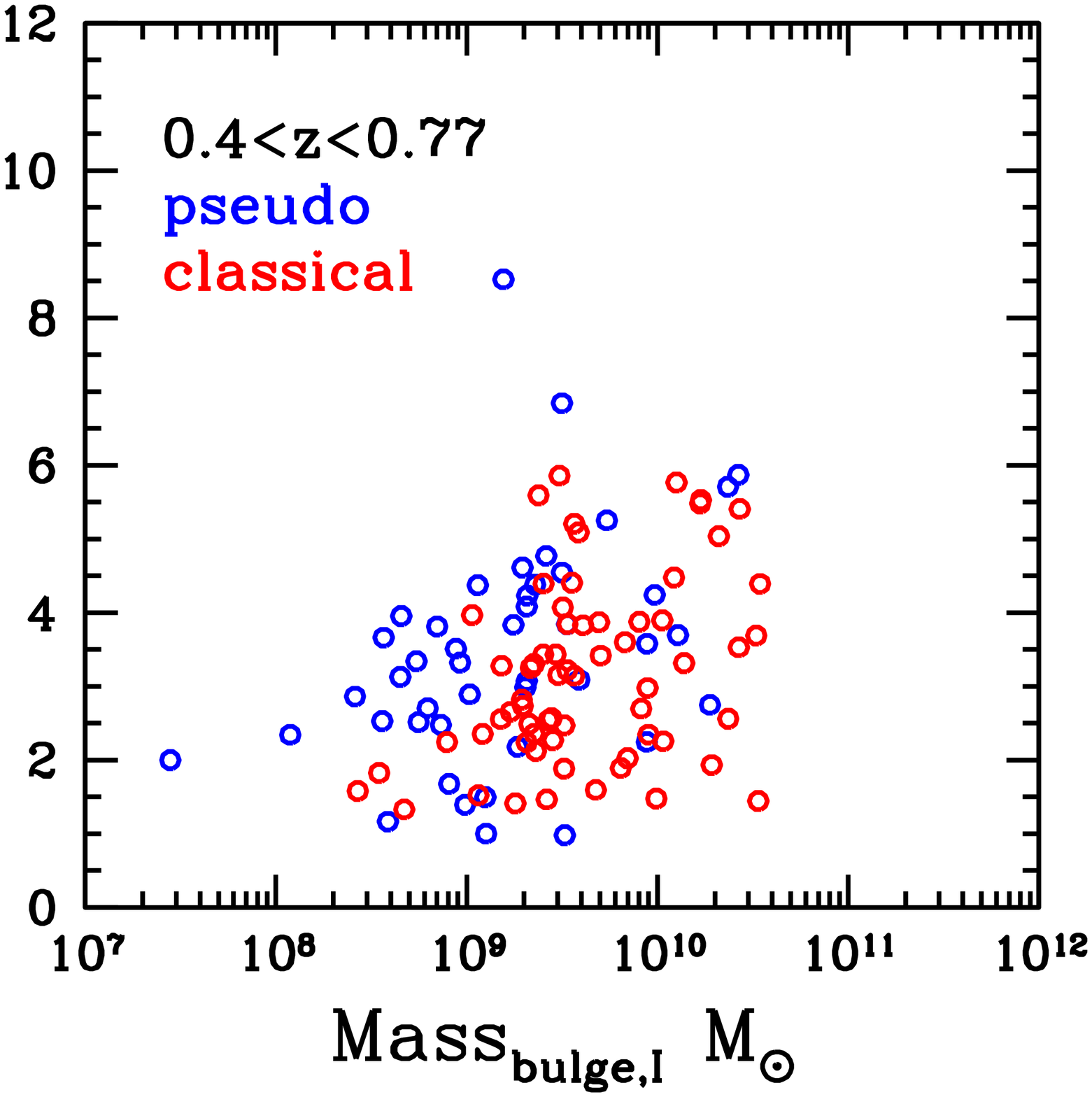}
      \includegraphics[width=55mm]{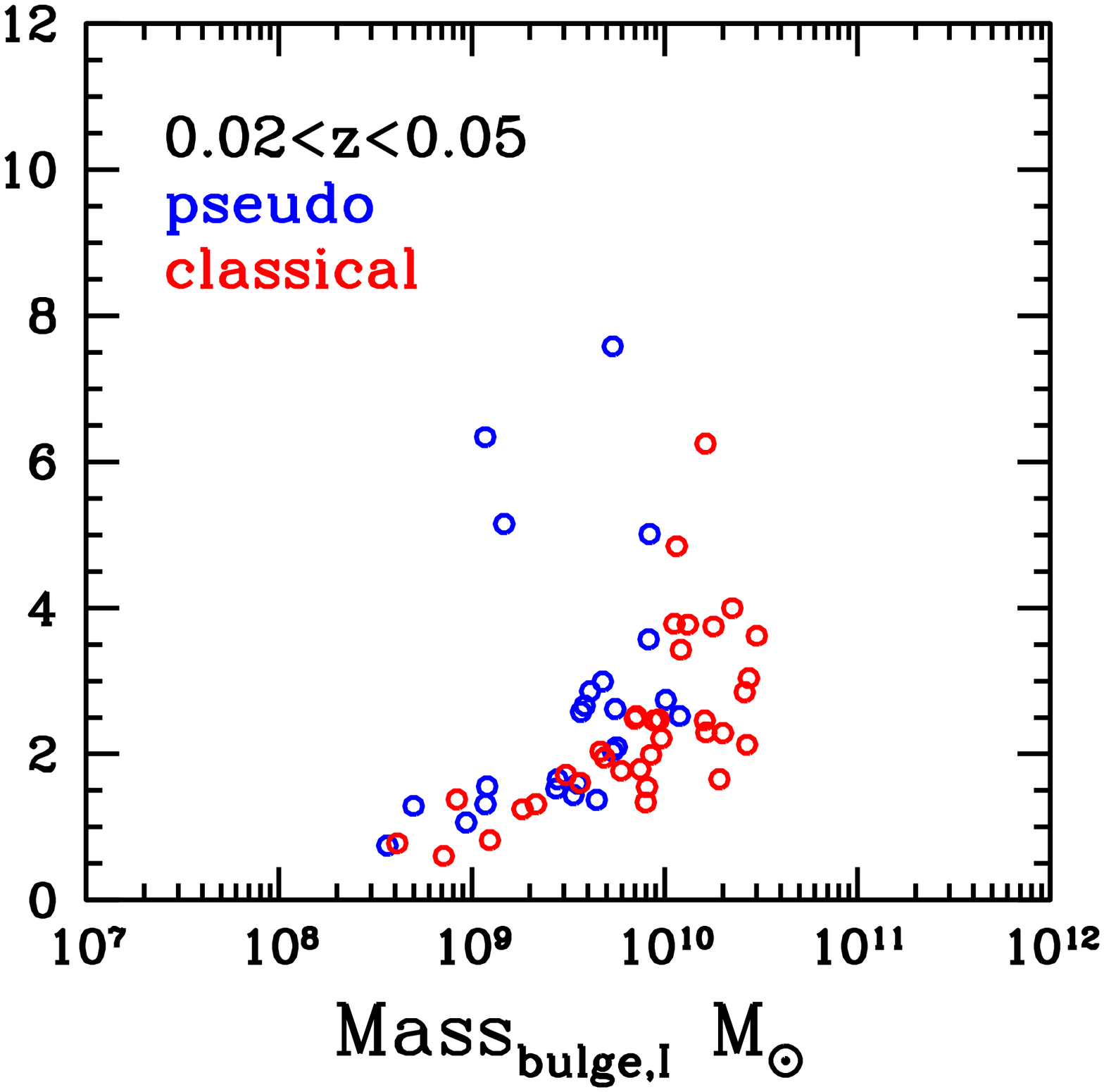}}
\caption{ The effective radius of the bulge is plotted against its absolute magnitude in the three redshift ranges in {\bf First row:} rest-frame {\it B}-band and in {\bf Second row:} rest-frame {\it I}-band. {\bf Third row:} The effective radius of the bulge is plotted against its stellar mass in the three redshift ranges in rest-frame {\it I}-band. In all the plots, pseudo bulges are marked with blue colour and classical bulges are marked with red colour.}
\label{fig:reb-vs-mb}
\end{figure*}

\section{Size - Magnitude relation}
\label{sec:size}
In Fig.~\ref{fig:reb-vs-mb}, we present the bulge absolute magnitude ($M_{bulge}$) versus size ($R_e$) relation for classical and pseudo bulges in the three redshift bins, in rest-frame {\it B} and {\it I}-band. It is interesting to note that both pseudo and classical bulges occupy distinct non-overlapping regions in the parameter space spanned by $R_e - M_{bulge}$ at all redshifts, in rest-frame {\it B}-band. In {\it I}-band also the distinctness is maintained, although not as clean as that observed for the optical. At any given redshift, for a given size, we find pseudo bulges to be fainter than classical bulges and for a given absolute magnitude, pseudo bulges are larger in size than classical bulges - this is in compliance with what has been found for the local bulges from SDSS survey \citep{Gadotti2009}. In other words, there appears to be some sort of invariance (albeit in qualitative sense) in the scaling relation all the way from $z\sim1$ to now - implying a different formation mechanism for classical and pseudo bulges. This is also in concordance with the results of the KS and AD-test, which quantitatively confirms the distinctness of the two bulge samples at all redshift ranges, Table.~\ref{ks-test}, consolidating the view that pseudo and classical bulges formed in a different way, although we see a similarity (in the sense of clumpiness) in the host galaxy morphology.

In the last panel of Fig.~\ref{fig:reb-vs-mb}, we plot bulge mass ($Mass_{bulge,I}$) versus effective radius ($R_{e,I}$), to study the bulge size-mass relation at three redshift ranges. Note that bulge mass has been obtained by multiplying the total-stellar-mass of the galaxy with $B/T$ of rest-frame {\it I}-band as it is considered a better tracer of bulge light. At high redshifts (0.77-1.0), pseudo bulges seem to support larger sizes than classical bulges for a given mass. At intermediate (0.4-0.77) and local (0.02-0.05) redshift range, classical bulges appear be more massive than pseudos for a given size. In our opinion, the implications of these findings need to be understood in greater detail and perhaps should be carried out on a bigger sample.


\begin{figure*}
\mbox{\includegraphics[width=55mm]{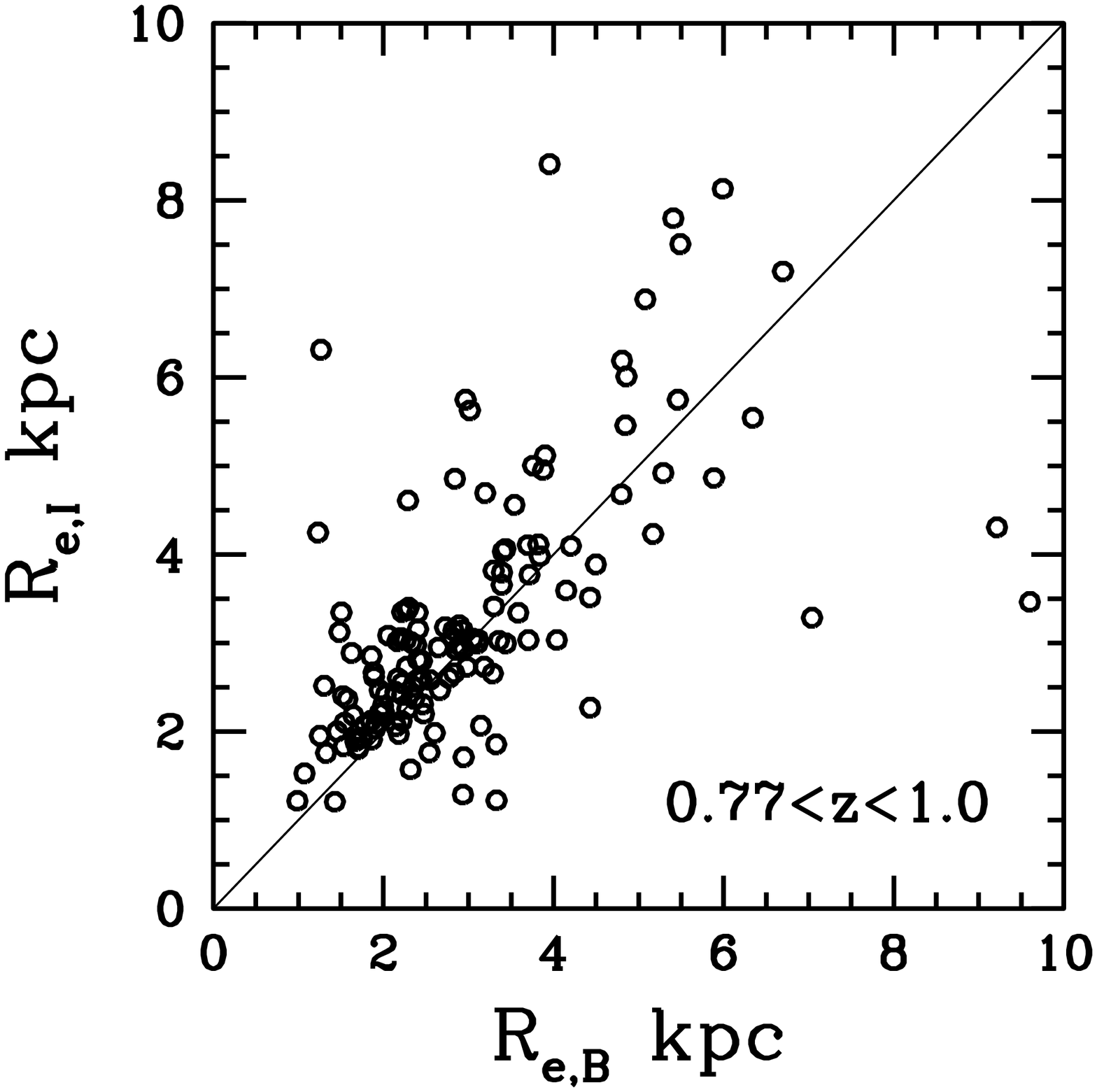}
      \includegraphics[width=55mm]{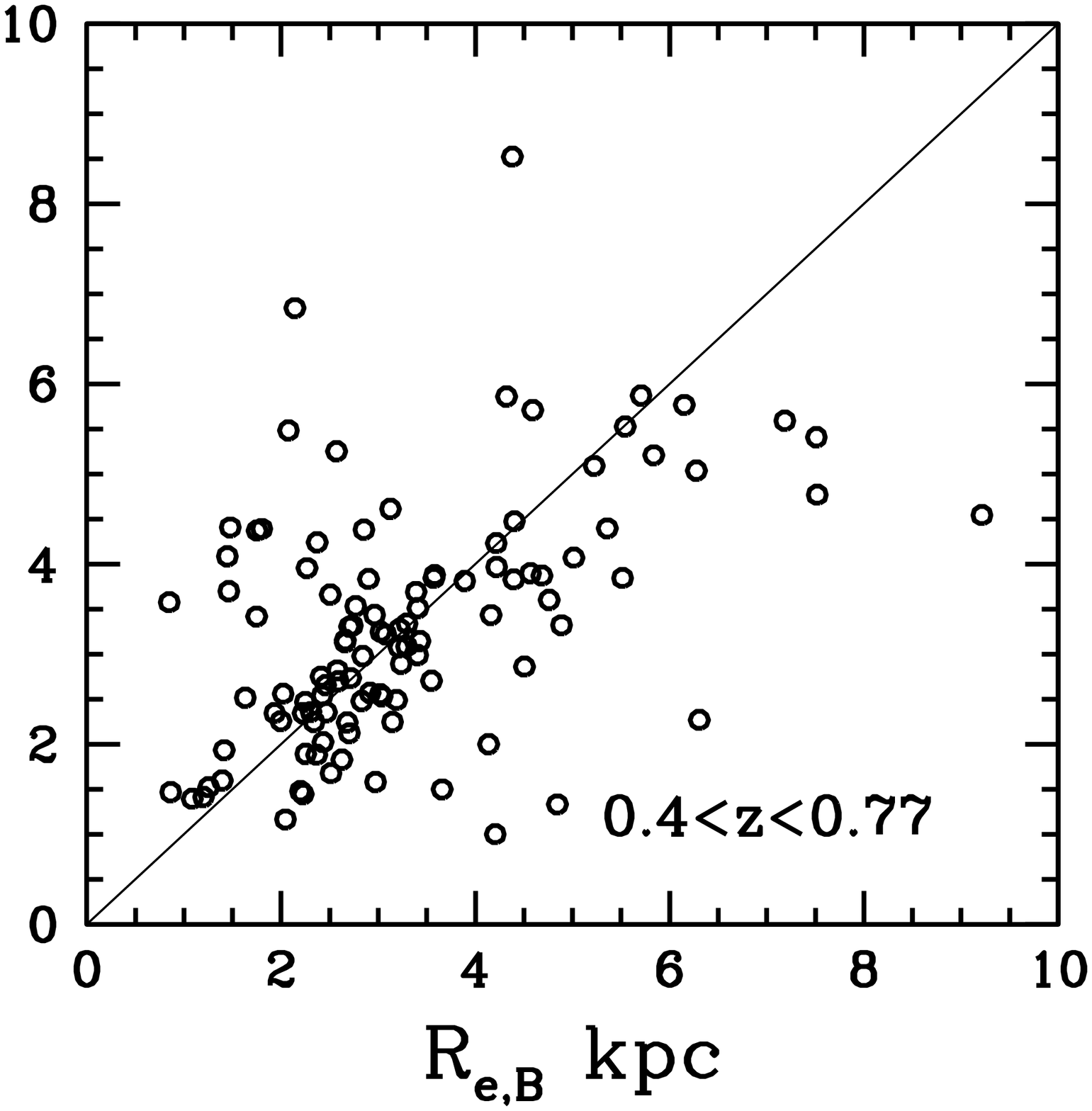}
      \includegraphics[width=55mm]{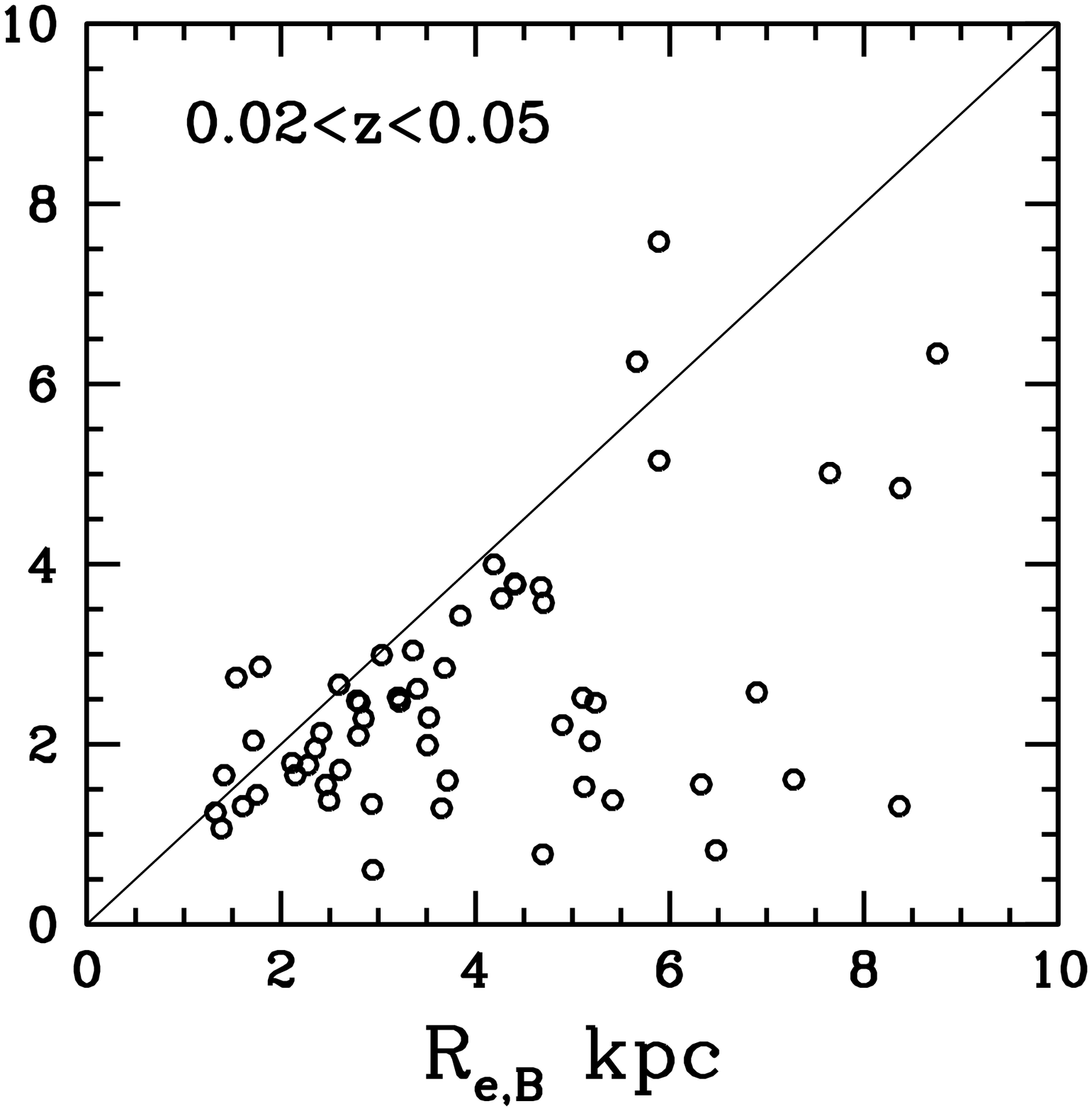}}
\mbox{\includegraphics[width=55mm]{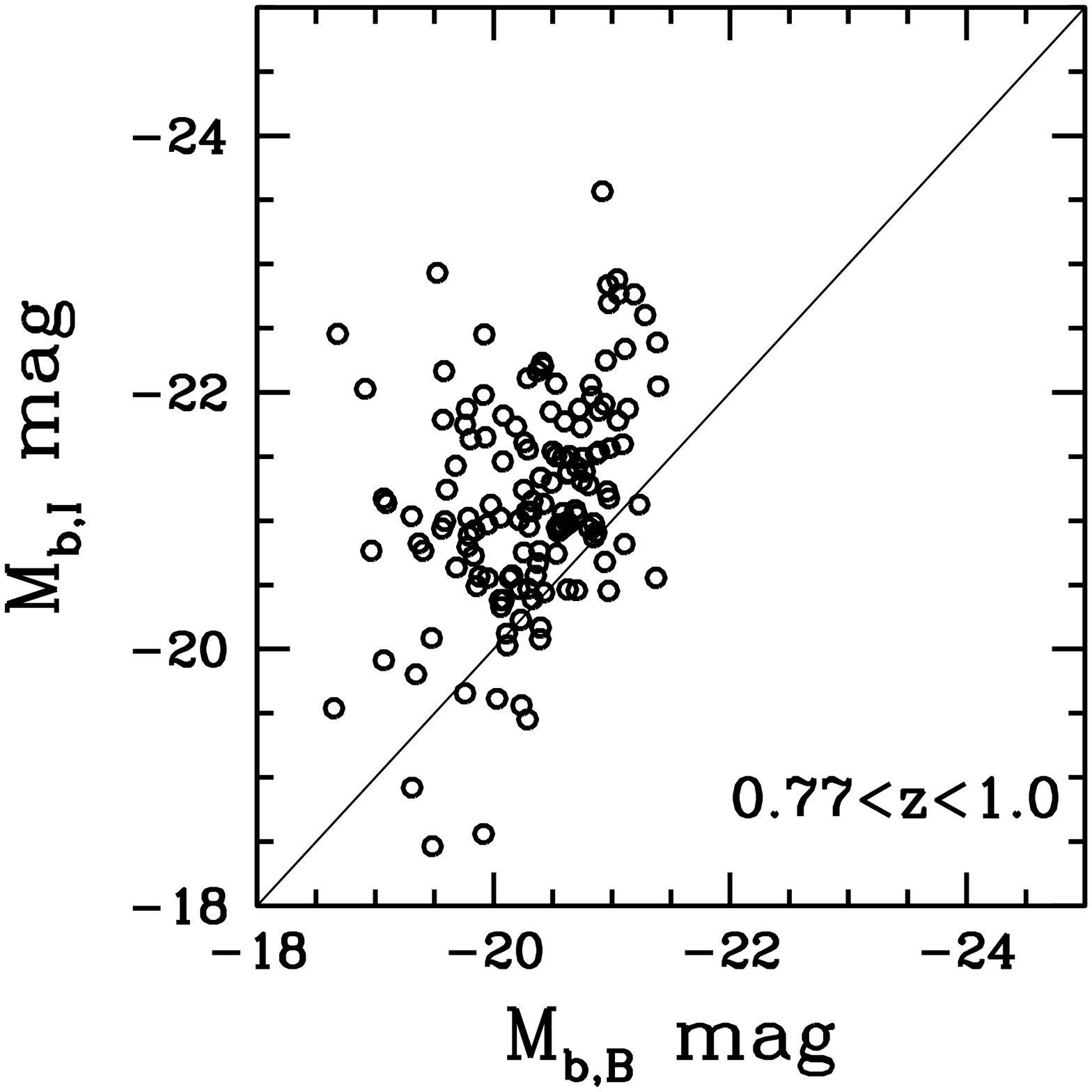}
      \includegraphics[width=55mm]{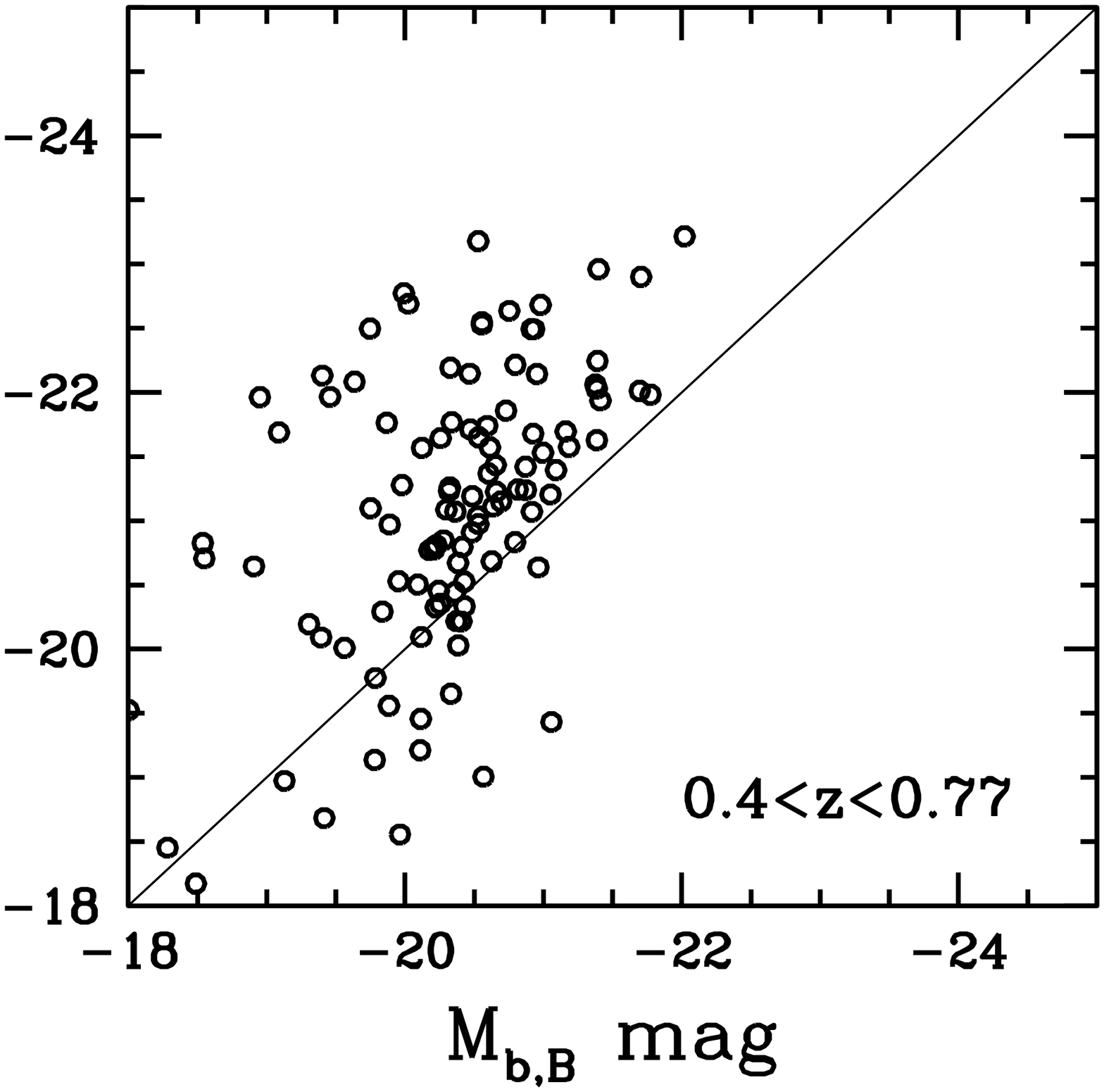}
      \includegraphics[width=55mm]{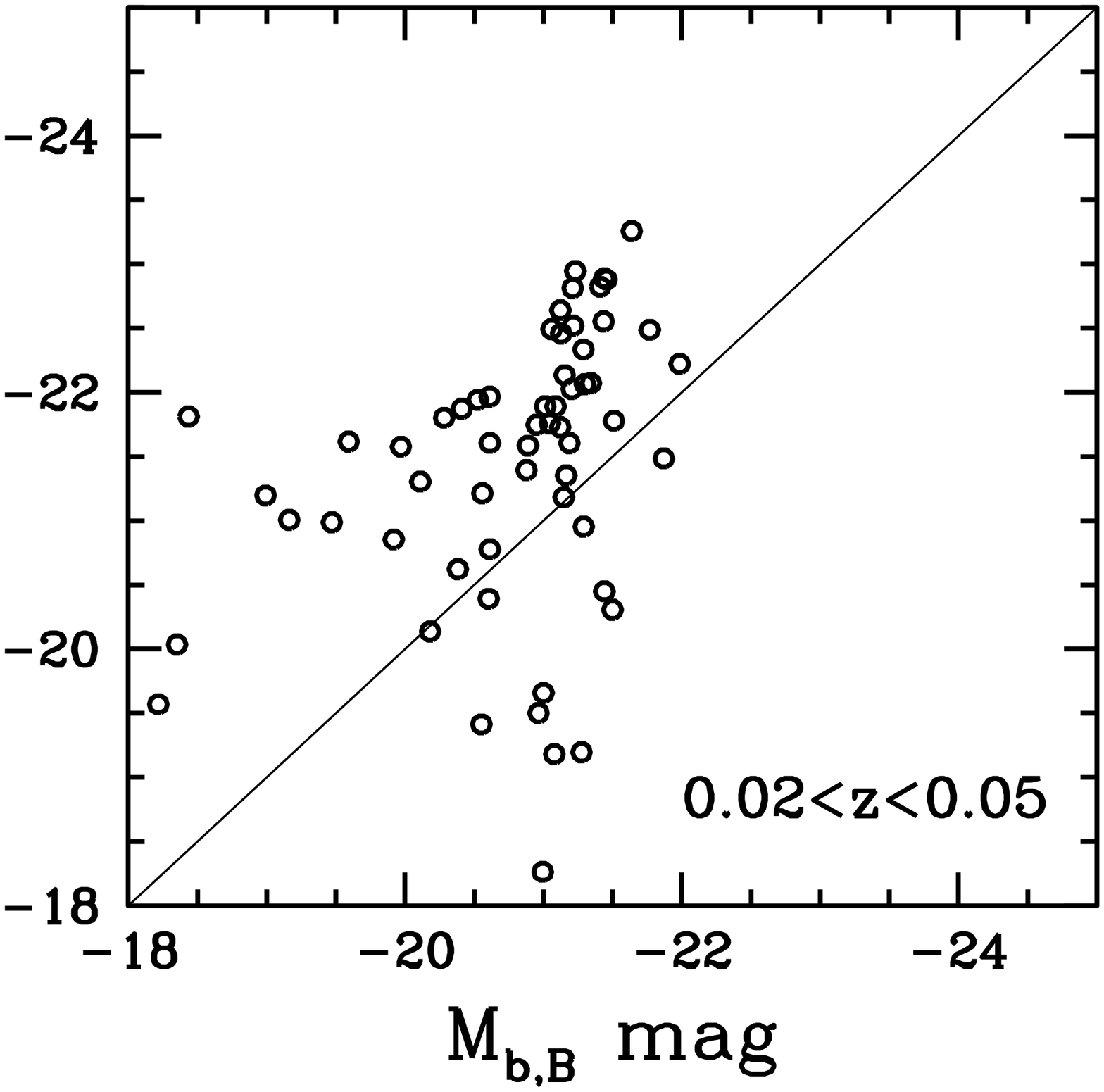}}
\mbox{\includegraphics[width=55mm]{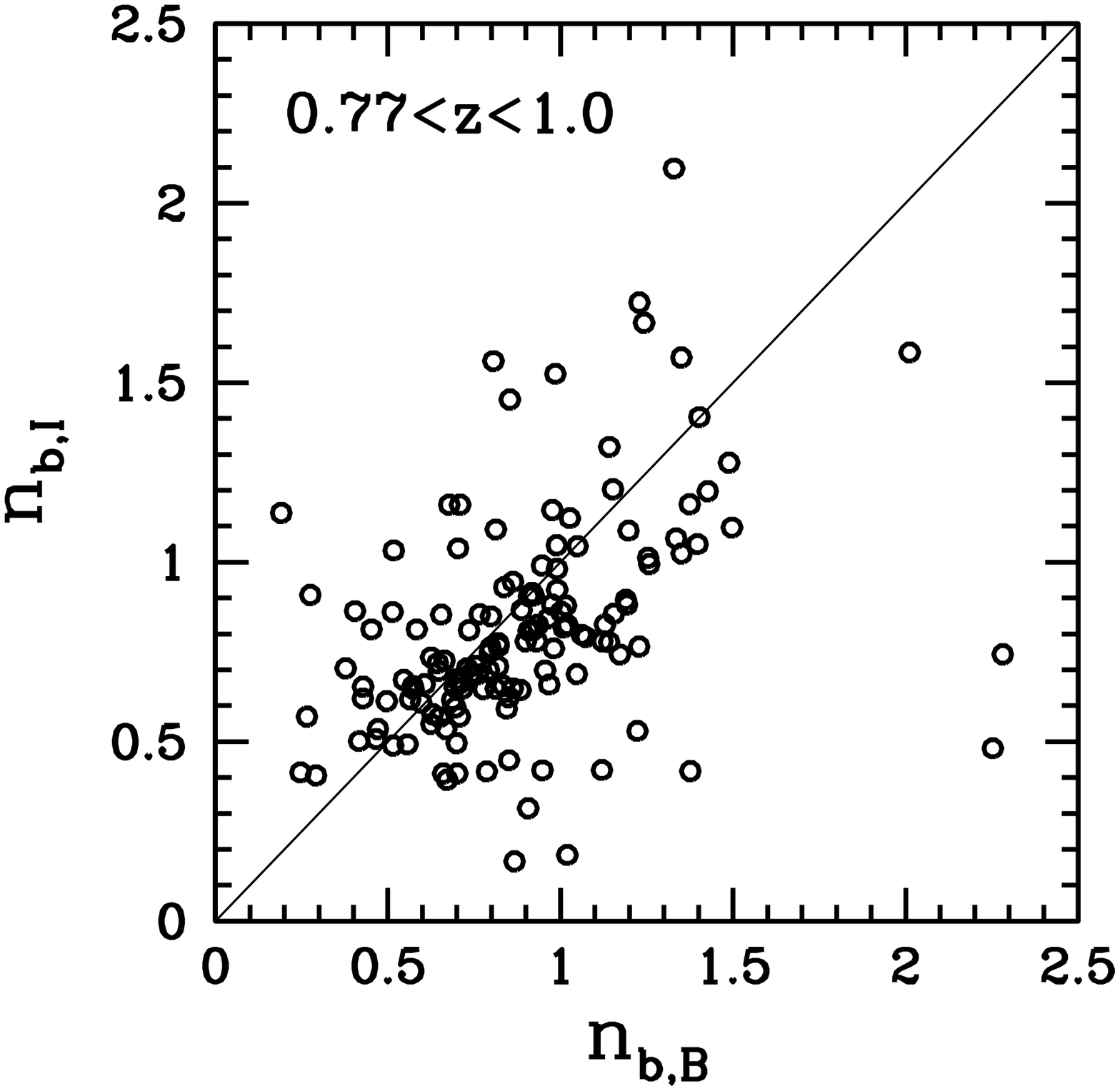}
      \includegraphics[width=55mm]{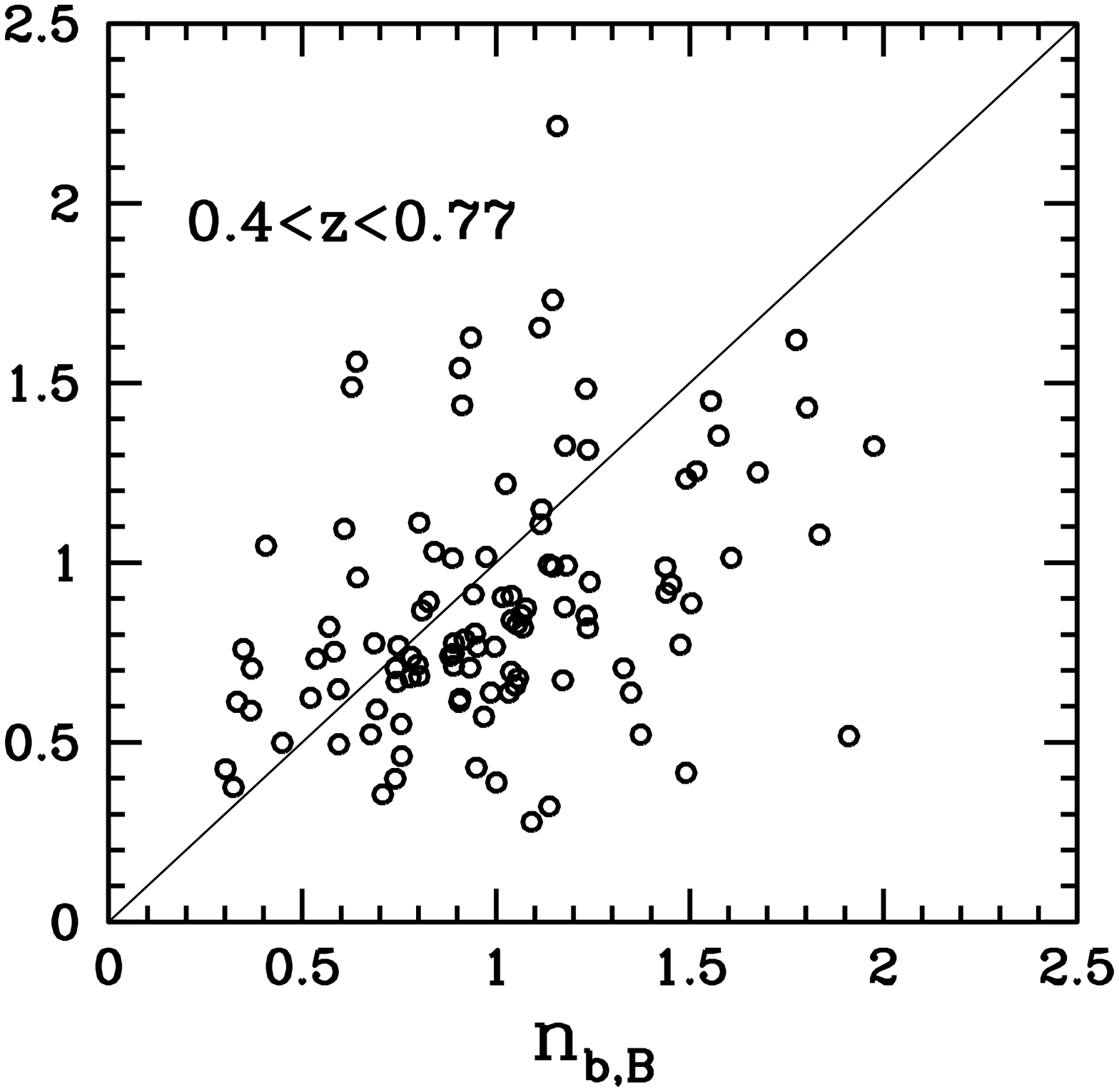}
      \includegraphics[width=55mm]{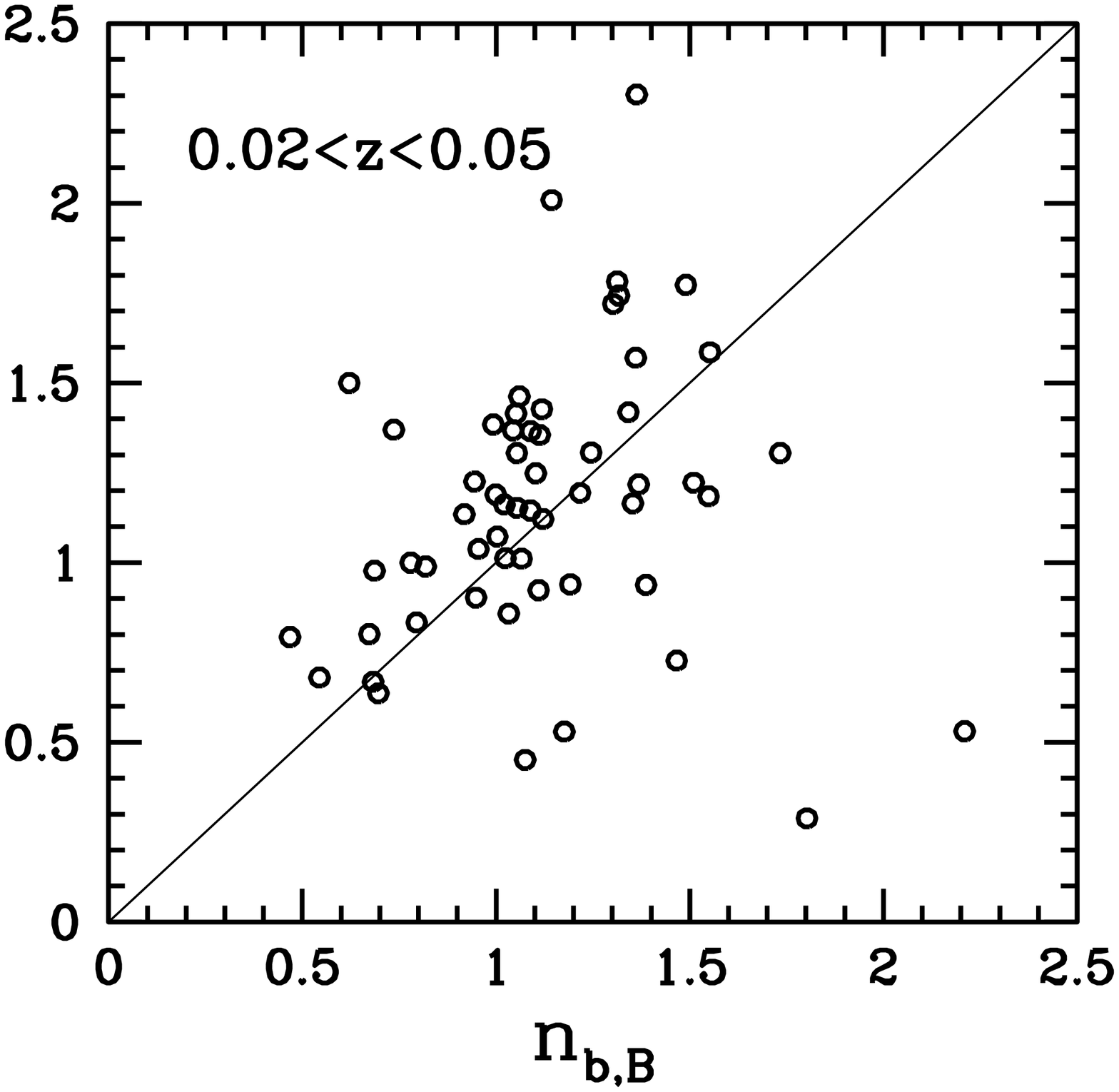}}
\mbox{\includegraphics[width=55mm]{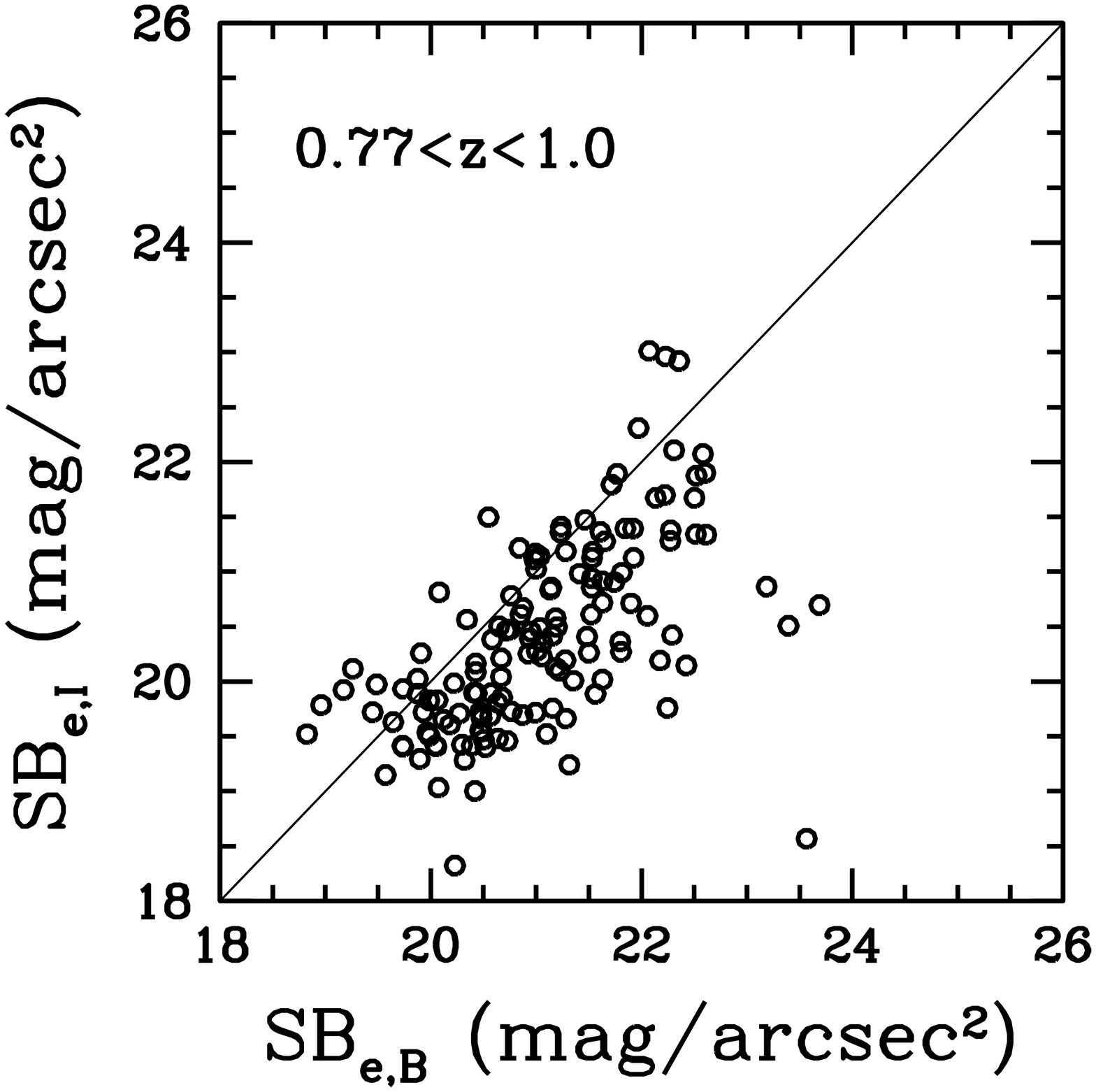}
      \includegraphics[width=55mm]{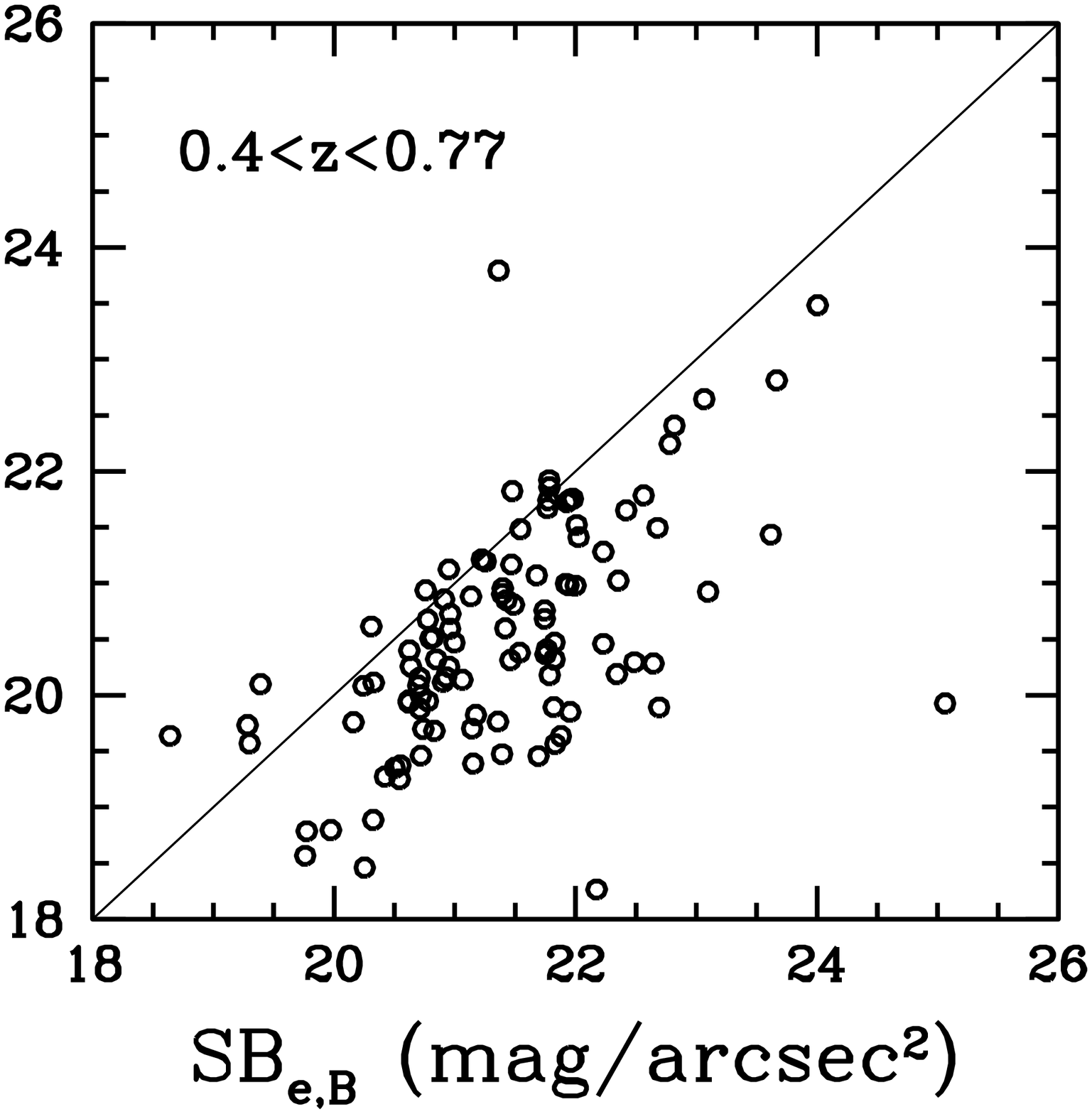}
      \includegraphics[width=55mm]{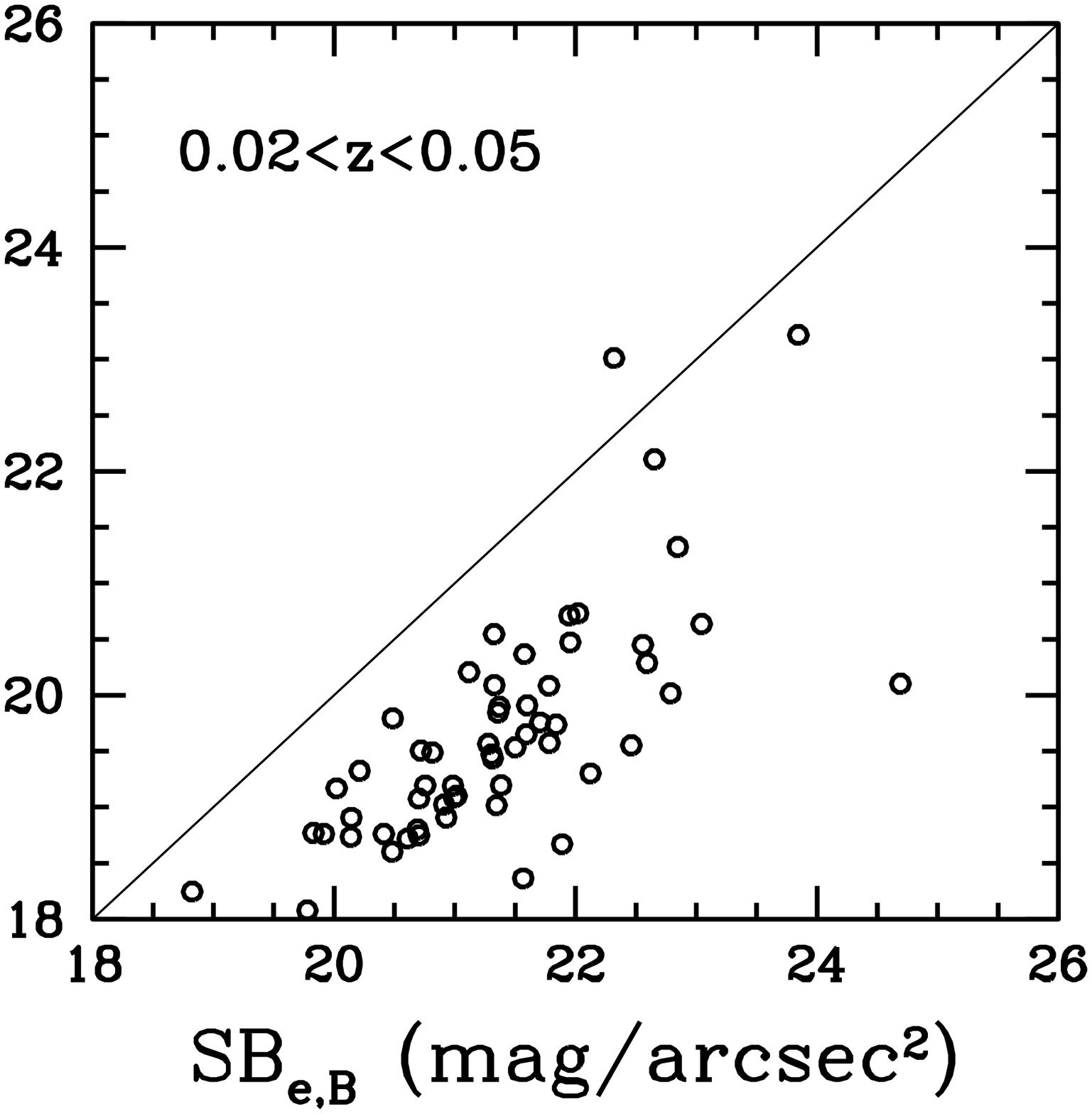}}
\caption{The measurements of the bulge, i.e, its effective radius, absolute magnitude, S\'ersic-index and effective surface brightness, in rest-frame {\it B}-band are plotted against their corresponding values in rest-frame {\it I}-band. The length of the two-axes is kept equal and a solid line referring to the middle of the two axes is marked in each plot, to facilitate comparison.}
\label{fig:comparison-plots}
\end{figure*}

\section{Comparison}
\label{sec:comparison}
In Fig.~\ref{fig:comparison-plots}, we compare the measured bulge parameters in rest-frame {\it B} with those in rest-frame {\it I}-band. The distribution of effective radius of the bulge in the two bands correlates at the higher redshift-ranges (0.77-1.0, 0.4-0.77), however, it shifts towards the optical as we reach the present epoch, affirming our earlier findings that the infrared size shrinks while the optical size expands with time. The bulges appear to be brighter in infrared for all the three redshift ranges, which is expected as the bulges are supposedly dominated with older stars. The bulge S\'ersic indices computed  in the two bands, show rather weak correlation at all redshift ranges. Becoming more luminous and shrinking to smaller sizes in the infrared, makes the bulges more compact at local redshifts, which reflects in the measurement of their S\'ersic-indices as well as effective surface brightness. The distribution of the latter is shifted towards the optical (thus, brighter in the infrared) in the highest redshift range (0.77-1.0) and shifts more and more towards that side at progressively lower redshifts. 


\section{Summary and discussion}
\label{sec:discuss}
We have studied the structure and morphology of bright ($M_B<-20$) disc galaxies since $z\sim1$ in rest-frame {\it B} and {\it I}-band using images from {\it HST} in CDFS and SDSS. The two-component structural decomposition of these galaxies into a photometric bulge and disc allowed us to track specifically the growth of bulges since then. By employing Kormendy relation (as suggested by \citet{Gadotti2009}), we further classified our photometric bulges into classical and pseudo bulges. Structural properties of these bulges have been subsequently examined as a function of redshift and a comparative study of the bulge properties between optical and infrared is presented. The main results from our study are summarized below.

1. Bulges in Optical: Both pseudo and classical bulges have steadily grown in terms of size, luminosity, S\'ersic index and bulge-total ratio, from $z\sim1$ to $z\sim0$. Classical bulges are found to be twice as luminous and $\sim$2.5 times brighter than their pseudo counterparts at all redshifts. Both bulge types double their luminosity with time and become increasingly dominant component of the galaxy. For galaxies with pseudo bulges, B/T rises from 10 to 26\%, and for those with classicals, it rises from 21 to 52\%, thus, emissivity of the galaxies shifts significantly to bulges.

2. Bulges in Infrared: For pseudos, there is no increase in luminosity and there is significant contraction ($\sim$1 Kpc) in size, suggesting that pseudo bulges grow/survive from gas readjustment and in-situ star formation, making older stellar population an increasingly insignificant contributor, from $z\sim1$ to $z\sim0$. Classical bulges, on the contrary, witness a considerable increase ($\sim$2.5 times) in their luminosity, thus, becoming equally dominated by young and older stellar population.

3. Bulge stellar mass: Mass of an average pseudo bulge is found to be about half of that of an average classical bulge, at all redshift ranges. Both bulge types witness a vast increase in their stellar mass. For pseudo bulges, it doubles and for classicals, it grows by $\sim$2.6 times reaching $\sim 0.88 (e+10) M_{\odot}$. The host disc mass has grown concurrently, thus, external processes have certainly played a significant role in bulge-disc evolution, since $z\sim1$. 

4. Host-discs in Optical: The luminosity distribution of pseudo and classical bulge host discs overlaps only at $z\sim1$. As we move to lower redshifts, classical bulge host-discs faint in comparison to pseudo bulge host-discs. This suggests that the loss of star forming material has been more severe for classical bulge host-discs, probably, in the form of clump migration. Based on visual inspection, similar fraction of pseudo and classical bulge host discs are observed to support clumps, at high redshifts.

5. Host-discs in Infrared: The luminosity distribution of pseudo and classical bulge host discs overlaps at all redshift ranges, from $z\sim1$ to $z\sim0$. Both become significantly luminous ($\sim1$ mag) with time, suggesting the ageing of existing stellar population.

6. Different evolutionary history: Pseudo and classical bulges differ from each other at 95\% confidence level in both KS and AD test, in optical, at all redshift ranges probed in this work. In infrared, although for the highest redshift range (0.77-1.0) they are similar, later, they also start to differ from each other. Even in the size-magnitude plane, the two bulge types occupy distinct non-overlapping regions, at all redshifts, suggesting different growth mechanisms.

Recent studies have revealed that bulges (or proto-bulges) are already in place by about redshift $z\sim1$ \citep{Huertas-Companyetal2015,MargalefBentaboletal2016,Tadakietal2017}. One of the primary goals in the area of galaxy formation and evolution has been to understand the growth of these bulges. Several recent studies have contributed in this regard, and the overall emerging picture of bulge growth in disc galaxies are in close agreement with each other. For example, \cite{Tascaetal2014} carried out the bulge-disc decomposition of galaxies taken from zCOSMOS survey in rest-frame {\it B}-band and found that bulge light has increased by $\sim30\%$ since $z\sim0.8$. Similarly, \cite{Langetal2014, Bruceetal2014} used CANDELS survey to study the bulge growth in the infrared ({\it H}-band) starting from redshift $z\sim2.5$, and reported a significant increase in the bulge-to-total ratio since then. Since they fixed the bulge S\'ersic index to $n=4$ in decomposition process, insight about which bulges are growing to what extent, was missing. In this regard, our analysis brings forth a picture of how classical and pseudo bulges, based on our decomposition and classification carried out in both rest-frame {\it B} and {\it I}-band, have grown and to what extent since $z\sim1$. This has produced some cognizance regarding the physical processes which seem to have contributed in this phase of evolution. 

One of the generic features of the high redshift galaxies is that they are clumpy \citep{Guoetal2015,Shibuyaetal2016}, especially in the optical band (see our Figure~\ref{fig:clump-cl} and Figure~\ref{fig:clump-pb}). Whether these clumps are accreted or formed in-situ as a result of disc gravitational instability \citep{Ceverinoetal2015, Shibuyaetal2016}, they are likely to migrate to the centre as a result of dynamical friction and join the ongoing bulge building as suggested by \cite{Elmegreenetal2008, InoueSaitoh2012}.

We report that both classical and pseudo bulges have grown and nearly doubled their stellar mass since $z\sim1$. Thus, clump migration and star-formation alone are not responsible for the growth; materials have to be accreted from outside. So along with a number of other studies based on local as well high-redshift galaxies \citep{Elmegreenetal2008,Weinzirletal2009,Hopkinsetal2010,Buitragoetal2013,Naabetal2014,Sachdevaetal2015}, we also advocate minor mergers as one of the active physical processes responsible for bulge growth. 

In addition, it might be possible that secular evolution also played an equally important role in the growth of classical and pseudo bulges since $z \sim 1$ \citep[e.g.,][]{ElicheMoraletal2006,LopezSanjuanetal2009,Hopkinsetal2010,Bournaudetal2014,Kaviraj2014}. At this point, it is worth getting back to the growth history of classical bulges in our sample. In rest-frame B-band, the classical bulges have grown by about $0.6$~mag while their host discs get fainter by about $1$~mag since $z\sim1$. In a similar study, \cite{Tascaetal2014} also reported that the classical bulge host discs became fainter while the bulge grew since $z\sim0.8$ in the rest-frame {\it B}-band. Could it be that the classical bulge grew at the cost of disc fading$?$ If this is indeed true, it might be an indication of the secular process in action - where disc materials help developing central concentration via redistribution of angular momentum which is known to occur due to non-axisymmetries present in the disc \citep{KormendyandKennicutt2004,SahaJog2014}. Our analysis suggests that a joint role played by inward-migration of star-forming clumps, minor mergers and secular processes is behind the overall growth of bulges since $z\sim1$. High resolution and deep imaging observations (from JWST) would be required to have further discernment along this line.

We are grateful to the anonymous refree for reading the draft with immense care and providing detailed comments for its refinement. 
 
\bibliographystyle{aasjournal}


\end{document}